\documentclass[aps,prb,reprint,superscriptaddress]{revtex4-1}

\usepackage{amsmath,amsfonts,amssymb,mathtools,bbold,braket,graphicx,hyperref}
\usepackage[utf8]{inputenc}
\usepackage{layouts}

\newcommand{\vect}[1]{\boldsymbol{#1}}

\newcommand{\id}{\mathbb{1}}

\begin{document}
  \title{Induced spin-orbit coupling in twisted graphene-TMDC heterobilayers: twistronics meets spintronics}
  \author{Alessandro David}
  \email{alessandro.david@uni-konstanz.de}
  \affiliation{Department of Physics, University of Konstanz, D-78464 Konstanz, Germany}
  \author{P\'eter Rakyta}
  \author{Andor Korm\'anyos}
  \email{andor.kormanyos@complex.elte.hu}
  \affiliation{Department of Physics of Complex Systems, E\"otv\"os Lor\'and University, Budapest, Hungary}
  \author{Guido Burkard}
  \email{guido.burkard@uni-konstanz.de}
  \affiliation{Department of Physics, University of Konstanz, D-78464 Konstanz, Germany}
  
  \begin{abstract}
    We propose an interband tunneling  picture to explain and  
    predict the interlayer twist angle dependence of the induced spin-orbit coupling in heterostructures of graphene and monolayer transition metal 
    dichalcogenides (TMDCs). We obtain a compact analytic formula for the 
    induced valley Zeeman and Rashba spin-orbit coupling in terms of the TMDC band structure parameters and interlayer tunneling matrix elements. 
    We parametrize the tunneling matrix elements with few parameters, which in our formalism are independent of the twist angle between the layers.
   We estimate the value of the tunneling parameters from existing DFT calculations at zero twist angle and we use them to predict the induced spin-orbit coupling at non-zero angles. Provided that the energy of the Dirac point of graphene is close to the TMDC conduction band, we expect a sharp increase of the induced spin-orbit coupling  around a twist angle of 18 degrees.
  \end{abstract}
  
  \maketitle
 
 \section{Introduction}
  
  Since its isolation, graphene\cite{novoselov_electric_2004, novoselov_two-dimensional_2005} has shown a plethora of interesting 
  phenomena\cite{castro_neto_electronic_2009}. Among others, long spin-relaxation times \cite{drogeler_spin_2016, singh_nanosecond_2016} and spin-diffusion lengths \cite{ingla-aynes_$24ensuremath-ensuremathmumathrmm$_2015} have been observed in graphene, making it a strong candidate 
  for spintronics applications \cite{han_graphene_2014}. However, the weak intrinsic spin-orbit coupling (SOC) of graphene hinders the control and tunability of possible spintronics devices. Moreover, the quantum spin Hall effect was initially predicted for graphene \cite{kane_quantum_2005}, but the low intrinsic SOC \cite{gmitra_band-structure_2009} has prevented the experimental verification of this prediction.

  A recent impetus to graphene spintronics has been given by van der Waals engineering \cite{geim_van_2013}, i.e., the fabrication of 
  heterostructures of different two-dimensional materials weakly bound by van der Waals forces.
  These heterostructures can posses functionalities that the individual constituent layers may not have. In order to increase the SOC in graphene, one of the most actively pursued directions is to interface it with materials that have strong intrinsic SOC, such as transition metal dichalcogenides (TMDCs) \cite{avsar_spinorbit_2014, wang_strong_2015, wang_origin_2016, yang_tunable_2016, yan_two-dimensional_2016, yang_strong_2017, ghiasi_large_2017, dankert_electrical_2017, volkl_magnetotransport_2017, zihlmann_large_2018, wakamura_strong_2018, leutenantsmeyer_observation_2018, omar_spin_2018, benitez_strongly_2018, safeer_room-temperature_2019}. TMDCs are expected to be good candidates for graphene spintronics for two reasons: i) it was shown that TMDC substrates do not degrade the mobility of
  graphene \cite{omar_spin_2018, kretinin_electronic_2014}, and ii) they host a strong intrinsic SOC of the order of 100 meV (10 meV) in their valence
  (conduction) band \cite{kormanyos_k_2015} and hence can potentially be suitable materials for proximity induced SOC. Indeed, the measurement of weak
  antilocalization (WAL) \cite{wang_strong_2015, wang_origin_2016, yang_tunable_2016, yang_strong_2017, volkl_magnetotransport_2017, zihlmann_large_2018, wakamura_strong_2018} and the beating of Shubnikov-de Haas oscillations (SdH) \cite{wang_origin_2016} proved that SOC is strongly enhanced in graphene/TMDC heterostructures. Details regarding the type and magnitude of the proximity induced SOC are less clear. Based on WAL measurement, Refs.~\citenum{yang_tunable_2016, yang_strong_2017} argued that the induced SOC in graphene is of Rashba type which is due to the inversion symmetry breaking effect of the substrate. The measurements of a large anisotropy of the in-plane and out-of-plane spin-relaxation times \cite{ghiasi_large_2017, benitez_strongly_2018} can be interpreted \cite{cummings_giant_2017} as an indication that a valley-Zeeman type SOC is also induced and its magnitude is comparable to the Rashba type SOC. This is consistent with the data extracted from SdH oscillations \cite{wang_origin_2016} and a similar conclusion was also reached in a more recent WAL measurement \cite{zihlmann_large_2018}. These measurements usually employed either bulk or few-layer TMDC substrate. On the other hand, Ref.~\citenum{wakamura_strong_2018} found that a monolayer TMDC substrate may induce strong Kane-Mele type SOC.
  
  On the theoretical side, density functional theory (DFT) calculations for aligned graphene/TMDC structures\cite{wang_strong_2015, kaloni_quantum_2014, gmitra_graphene_2015, gmitra_trivial_2016, singh_proximity-induced_2018}  showed that SOC can be induced in graphene. Direct comparison between these theoretical results and the measurements is not straightforward. Firstly, the DFT bandstructure calculations are usually fitted with model Hamiltonians for graphene in order to extract the SOC constants and the corresponding energy scales. However, most measurements yield information on spin-relaxation times. Therefore further information about intervalley scattering times as well as the dominant spin-relaxation mechanisms is needed in order to interpret the observations in terms of SOC energy scales. Secondly, while most measurements used few-layer TMDCs as substrates, the DFT calculations assumed monolayer TMDCs. It is not entirely clear if the differences in the band structure of monolayer and bulk TMDCs can influence the induced SOC. Thirdly, in contrast to the theoretical calculations,  in the experiments the layers were not intentionally aligned and in general there is most likely to be a twist angle between them, as observed in Ref.~\citenum{pierucci_band_2016}.
  (We note that Refs.~\citenum{wang_electronic_2015, felice_angle_2017} performed calculations for a few twist angles where the graphene and TMDC layers form approximately commensurate structures, but the SOC was not taken into account.) The tight-binding (TB) models of 
  Refs.~\citenum{alsharari_mass_2016, alsharari_topological_2018} considered aligned structures  or small twist angles.
  Only very recently was the TB methodology extended to the calculation of induced SOC for arbitrary twist angle between graphene and the TMDC substrate \cite{li_twist-angle_2019}. 
  
  Here we use an approach that  describes the induced SOC in terms of virtual band-to-band tunneling between graphene and the monolayer TMDC substrate. This  perturbative approach is motivated by previous DFT calculations \cite{wang_strong_2015, kaloni_quantum_2014, gmitra_graphene_2015, gmitra_trivial_2016, singh_proximity-induced_2018, wang_electronic_2015, felice_angle_2017}
  which show that the linear dispersion of graphene close to the Dirac point is  preserved  because 
  the interaction between the layers is  rather weak. In real space, we take into account  tunneling processes between graphene and 
  the closest layer of chalcogen atoms in the TMDC. This approximation allows to obtain a simple 
  and effective parametrization of the interlayer tunneling using  just two real parameters. We show how these parameters 
  can  be applied to describe tunneling for all twist angles. 
  We then calculate the induced valley Zeeman and Rashba type SOC in graphene as a function of interlayer twist angle and demonstrate the close relation between the 
  intrinsic properties of the substrate and the induced SOC in graphene. As a concrete example we consider graphene on monolayer 
  MoS$_2$, but the same approach can be used for other semiconductor monolayer TMDC where the Dirac point of graphene is 
  in the band gap of the substrate. The possibility to tune the strength of the induced SOC in graphene by changing the interlayer twist angle links graphene spintronics with the newly emerging field of twistronics \cite{bistritzer_moire_2011, carr_twistronics:_2017,   cao_unconventional_2018, ribeiro-palau_twistable_2018}.
  
  This paper is organized as follows. In Sec.~\ref{sec:heterostructure} we present the details of the heterostructure. In Sec.~\ref{sec:tunneling} we describe the tunneling between the two layers and we introduce the idea of tunneling to a band. We construct a Hamiltonian for the Dirac points of graphene in Sec.~\ref{sec:effectiveHam} and we indicate how valley Zeeman and Rashba type SOC are induced in graphene by the TMDC substrate in Sec.~\ref{sec:valley-zeeman} and Sec.~\ref{sec:rashba}, respectively. We present and discuss our result in Sec.~\ref{sec:discussion} and we draw our conclusions in Sec.~\ref{sec:conclusion}.

\section{Twisted Heterostructure}\label{sec:heterostructure}

  \begin{figure}
    \centering
    \includegraphics[width=\columnwidth]{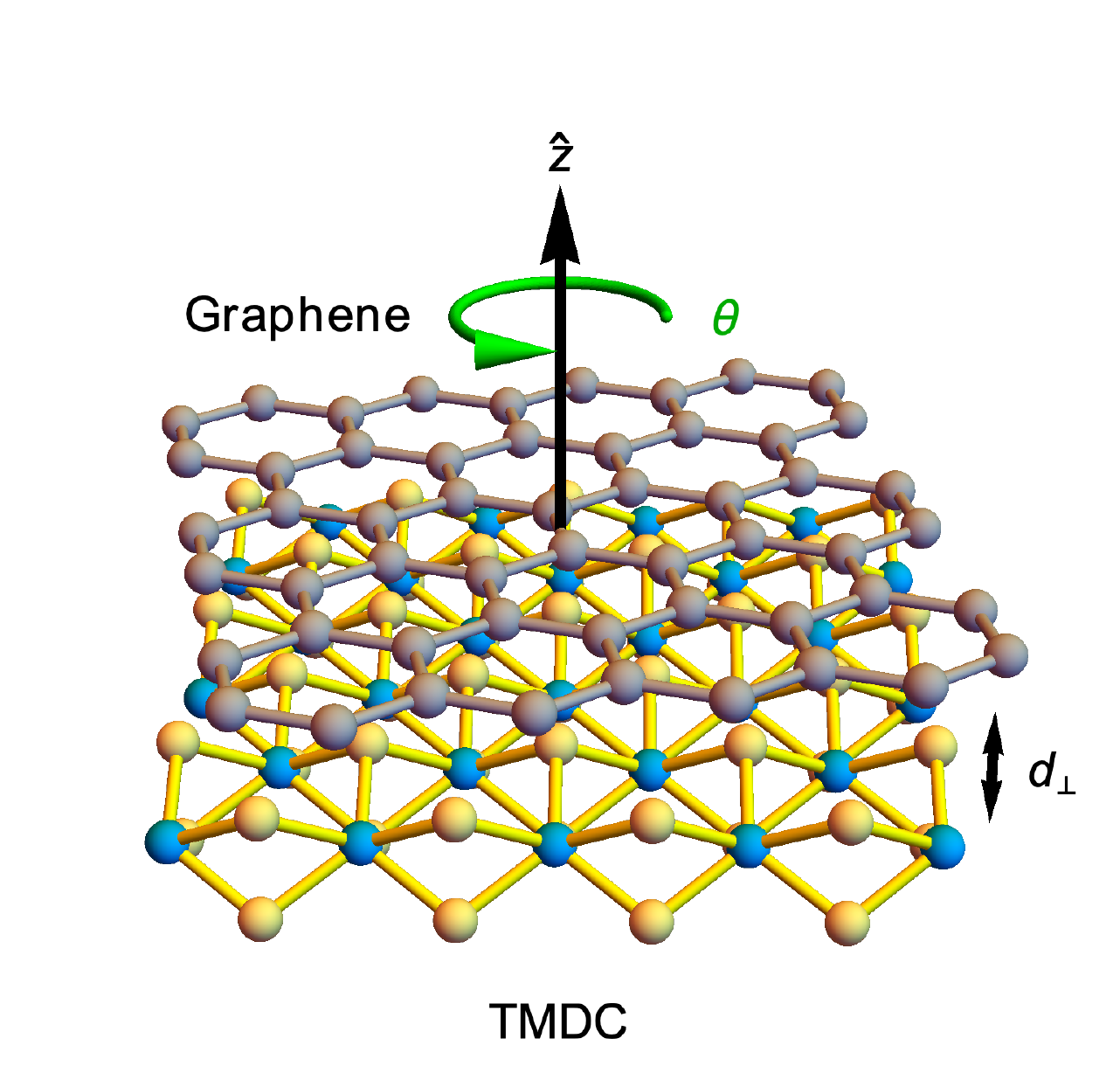}
    \caption{3D view of graphene on top of monolayer TMDC. Here $\theta$ is the twist angle between graphene and the TMDC layer, while $d_\perp$ is the perpendicular distance between graphene and the upper (closest) chalcogen layer of the TMDC.}
    \label{fig:heterostructure}
  \end{figure}
  
  Graphene \cite{novoselov_electric_2004, novoselov_two-dimensional_2005, castro_neto_electronic_2009} and monolayer TMDCs \cite{mak_atomically_2010, splendiani_emerging_2010, kormanyos_k_2015} share the same 2D hexagonal structure given by two triangular sublattices, $A$ and $B$. 
  For graphene the lattice constant is $a_G = 2.46$ Å and the two sublattices are occupied by carbon atoms. Conduction and valence band of graphene show conic dispersion relations at the two inequivalent corners of the Brillouin zone, $\vect{K}_\tau = \tau \vect{K} = 4\pi/3a_G (\tau, 0)$, where $\tau = \pm 1$, also known as Dirac points. A two-band nearest-neighbor tight-binding (TB) model that takes into account only one $p_z$ orbital per carbon atom leads to the Hamiltonian\cite{castro_neto_electronic_2009}
  \begin{equation}\label{eqn:graphene}
    h_{\tau\vect{K}}^\text{gr} (\vect{k}) =
      \hbar v_F \left ( \tau k_x \sigma_x + k_y \sigma_y \right ),
  \end{equation}
  where $|\vect{k}| \ll |\vect{K}|$, $\sigma_{x}$, $\sigma_{y}$ are Pauli matrices for the sublattice pseudospin and $v_F$ is the Fermi velocity of the electrons.  
  Monolayer TMDCs have larger lattice constants than graphene ($a_T=3.1 \div 3.3$ \AA), therefore smaller Brillouin zones. 
  The metal atoms occupy the $A$ sublattice sites, while the chalcogen atoms are found on the $B$ sublattice sites but vertically shifted by $\pm d_\text{X--X}/2$, where $d_\text{X--X}$ is the chalcogen-chalcogen distance\cite{kormanyos_k_2015}. 
  We consider a heterobilayer van der Waals structure formed by graphene deposited on top of monolayer TMDC. The graphene layer is 
  separated by $d_\perp$ from the topmost TMDC chalcogen layer (see Fig.~\ref{fig:heterostructure}). Because of the lattice constant difference between graphene and the TMDC 
  they do not form a commensurate structure.  In general, 
  the graphene lattice vectors can be rotated by angle $\theta$ with respect to the TMDC lattice vectors and 
  the $A$ sublattice of graphene may be  shifted horizontally with respect to the $A$ sublattice of the TMDC by vector $\vect{r}_0$. 
  (The vector $\vect{r}_0$ is contained in the first (rotated) unit cell of graphene.) 
  In the rest of the paper, we use the following notations: primed quantities are related to the TMDC and every vector $\vect{r}$ that is rotated by an angle $\theta$ with respect to its original definition is indicated by $\vect{r}^{\theta} = R(\theta)\vect{r}$, where $R$ is the rotation operator around the $z$-axis. The sublattice sites are found at the positions 
  $\vect{R}_{X}^{\theta} = n_1 \vect{a}_1^\theta + n_2 \vect{a}_2^\theta + \vect{\tau}_{X}^\theta + \vect{r}_0$, $\vect{R}_{X'} = n'_1 \vect{a}'_1 + n'_2 \vect{a}'_2 + \vect{\tau}_{X'}$, where $X = A, B$ and  $X' = A', B'$ refer to the graphene and TMDC sublattice, respectively. Here, $\vect{a}_{1,2}$ ($\vect{a}'_{1,2}$) are the lattice vectors of graphene (TMDC) and $\vect{\tau}_X$ ($\vect{\tau}_{X'}$) indicates the position of sublattice $X$ ($X'$) in the unit cell. See Appendix~\ref{sec:definitions} for the explicit definitions used in this work.
  
  \section{Interlayer tunneling}\label{sec:tunneling}
  
  Looking at the \emph{ab initio} calculations of Ref.~\citenum{wang_strong_2015, gmitra_trivial_2016}, the Dirac point of graphene is located 
  inside the TMDC band gap and its linear dispersion is mostly unaffected. However, modifications of the graphene bands very close to the 
  Dirac point indicate spin-orbit splittings and possibly the presence of a spin-independent band gap opening as well. 
  We will  use  perturbation theory to give a microscopic description of the induced spin-splitting  of the graphene bands. 
  
  The total Hamiltonian has three parts, describing the isolated eigenstates of graphene and TMDC and the interlayer tunneling respectively, $H_\text{tot} = H_\text{gr} + H_\text{tmdc} + H_\text{T}$.
  The  theory for interlayer interactions in incommensurate atomic layers\cite{koshino_interlayer_2015} gives a compact analytic form, 
   in momentum space, for the interlayer tunneling matrix elements 
   $U_{XX'} (\vect{k}, \vect{k}') = \; _\text{gr}\bra{X, \vect{k}^{\theta}} H_\text{T} \ket{X', \vect{k}'}_\text{tmdc}$ 
   between unperturbed graphene and TMDC states.
    Here,  $X$ and $X'$ run over the sublattice indices and, in general, also over all the atomic orbitals located on the same sublattice site. 
    If there is only one atomic orbital per lattice site, the 
    Bloch states read $\ket{X, \vect{k}^{\theta}}_\text{gr} = N^{-1/2} \sum_{\vect{R}_{X}^\theta} e^{i \vect{k}^{\theta} \cdot \vect{R}_{X}^\theta} \ket{\vect{R}_{X}^\theta}$ and $\ket{X', \vect{k}'}_\text{tmdc} = N'^{-1/2} \sum_{\vect{R}_{X'}} e^{i \vect{k}' \cdot \vect{R}_{X'}} \ket{\vect{R}_{X'}}$ and the theory gives\cite{bistritzer_transport_2010, bistritzer_moire_2011,koshino_interlayer_2015}
  \begin{multline}
  \label{eqn:umklapp}
    U_{XX'} (\vect{k}, \vect{k}') =
      \sum_{\vect{G}, \vect{G}'}
        \delta_{\vect{k}^{\theta} + \vect{G}^{\theta}, \vect{k}' + \vect{G}'} \;
        t_{X'} (\vect{k}' + \vect{G}') \\
        \times e^{i \vect{G}^{\theta} \cdot (\vect{\tau}_X^{\theta}+\vect{r}_0)
        -i \vect{G}' \cdot \vect{\tau}_{X'}},
  \end{multline}
  where $\vect{G}$, $\vect{G}'$ are reciprocal lattice vectors of graphene and TMDC, respectively. 
  The term  $\delta_{\vect{k}^{\theta} + \vect{G}^{\theta}, \vect{k}' + \vect{G}'}$  expresses quasi-momentum conservation. 
  In the derivation of Eq.~\eqref{eqn:umklapp} the Slater-Koster two-center approximation \cite{slater_simplified_1954} 
  has been used, whereby $\bra{\vect{R}_{X}^{\theta}} H_\text{T} \ket{\vect{R}_{X'}} = 
  T_{XX'} (\vect{R}_{X}^{\theta}-\vect{R}_{X'})$ and
  the tunneling strength in momentum space, $t_{X'} (\vect{q})$, is the Fourier transform of $T_{XX'} (\vect{R})$. As we consider only one $p_z$ orbital per carbon atom and we adopt the Slater-Koster approximation, $t_{X'} (\vect{q})$ is insensitive to the graphene sublattice index $X$ (see Appendix \ref{sec:slaterkoster}).
  
  Considering now the graphene on monolayer TMDC heterostructure, in real space an electron from graphene may tunnel to any of 
  the three layers of atoms of the TMDC.  However, the probability to reach the second or the third atomic layers of the monolayer TMDC 
  is exponentially suppressed with respect to reaching the first, closest one. Therefore, to describe the tunneling we consider 
  only the first (upper) chalcogen layer that is closer to graphene. 
  In contrast to graphene, monolayer TMDCs have a rather complicated band structure.  Since DFT calculations indicate that 
  the Dirac point of graphene is found inside the band gap of the TMDC, we expect that the most important bands of the TMDC are those 
   nearest in energy,  namely the conduction and the valence bands. These bands are mainly formed by metal atom $d$ orbitals, 
   but the weights of  chalcogen atom $p$ orbitals are non-zero\cite{kormanyos_k_2015}. 
   It follows that the nearest chalcogen layer approximation for tunneling can be used in combination with the 
   band description of the TMDC. Accordingly, we need to extend the theory of Ref.~\citenum{koshino_interlayer_2015} 
   to consider tunneling not from atomic orbital to atomic orbital but from orbital to an energy band.
  
  The state of an electron in band $b$ of the TMDC, can be written as a linear combination of single orbital Bloch states, $\ket{b,\vect{k}'}_\text{tmdc} = \sum_{X'} c_{bX'} (\vect{k}') \ket{X', \vect{k}'}_\text{tmdc}$. Here the complex amplitudes $c_{bX'} (\vect{k}')$ are different for each band $b$. In our approximation, when computing the interlayer tunneling matrix, this sum runs over the three $p$ orbitals of the nearest chalcogen layer, hence $\vect{\tau}_{X'} = \vect{\tau}_{B'}$ in Eq.~\eqref{eqn:umklapp}. We introduce the interlayer tunneling matrix element between orbital $X$ of graphene and band $b$ of the TMDC as $U_{Xb} (\vect{k}, \vect{k}') = \; _\text{gr}\bra{X, \vect{k}^{\theta}} H_\text{T} \ket{b, \vect{k}'}_\text{tmdc}$. As a consequence, $t_{X'} (\vect{k}' + \vect{G}')$ in Eq.~\eqref{eqn:umklapp} is replaced by $t_{b} (\vect{k}' + \vect{G}')$, the band tunneling strength,
  \begin{equation}\label{eqn:tunnstrength}
    t_{b} (\vect{k}' + \vect{G}') = \sum_{X'} c_{bX'} (\vect{k}') \, t_{X'} (\vect{k}' + \vect{G}').
  \end{equation}
  
  \section{Bilayer Hamiltonian}\label{sec:effectiveHam}
  
  \begin{figure}
    \centering
    \includegraphics[width=.9\columnwidth]{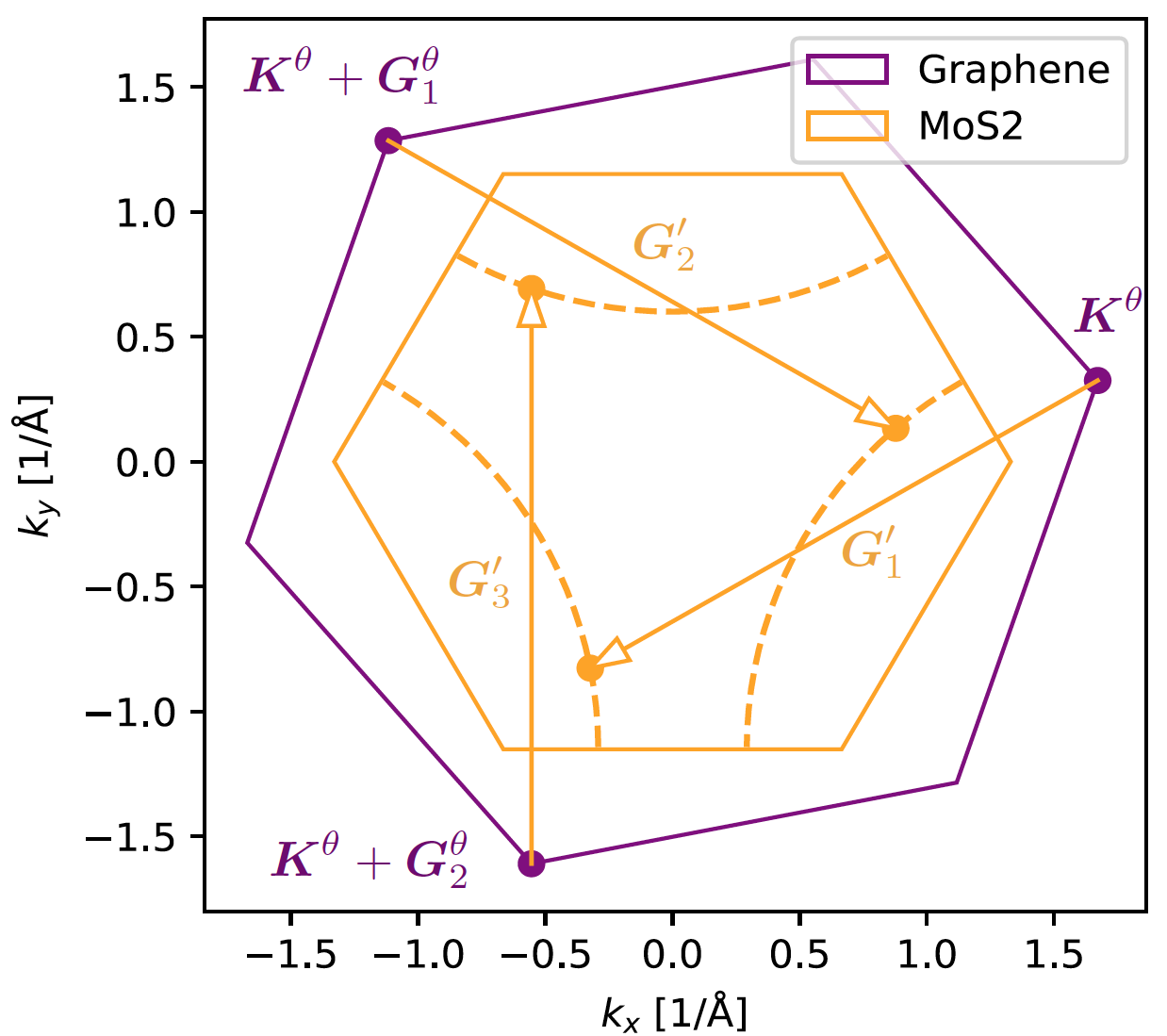}
    \caption{Backfolded TMDC BZ vectors satisfying the quasi-momentum conservation of Eq.~\eqref{eqn:backfolded} for the rotated Dirac point of graphene $\vect{K}^\theta$. The dashed lines indicate the full paths of the backfolded vectors in the range of twist angles $\theta \in [0, \pi/3]$. Moreover, $\vect{G}_{1,2}^\theta$ are rotated reciprocal lattice vectors of graphene, while $\vect{G}'_{1,2,3}$ are reciprocal lattice vectors of the TMDC. As an example, here we have shown in orange the BZ of MoS$_2$ (with lattice constant $a_T = 3.15$ \AA).}
    \label{fig:geometry}
  \end{figure}
  
  We expect $|t_{b}(\vect{q})|$ to decay very fast in $|\vect{q}|$ \cite{bistritzer_transport_2010, bistritzer_moire_2011, koshino_interlayer_2015}, therefore we consider only vectors $\vect{k}'$ in the TMDC BZ that respect the quasi-momentum conservation of Eq.~\eqref{eqn:umklapp}, i.e. $\tau\vect{K}^{\theta} + \vect{G}^{\theta} = \vect{k}' + \vect{G}'$, and such that $|\vect{k}' + \vect{G}'|$ is minimum. We find that these two conditions are satisfied for three distinct points $\tau\vect{k}'_j$, $j=1,2,3$, of the TMDC BZ, for a fixed value of $\tau$. This is similar to what happens for rotated bilayer graphene \cite{bistritzer_moire_2011}. When $\theta \in [0, \pi/3]$, for our choice of reciprocal lattice vectors, these three points are
  \begin{equation}\label{eqn:backfolded}
    \begin{aligned}
      \tau\vect{k}'_1 & = \tau (\vect{K}^\theta - \vect{b}'_1), \\
      \tau\vect{k}'_2 & = \tau (\vect{K}^\theta + \vect{b}_2^\theta
        - \vect{b}'_2), \\
      \tau\vect{k}'_3 & = \tau (\vect{K}^\theta - \vect{b}_1^\theta
        + \vect{b}'_1 + \vect{b}'_2),
    \end{aligned}
  \end{equation}
  where $\vect{b}_{1,2}$ ($\vect{b}'_{1,2}$) are the reciprocal lattice vectors of graphene (TMDC). (See Fig.~\ref{fig:geometry} and Appendix~\ref{sec:definitions}.) Then one can show (see Appendix~\ref{sec:slaterkoster} and Appendix~\ref{sec:symmetryorbitalweights})  that the band tunneling strength in Eq.~\eqref{eqn:tunnstrength} can be parametrized by two real numbers, $t_\parallel$ and $t_\perp$,
  \begin{multline}\label{eqn:bandtunn}
    t_b (\tau\vect{K}^\theta) =
      i \tau [c_{bx} (\tau\vect{k}'_1) \cos\theta
        + c_{by} (\tau\vect{k}'_1) \sin\theta] \;
        t_\parallel \\
      + c_{bz} (\tau\vect{k}'_1) \; t_\perp,
  \end{multline}
  where the connection between the Dirac point $\tau\vect{K}^\theta$ and first backfolded point $\tau\vect{k}'_1$ is given in Eq.~\eqref{eqn:backfolded}. We estimate $t_\parallel \approx t_\perp \approx 100$ meV, see Appendix \ref{sec:estimation} for details. In order to compute the band tunneling strength for all twist angles $\theta$, Eq.~\eqref{eqn:bandtunn} requires the knowledge of the orbital amplitudes $c_{bp}(\tau\vect{k}'_j)$, $p=x,y,z$, which are intrinsic properties of the TMDC. We have obtained their values for MoS$_2$ from the tight-binding model of Ref.~\citenum{fang_ab_2015}.
  
  One can then set up a bilayer Hamiltonian valid for a neighborhood of the Dirac point $\tau\vect{K}$ that describes the hybridization with the TMDC,
  \begin{equation}\label{eqn:effham}
    \mathcal{H} = 
    \left (
    \begin{array}{c|cccc}
    h_{\tau\vect{K}}^{\text{gr},\theta} (\delta\vect{k}) & 
    T_{\tau\vect{k}'_1} &
    T_{\tau\vect{k}'_2} &
    T_{\tau\vect{k}'_3} \\
    \hline
    T_{\tau\vect{k}'_1}^\dagger & 
    h_{\tau\vect{k}'_1}^\text{tmdc} (\delta\vect{k})
    & 0 & 0 \\
    T_{\tau\vect{k}'_2}^\dagger & 0 &
    h_{\tau\vect{k}'_2}^\text{tmdc} (\delta\vect{k})
    & 0 \\
    T_{\tau\vect{k}'_3}^\dagger & 0 & 0 &
    h_{\tau\vect{k}'_3}^\text{tmdc} (\delta\vect{k})
    \end{array}
    \right ).
  \end{equation}
   Here $\delta\vect{k}$ is a small displacement, $|\delta\vect{k}|\ll |\vect{K}|$, from the backfolded vectors $\tau\vect{k}'_j$. 
   The displacement from the Dirac point is therefore $\delta\vect{k}^{\alpha=-\theta}$ in graphene’s coordinate system. 
   The rotated graphene Hamiltonian reads
  \begin{equation}\label{eqn:rotatedgraphene}
    h_{\tau\vect{K}}^{\text{gr},\theta} (\delta\vect{k}) =
    \hbar v_F \tau |\delta\vect{k}|
    \begin{pmatrix}
      0 & e^{-i \tau (\varphi_{\delta\vect{k}} - \theta)} \\
      e^{i \tau (\varphi_{\delta\vect{k}} - \theta)} & 0
    \end{pmatrix} \otimes \id_S,
  \end{equation}
  with $\varphi_{\delta\vect{k}} = \arctan(\delta\vect{k}_x / \delta\vect{k}_y)$ and $\id_S$ is the identity matrix for the spin degree of freedom. Moreover, $h_{\tau\vect{k}'_j}^\text{tmdc} (\delta\vect{k})$ describes the Hamiltonian of the TMDC at a vector $\delta\vect{k}$ distance from $\tau\vect{k}'_j$. $h_{\tau\vect{k}'_{2,3}}^\text{tmdc}(\delta\vect{k})$ can be obtained from $h_{\tau\vect{k}'_1}^\text{tmdc}(\delta\vect{k})$ because the points $\tau\vect{k}'_j$ have $C_3$ symmetry with respect to the $\Gamma$ point of the TMDC BZ. Therefore
  \begin{equation}\label{eqn:tmdsymmetry}
    \begin{aligned}
      h_{\tau\vect{k}'_2}^\text{tmdc} (\delta\vect{k}) & =
        h_{\tau\vect{k}'_1}^\text{tmdc} (\delta\vect{k}^{\alpha=-2\pi/3}), \\
      h_{\tau\vect{k}'_3}^\text{tmdc} (\delta\vect{k}) & =
        h_{\tau\vect{k}'_1}^\text{tmdc} (\delta\vect{k}^{\alpha=+2\pi/3}).
    \end{aligned}
  \end{equation}
  In the simplest case  $h_{\tau\vect{k}'_j}^\text{tmdc} (\delta\vect{k})$ contains the dispersion of those bands that we take into account, i.e., 
  valence and the conduction band. 
  In our case $h_{\tau\vect{k}'_j}^\text{tmdc} (\delta\vect{k})$ also includes the effects of the intrinsic SOC of the TMDC 
  on the band structure. 
  The dispersion of the bands can be obtained, e.g., using the $k \cdot p$ method (see Appendix \ref{sec:schriefferwolff}) 
  or taken from TB calculations. 
  Finally, the tunneling from the $\tau\vect{K}^\theta$ point of graphene to the $\tau\vect{k}'_j$ points of the TMDC BZ is given 
  by $T_{\tau\vect{k}'_j}$. In our approximation, the tunneling matrices $T_{\tau\vect{k}'_j}$ do not depend on 
  the value of the small wave vector $\delta\vect{k}$.
  Using Eq.~\eqref{eqn:umklapp} and Eq.~\eqref{eqn:tunnstrength}, for each band $b$ of the TMDC that we take into account 
  in $h_{\vect{k}'_j}^\text{tmdc} (\delta\vect{k})$  the corresponding column of the tunneling matrix $T_{\tau\vect{k}'_j}$ reads 
  \begin{equation}\label{eqn:tunnelingmatrix}
    (T_{\tau\vect{k}'_j})_b = e^{-i \tau\vect{G}'_j \cdot \vect{\tau}_{X'}}
    e^{i \tau\vect{G}_j^{\theta} \cdot \vect{r}_0} t_{b} (\tau\vect{K}^{\theta})
    \begin{pmatrix}
      1 \\
     e^{i \tau\phi_j}
    \end{pmatrix},
  \end{equation}
  where $\vect{G}_j = 0, \vect{b}_2, -\vect{b}_1$ and $\vect{G}'_j = \vect{b}'_1, \vect{b}'_2, -\vect{b}'_1-\vect{b}'_2$ for $j=1,2,3$, moreover $\phi_j = \vect{G}_j \cdot \vect{\tau}_B = 0, 2\pi/3, -2\pi/3$.  
  We assume that $T_{\tau\vect{k}'_j}$ preserves the spin degree of freedom and therefore it is diagonal in the spin space.
  
  \section{Valley-Zeeman SOC}\label{sec:valley-zeeman}
  
  In order to gain further understanding of how the intrinsic properties of the monolayer TMDC determine the induced valley-Zeeman type SOC, we apply a Schrieffer-Wolff transformation \cite{schrieffer_relation_1966, bravyi_schriefferwolff_2011} to Eq.~\eqref{eqn:effham} to derive an effective graphene Hamiltonian. Following Ref.~\citenum{winkler_spin--orbit_2003}, within second-order the perturbation reads
  \begin{equation}\label{eqn:deltaHgr}
    \delta H^{\text{gr},\tau}_{Xs,X's} = \sum_{j,b}
      \frac{(T_{\tau\vect{k}'_j})_{X,b} \,
        (T_{\tau\vect{k}'_j}^\dagger)_{b,X'}}
        {E_D^\text{gr} - E_{bs}^\text{tmdc}(\tau\vect{k}'_j+\delta\vect{k})},
  \end{equation}
  where $X,X'=A,B$ refers to the graphene sublattices, $s=\uparrow,\downarrow$ is the spin index, $j=1,2,3$ and $b$ is the band index. Moreover, $E_D^\text{gr}$ is the energy of the Dirac point that we fix, without the loss of generality, to $E_D^\text{gr}=0$, while $E_{bs}^\text{tmdc}(\tau\vect{k}'_j+\delta\vect{k})$ is the energy of the TMDC band $b$, spin index $s$, at the BZ point $\tau\vect{k}'_j+\delta\vect{k}$. We remark that Eq.~\eqref{eqn:deltaHgr} does not describe spin-flip processes ($\delta H^{\text{gr},\tau}_{X \uparrow, X' \downarrow} = 0$) because the tunneling matrices of Eq.~\eqref{eqn:tunnelingmatrix} are spin-preserving. One can make use of the threefold rotational symmetry to simplify Eq.~\eqref{eqn:deltaHgr} (see Appendix \ref{sec:schriefferwolff}). Expanding $E_{bs}^\text{tmdc}$ up to linear terms in $\delta\vect{k}$, it turns out that the diagonal matrix elements, $\delta H_{Xs, Xs}^{\text{gr},\tau}$, are $\delta\vect{k}$-independent,
  \begin{equation}\label{eqn:diagonal}
    \delta H^{\text{gr},\tau}_{Xs,Xs} = -3\sum_b \frac{|t_b(\tau\vect{K}^{\theta})|^2}{E_b(\vect{k}'_1) + s\tau \Delta_{0,b}(\vect{k}'_1)},
  \end{equation}
  where  $E_b(\vect{k}'_1)$ is the energy of the TMDC band $b$ (ignoring SOC) at $\vect{k}'_1$, computed with respect to the Dirac point of graphene and $\Delta_{0,b}(\vect{k}'_1)$ is the spin splitting of band $b$ at $\vect{k}'_1$ due to the diagonal part of the intrinsic SOC of the TMDC \cite{kormanyos_k_2015}. Neglecting a constant shift, Eq.~\eqref{eqn:diagonal} can be rewritten as $H_\text{VZ} = \lambda_\text{VZ} \, \tau s_z$, where $s_z$ is a Pauli matrix for spin. The Hamiltonian term $H_\text{VZ}$ describes the induced valley Zeeman SOC and the constant $\lambda_\text{VZ}$ is given by
  \begin{equation}\label{eqn:lambdavalleyzeeman}
    \lambda_\text{VZ} = 3 \sum_b \frac{|t_b(\tau\vect{K}^{\theta})|^2 \Delta_{0,b}(\vect{k}'_1)}{E_b^2(\vect{k}'_1) - \Delta_{0,b}^2(\vect{k}'_1)}.
  \end{equation}
  This is the first important result of our work. It shows explicitly how $\lambda_\text{VZ}$ depends on the intrinsic properties of the TMDC substrate and the twist angle  $\theta$ between the layers. The latter determines the wavenumber $\vect{k}'_1$ and affects the tunneling strength $t_b(\tau\vect{K}^{\theta})$ through Eq.~\eqref{eqn:bandtunn}.
  
  The off-diagonal matrix elements $\delta H_{As, Bs}^{\text{gr},\tau} (\delta\vect{k})$  in  Eq.~\eqref{eqn:deltaHgr} are $\delta\vect{k}$-dependent,
  \begin{multline}\label{eqn:offdiagonal}
    \delta H_{As,Bs}^{\text{gr},\tau}(\delta\vect{k}) = \\
      \frac{3}{2}
      \left (
        \sum_b 
        \frac{w_{bs\tau,\tau}(\vect{k}'_1) |t_b(\tau\vect{K}^{\theta})|^2}
        {E_{bs\tau}^2(\vect{k}'_1)}
      \right ) [\tau\delta\vect{k}_x - i \delta\vect{k}_y],
  \end{multline}
  where $E_{bs\tau}(\vect{k}'_1)=E_b(\vect{k}'_1) + s\tau \Delta_{0,b}(\vect{k}'_1)$ and $w_{bs\tau,\tau}(\vect{k}'_1)$ is a complex quantity related to the local slope of the TMDC band $b$ (see Appendix~\ref{sec:schriefferwolff}). Eq.~\eqref{eqn:offdiagonal} gives a  correction to the Fermi velocity of pristine graphene. The proximity corrected Fermi velocity is
  \begin{equation}
    \widetilde{v}_F = \left |
      v_F + e^{i\tau\theta} \frac{3}{2\hbar}
      \sum_b
      \frac{w_{bs\tau,\tau}(\vect{k}'_1) |t_b(\tau\vect{K}^{\theta})|^2}
      {E_{bs\tau}^2(\vect{k}'_1)}
    \right |.
  \end{equation}
  We have numerically computed this correction for a pristine graphene Fermi velocity $v_F = 10^6$ m/s, using MoS$_2$ as the TMDC compound. The correction we find is in the order of $\pm 0.2$\% depending on the twist angle. In general the value of $v_F$ is more sensitive to the dielectric constant of the environment \cite{hwang_fermi_2012}, therefore we will not discuss this effect further. 
  
  \section{Rashba type SOC}\label{sec:rashba}
  
  As already mentioned, WAL measurements suggest that a Rashba-type SOC is also
  induced in graphene. Traditionally, the Rashba SOC in graphene was understood
  in terms of a symmetry breaking effect of a perpendicular electric field
  \cite{min_intrinsic_2006, gmitra_band-structure_2009, han_graphene_2014}.
  More generally, one can expect that Rashba-type SOC is induced when
  structural asymmetry is present in the heterostructure. Indeed, the DFT
  calculation of Ref.~\citenum{gmitra_trivial_2016} indicated that even for
  zero external electric field a finite Rashba SOC is induced in graphene. To
  our knowledge, the microscopic mechanisms giving rise to the induced
  Rashba SOC has not yet been discussed. We show that an important contribution
  comes from virtual interlayer tunneling processes that are facilitated by
  the off-diagonal spin-flipping elements of the intrinsic SOC matrix of the
  monolayer TMDC, indicated by $(H_\text{soc})_{b \uparrow, b' \downarrow}$
  and $(H_\text{soc})_{b \downarrow, b' \uparrow}$. Such off-diagonal matrix
  elements are allowed between pairs of bands if one of the bands is symmetric (even)
  and the other one is antisymmetric (odd) with respect to reflection on the
  horizontal mirror plane of the TMDC (see, e.g.,
  Ref.~\citenum{kormanyos_spin-orbit_2014} for further discussion of the SOC in
  monolayer TMDCs). In third order perturbation theory one finds the following
  matrix elements \cite{winkler_spin--orbit_2003},
  \begin{multline}\label{eqn:Rashba-gen}
   (\delta H^{\text{gr},\tau}_\text{R})_{X \uparrow, X' \downarrow} = \\
   \sum_{j,b,b'}
    \frac{(T_{\tau\vect{k}'_j})_{X,b}
    (H_\text{soc})_{b \uparrow, b' \downarrow}
      (T_{\tau\vect{k}'_j}^\dagger)_{b',X'}}
      {[E_D^\text{gr} - E_{b}^\text{tmdc}(\tau\vect{k}'_j)]
        [E_D^\text{gr} - E_{b'}^\text{tmdc}(\tau\vect{k}'_j)]}
  \end{multline}
  and $(\delta H^{\text{gr},\tau}_\text{R})_{X \downarrow, X' \uparrow}$ is
  analogously defined. Here $b \neq b'$ is the band index and in the
  denominator we have neglected the dependence of the TMDC band energies
  $E_{b}^\text{tmdc}(\tau\vect{k}'_j)$ on the intrinsic SOC (c.f.,
  Eq.~\eqref{eqn:deltaHgr}) because it would lead to higher order effects.
  The matrix elements $(H_\text{soc})_{b \uparrow, b' \downarrow}$ can be
  calculated using the TB model of Ref.~\citenum{fang_ab_2015}, while the
  tunneling matrices $(T_{\tau\vect{k}'_j})_{X,b}$ and
  $(T_{\tau\vect{k}'_j}^\dagger)_{b',X'}$ can be obtained in the same way as
  explained in Sec.~\ref{sec:effectiveHam}. As we show in Appendix~\ref{sec:RashbaSOC},
  each pair of even and odd bands leads to a Rashba SOC strength 
   \begin{equation}\label{eqn:Rashba-strength}
      \lambda_{\text{R},eo} = \frac{6 \gamma_d
      |{T}_{e,o}^{}(\vect{K}^\theta)| |\Lambda_1(\vect{k}'_1)|}
      {\left(E^\text{gr}_{D}-E_{e}^\text{tmdc}(\vect{k}'_1)\right) 
      \left(E^\text{gr}_{D}-E_{o}^\text{tmdc}(\vect{k}'_1)\right)}
  \end{equation}
   and to a complex phase factor  $e^{i\vartheta_{eo}}$, where
   $\vartheta_{eo}=\mathrm{Arg}[\Lambda_1(\vect{k}'_1)].$ Here $\gamma_d$
   is the atomic SOC strength of the metal atoms' $d$ orbitals of the TMDC,
  $T_{e,o} (\vect{K}^\theta) = t_e (\vect{K}^\theta) t_o^*(\vect{K}^\theta)$,
  with $t_b$ defined in
  Eq.~\eqref{eqn:bandtunn} and  $\Lambda_1$ is a complex quantity formed by the SOC matrix elements of the TMDC.
  We give  the explicit definition of $\Lambda_1$ as well as the details of the calculations
  leading to Eq(\ref{eqn:Rashba-strength}) in Appendix~\ref{sec:RashbaSOC}. 
  To obtain the total Rashba SOC strength one has to sum over all possible pairs of even and odd bands,
  including the complex phase factors $e^{i\vartheta_{eo}}$.
  Therefore one has
  $
  \lambda_{\text{R},\text{tot}} = |\lambda_{\text{R},e_1 o_1}e^{i\vartheta_{e_1 o_1}} +
  \lambda_{\text{R},e_2 o_2}e^{i\vartheta_{e_2 o_2}} + \ldots|
  $
  and
  $
  \vartheta_\text{tot}=\mathrm{Arg}[\lambda_{\text{R},e_1 o_1}e^{i\vartheta_{e_1 o_1}} +
  \lambda_{\text{R},e_2 o_2}e^{i\vartheta_{e_2 o_2}} + \ldots]$.
  In the end one finds that the induced  Rashba type SOC in graphene
  reads  $H_\text{R} = \lambda_{\text{R,tot}} e^{-i \vartheta_\text{tot}s_z/2}
  (\tau \sigma_x s_y - \sigma_y s_x) e^{i \vartheta_\text{tot} s_z / 2}$, where
  $s_x$, $s_y$ are spin Pauli matrices.  As one can see from 
  Eq.~\eqref{eqn:Rashba-strength} the induced Rashba type
  SOC, similarly to the induced valley Zeeman SOC,   is a
  second order process in the interlayer tunneling, but in addition it
  involves a spin-flip process within the monolayer TMDC.  We show the results of our numerical
  calculations for $\lambda_{\text{R}}$  in Fig.~\ref{fig:induced-rashba}.
  Finally, the total effective graphene Hamiltonian reads 
  $
  H_\text{G} (\delta\vect{k}) = h_{\tau\vect{K}}^{\text{gr},\theta} (\delta\vect{k}) +
  H_\text{VZ} + H_\text{R}, 
  $
  see Eq.~\eqref{eqn:rotatedgraphene} and
  Eq.~\eqref{eqn:diagonal} for the first two terms and Eq.~(\ref{eqn:Rashba-strength}) for  $H_\text{R}$. 
  
  \section{Discussion}\label{sec:discussion}
  
  \begin{figure}
    \centering
    \includegraphics[width=\columnwidth]{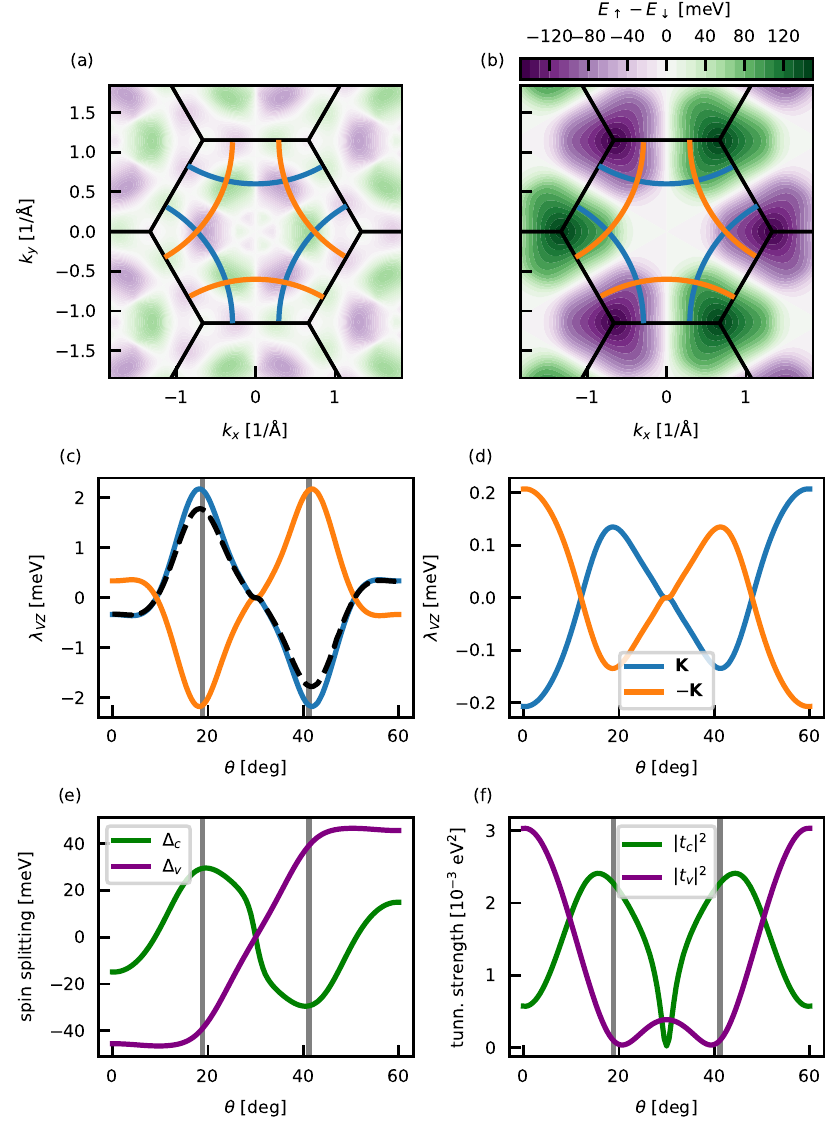}
    \caption{\emph{(a) and (b)}. Spin splitting in conduction, (a), and valence band, (b), of the TMDC. Blue (orange) arcs indicate the paths of the three backfolded vectors $\vect{k}'_j$ ($-\vect{k}'_j$) for Dirac point $\vect{K}$ ($-\vect{K}$). \emph{(c)}. Valley Zeeman spin-orbit strength induced in graphene when the Dirac point energy is close to the TMDC conduction band edge ($f_G = 1$). The blue (orange) line shows the result of second-order perturbation theory for Dirac point $\vect{K}$ ($-\vect{K}$), as derived in Eq.~\eqref{eqn:lambdavalleyzeeman}. The dashed black line is obtained from the exact diagonalization of the bilayer Hamiltonian, Eq.~\eqref{eqn:effham}, for $\vect{K}$. \emph{(d)}. Same as (c) but in the case when the Dirac point energy is in the middle of the TMDC band gap ($f_G = 0.55$) and with a larger TMDC band gap of $E_G = 2.0$ eV in order to reproduce the case of Ref.~\citenum{pierucci_band_2016}. \emph{(e)}. Spin-orbit splitting in TMDC encountered by the backfolded vectors of $\vect{K}$ along the paths in (a) (green line) and (b) (purple line). \emph{(f)}. Tunneling strength squared for a tunneling process from graphene to the conduction (green line) or the valence band (purple line) of the TMDC. The gray vertical lines in (c), (e) and (f) highlight the angles where the backfolded vectors $\vect{k}'_j$ get as close as possible to the maximum of the spin-splitting in the conduction band of the TMDC.}\label{fig:valleyzeemanone}
  \end{figure}
  
  \begin{figure}
    \centering
    \includegraphics[width=\columnwidth]{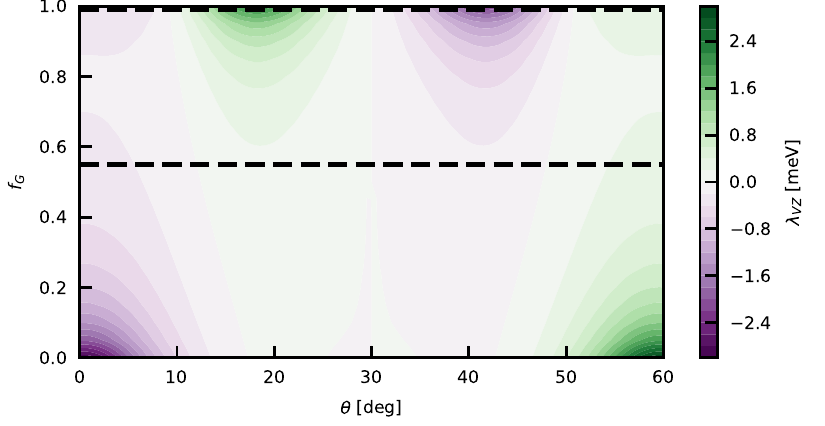}
    \caption{Induced valley Zeeman SOC as a function of the twist angle $\theta$ and the parameter $f_G$ that indicates how close the Dirac point lies to the conduction ($f_G = 1$) or to the valence band ($f_G = 0$). The dashed black lines indicate the two values of $f_G = 1$ and $f_G = 0.55$ used in Fig.~\ref{fig:valleyzeemanone}(c) and (d) respectively.}
    \label{fig:valleyzeemantwo}
  \end{figure}
  
  In order to show explicitly how the twist angle $\theta$ between the layers affects the induced SOC in graphene, we need the band structure of the TMDC substrate and the weights $c_{b,x,y,z}(\tau\vect{k}'_1)$ for all backfolded points $\vect{k}'_1$ in the BZ along the path shown in Fig.~\ref{fig:geometry}. As a concrete example, we take monolayer MoS$_2$ (lattice constant $a_T = 3.15$ Å \cite{kormanyos_k_2015}) and we extract these values from the TB model of Ref.~\citenum{fang_ab_2015}. The \emph{ab initio} calculations from Ref.~\citenum{gmitra_trivial_2016} show the Dirac point very close to the conduction band of MoS$_2$, while experimental results reported in Ref.~\citenum{pierucci_band_2016} indicate that the Dirac point should be found in the middle of the MoS$_2$ band gap. Because of these discrepancies, we treat the energy of the Dirac point of graphene within the band gap of the TMDC as a parameter in our theory. We parametrize this energy by a number $f_G \in [0, 1]$ whose value is a linear function of the position of the Dirac point in the TMDC band gap. When $f_G = 0$, the Dirac point is aligned with the TMDC valence band edge, while for $f_G = 1$ the Dirac point has the same energy as the TMDC conduction band edge.
  
  According to Eq.~\eqref{eqn:lambdavalleyzeeman}, the strength of the induced valley Zeeman SOC has three main contributions from each band $b$: i) it is proportional to the magnitude square of the tunneling strength $|t_b|^2$ and ii) to the spin splitting $\Delta_{0,b}$, while iii) it is inversely proportional to the energy difference  $E_b^2 - \Delta_{0,b}^2$. In our numerical calculations of $\lambda_\text{VZ}$, shown in Fig.~\ref{fig:valleyzeemanone}(c) and (d), we take into account two bands, the conduction ($b=c$) and the valence ($b=v$) bands (CB and VB). We plot $\Delta_{0,c}$ and $\Delta_{0,v}$ in Fig.~\ref{fig:valleyzeemanone}(a),(b) for the whole BZ of monolayer MoS$_2$ and in Fig.~\ref{fig:valleyzeemanone}(e) along the path of the $\vect{k}'_j$ points. Again along this path, we report the values of $|t_c|^2$ and $|t_v|^2$ in Fig.~\ref{fig:valleyzeemanone}(f).
  
  First we consider the case of the Dirac point close to the conduction band ($f_G \approx 1$) as reported by DFT calculations \cite{gmitra_trivial_2016}. Using Eq.~\eqref{eqn:lambdavalleyzeeman}, the calculated  $\lambda_\text{VZ}$ is plotted in Fig.~\ref{fig:valleyzeemanone}(c). One can see that starting from a small negative value at $\theta \gtrsim 0^{\circ}$, $\lambda_\text{VZ}$ vanishes for $\theta\approx 10^{\circ}$  and then increases to $2$ meV just before $\theta=20^{\circ}$. Then  $\lambda_\text{VZ}$ goes back to zero at $\theta= 30^{\circ}$ and the dependence is reflected with opposite sign between $\theta=30^{\circ}$ and $\theta=60^{\circ}$. To understand these features, note that Eq.~\eqref{eqn:deltaHgr} and Eq.~\eqref{eqn:lambdavalleyzeeman} suggest that when the Dirac point is very close to the CB (VB), the contribution from the VB (CB) to $\lambda_\text{VZ}$ is suppressed by the large value of $E_v^2(\vect{k}'_1)$ ($E_c^2(\vect{k}'_1)$). Hence, for $f_G \approx 1$, the behavior of $\lambda_\text{VZ}$ over $\theta \in [0, \pi/3]$ is qualitatively well explained by the contribution of the CB and the VB can be neglected.
  
  The reason for the vanishing  $\lambda_\text{VZ}$ for $\theta\approx 10^{\circ}$ and $\theta= 30^{\circ}$ is that also the TMDC CB spin-splitting goes to zero and changes sign at these angles. The zero spin splitting at $\theta= 30^{\circ}$ appears because the backfolded points $\vect{k}'_j$ lie on the $\Gamma$--$M$ line which by symmetry has no spin splitting \cite{kormanyos_k_2015}. In the case of $\theta\approx 10^{\circ}$, the backfolded points $\vect{k}'_j$ encounter a spin-splitting inversion  of the TMDC conduction band (see Fig.~\ref{fig:valleyzeemanone}(a)), i.e., the spin-split conduction bands cross  along certain low symmetry lines in the BZ. The peak around $\theta= 20^{\circ}$ is expected for multiple reasons. Close to $\theta= 20^{\circ}$ both spin splitting $\Delta_{0,c}(\vect{k}'_1)$ and tunneling strength $t_{c}(\vect{K}^{\theta})$ reach their largest absolute values (see green lines of Fig.~\ref{fig:valleyzeemanone}(e),(f)). For $\Delta_{0,c}(\vect{k}'_1)$ this happens because the backfolded points $\vect{k}'_j$ in the TMDC BZ get very close to the $Q$ valley of the CB, in the middle of the $\Gamma$--$K$ line, which has large spin splitting (see Fig.~\ref{fig:valleyzeemanone}(a)) \cite{kormanyos_k_2015}. The tunneling strength peak instead comes from a larger local weight of the $p_z$ orbitals (larger magnitude of orbital amplitudes $c_{cz} (\tau\vect{k}'_1)$ in Eq.~\eqref{eqn:bandtunn}). Additionally, the energy distance between the Dirac point of graphene and the bottom of the $Q$ point, which is a valley of the CB, is also smaller than for other $\vect{k}'_1$ points in the BZ. We have checked that the above comments remain valid even if we add in the calculation the first band above the conduction band (CB+1). Including this higher band does not change qualitatively the values of $\lambda_\text{VZ}$.
   
  To confirm the behavior predicted by second order perturbation theory, we have computed $\lambda_\text{VZ}$ at $\delta\vect{k}=0$ from exact diagonalization of the bilayer Hamiltonian in Eq.~\eqref{eqn:effham}. Only the CB and the VB were taken into account in $h_{\tau\vect{k}'_j}^\text{tmdc}$. The result is shown in Fig.~\ref{fig:valleyzeemanone}(c) by a dashed black line. The agreement is very close except for the largest absolute values where the second order perturbation results deviates by around 10\%. In these regions the Dirac points are quite near in energy to the CB of the TMDC and the small parameter $|t_b|/(E_b \pm \Delta_{0,b})$ increases up to $0.16$.
     
  
  It is known that DFT calculations (and TB models fitted to DFT calculations) underestimate the band gap of the TMDC. Indeed, the ARPES experiment of Ref.~\citenum{pierucci_band_2016} reports a larger band gap of 2.0 eV. Moreover, according to Ref.~\citenum{pierucci_band_2016}, in graphene/TMDC bilayers, the Dirac point of graphene is found in the middle of the TMDC band gap ($f_G \approx 0.55$). For these reasons we have computed the induced valley Zeeman SOC in Eq.~\eqref{eqn:lambdavalleyzeeman} for these alternative parameters (CB and VB dispersions were taken from the TB model as before). The results are plotted in Fig.~\ref{fig:valleyzeemanone}(d). Here, the contribution from the VB is larger close to $\theta = 0^\circ$ and $\theta = 60^\circ$ (see purple lines in Fig.~\ref{fig:valleyzeemanone}(e),(f)) while it fades away around $\theta = 20^\circ$ and $\theta = 40^\circ$ where the CB contribution is more significant (see green lines in Fig.~\ref{fig:valleyzeemanone}(e),(f)). Nevertheless, the values for $\lambda_\text{VZ}$ predicted in Fig.~\ref{fig:valleyzeemanone}(d) are one order of magnitude lower than those in Fig.~\ref{fig:valleyzeemanone}(c) (Dirac point close to CB). They are indeed suppressed by the large distance of the Dirac point from both CB and VB.
 %
 We show in Fig.~\ref{fig:valleyzeemantwo} the value of $\lambda_\text{VZ}$ computed from Eq.~\eqref{eqn:lambdavalleyzeeman} for all values of $f_G$ between 0 and 1. The dashed black lines indicates the two cuts at $f_G = 1$ (Fig.~\ref{fig:valleyzeemanone}(c)) and $f_G = 0.55$ (Fig.~\ref{fig:valleyzeemanone}(d)). One can observe that close to the VB ($f_G \approx 0$) the induced valley Zeeman SOC is comparable to the values obtained close to the CB. However close to the VB the highest spin-orbit strengths appear close to $\theta = 0^\circ$ and $\theta = 60^\circ$.

  \begin{figure}
    \centering
    \includegraphics[width=\columnwidth]{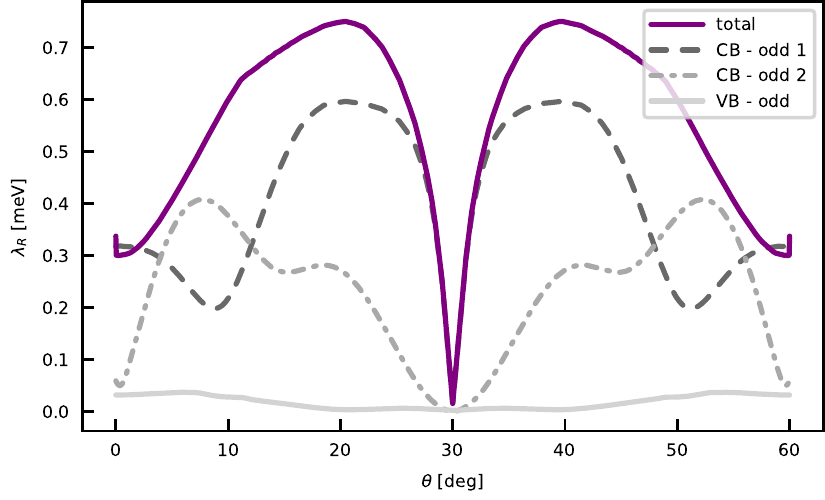}
    \caption{Magnitude of the induced Rashba type SOC as a function of the twist angle $\theta$ for $f_G = 1$. The purple line shows the total Rashba type SOC, the gray lines indicate separately the contribution related to two asymmetric bands above the conduction band and an asymmetric band below the valence band, respectively.}
    \label{fig:induced-rashba}
  \end{figure}
  
  In Fig.~\ref{fig:induced-rashba} we show the induced Rashba SOC as a function of the twist angle $\theta$ between the layers. In these calculations we again considered MoS$_2$ as a concrete example and used $f_G = 1$. The gray lines indicate the separate contributions to Eq.~\eqref{eqn:Rashba-gen} of three pairs of symmetric-antisymmetric bands. In particular, we consider the interaction of the symmetric CB with two asymmetric bands higher in energy and the interaction of the symmetric VB with one asymmetric band lower in energy. The purple line represents the total sum of the three gray contributions taking into account the complex phases associated with them, see Appendix~\ref{sec:RashbaSOC} for details of the  calculation. One can see that the twist angle can considerably change the value of the SOC strength $\lambda_\text{R}$. In particular, a twofold increase of $\lambda_\text{R}$ can be observed at $\theta\approx 20^{\circ}$ with respect to the $\theta = 0^{\circ}$ case. This is a somewhat smaller increase than in the case of $\lambda_\text{VZ}$, nevertheless it shows that $\lambda_\text{R}$ is tunable by the twist angle. The increase of $\lambda_\text{R}$ close to $20^{\circ}$ can partially be explained by the fact that one of  the asymmetric bands, whose energy appears in the denominator of Eq.~\eqref{eqn:Rashba-gen}, is quite close to the conduction band in the vicinity of the $Q$ point. Comparing Fig.~\ref{fig:valleyzeemanone}(c) and Fig.~\ref{fig:induced-rashba} one can see that for $\theta\approx 0$ the values of $\lambda_\text{VZ}$ and $\lambda_\text{R}$ are comparable, while for $\theta\approx 20^{\circ}$ the valley Zeeman SOC dominates the Rashba type SOC. One can also see that $\lambda_\text{R}$ drops to a small but non-zero value for $\theta=30^{\circ}$. This can be qualitatively understood by looking at Fig.~\ref{fig:valleyzeemanone}(f) which shows that the tunneling to the conduction band has a sharp minimum for this angle. 
  
  Finally, we note that Ref.~\citenum{li_twist-angle_2019} studied the same graphene/monolayer TMDC heterostructures using a  TB model to describe both graphene and the monolayer TMDC and setting  up a TB parametrization for the inter-layer coupling. This  approach, in principle, takes into account the coupling between all bands of the monolayer TMDC and graphene but also necessitates a number of new TB parameters to describe the interlayer coupling. For graphene/monolayer MoS$_2$ our results are, both for the induced valley Zeeman and the Rashba type SOC, qualitatively similar to Ref.~\citenum{li_twist-angle_2019}, which indicates that our approach captures the most important ingredients contributing to the induced SOC. However, the vanishing and sign change of $\lambda_\text{VZ}$ at $\theta\approx 10^{\circ}$ was not predicted in Ref.~\citenum{li_twist-angle_2019}. As explained above, we identified the band structure feature of the monolayer MoS$_2$ that gives rise to this behavior of $\lambda_\text{VZ}$ and we believe that it is not an artifact of our approach. This feature should appear in graphene/TMDC bilayers for other semiconductor monolayer TMDC compounds, not only for MoS$_2$. Regarding the induced Rashba SOC, for $\theta=0^\circ$ our result is in good qualitative agreement with Ref.\citenum{gmitra_trivial_2016}, where $\lambda_\text{R}$ was extracted from DFT calculations on commensurate graphene-TMDC supercells. 
  
  \section{Conclusions}\label{sec:conclusion}
  
  In this paper we have presented the analytic twist angle dependence of the induced spin-orbit coupling in graphene from the van der Waals interaction with monolayer TMDC. This fills the gap between experimental and theoretical works on twisted graphene-TMDC heterobilayers. While experiments most likely have a twist angle between the layers of the heterostructure, often unaccounted for in the analyses of the results and different from sample to sample, theory only considered zero or small twist angles. Here we have shown that the induced SOC may vary significantly and even vanish as a function of the twist angle and of the position of the Dirac point in the TMDC band gap, therefore the knowledge of both $\theta$ and $f_G$ is important in order to compare experiments performed with different samples. The largest values of the induced valley Zeeman type SOC  are $\sim 2$ meV when the Dirac point of graphene is close to the conduction band of the TMDC. In comparison, the intrinsic spin-orbit coupling of isolated graphene is expected to be in the order of $24$ $\mu$eV \cite{gmitra_band-structure_2009}. This indicates that, by juxtaposing monolayer TMDCs and by engineering the twist angle between the two layers, the induced SOC in graphene can be two orders of magnitude larger than the intrinsic one. We also identified a microscopic mechanism that gives rise to an induced Rashba type SOC and we have found that it can also be significantly enhanced as a function of the twist angle. 
  
  The use of a band-to-band tunneling picture was fundamental to reach our results. This framework simplifies the study of heterobilayers where the band structure of the individual constituent layers is well known and understood. Similarly to Ref.~\cite{li_twist-angle_2019}, it can also be used if the lattice constants of the individual layers are incommensurate. Moreover, as the complexity of the material increases and the number of orbitals involved in its valence and conduction bands becomes large, an orbital-to-orbital tunneling picture to describe interlayer tunneling would require a tight binding model with many parameters. In graphene/TMDC heterostructures, by using the nearest chalcogen layer approximation and the Fourier transform of the Slater-Koster matrix elements, the interlayer tunneling parametrization was reduced to just two overlap integrals. The bands of the isolated layers can be approximated by $k \cdot p$ theory which helped to obtain  the induced SOC  by applying quasi-degenerate perturbation theory. Using this approach we were able to separate the contribution from the different bands and analyse the behaviour of the induced valley Zeeman and Rashba type SOC as a function of the interlayer twist angle. Our approach makes the role of the intrinsic  properties of the substrate more apparent and, therefore, it might be used to screen potential substrate materials for desired induced SOC properties in van der Waals heterostructures. We assumed perfectly ballistic layers  in our work. An interesting extension would be to study the induced SOC in the presence of disorder effects. This may  affect the interpretation of WAL measurements, as  the interplay between spin, valley  and disorder physics  yields a rich behavior of the quantum correction to the conductivity \cite{ilic_weak_2019}.
  
  \section{Acknowledgements}
  
  We acknowledge funding from the DFG through SFB767, from CAP Konstanz and from FlagERA through iSpinText. P. R. and A. K. were supported by NKFIH within the Quantum Technology National Excellence Program (Project No. 2017-1.2.1-NKP-2017-00001) and by OTKA NN 127903 (Topograph FlagERA project). P. R. also acknowledges the funding from OTKA PD123927 and K123894. We would like to thank A. Pearce for technical assistance with Fig.~\ref{fig:heterostructure} and V. Shkolnikov for helpful discussions.
  
  \appendix
  
  \section{Definition of lattice vectors}\label{sec:definitions}
   
  The basis vectors for the hexagonal lattice are $\vect{a}_{1,2} = a(\pm 1/2, \sqrt{3}/2)$ with lattice constant $a = a_G$ ($a = a_T$) for graphene (TMDC). The $B$ sublattice is shifted by $\vect{\delta} = a/\sqrt{3}(0,1)$. The reciprocal lattice vectors $\vect{b}_{1,2}$ follow the relation $\vect{a}_i \cdot \vect{b}_j = 2\pi\delta_{ij}$, where $\delta_{ij}$ is the Kronecker delta, and are explicitly given by $\vect{b}_{1,2} = 4\pi / a\sqrt{3} (\pm \sqrt{3}/2,1/2)$. In the heterobilayer studied in this paper, graphene is on top of the TMDC layer, separated from the topmost TMDC chalcogen layer by $d_\perp$ (see Fig.~\ref{fig:heterostructure}). The positions of the atoms in the unit cell are given by $\vect{\tau}_X$ for graphene and by $\vect{\tau}_{X'}$ for the TMDC, with $X=A,B$ and $X'=A',B'_1,B'_2$, where $B'_1$ ($B'_2$) indicate the upper (lower) chalcogen atom site. We fix the origin of our coordinate system above a metal atom in the TMDC, but in the same plane as the upper chalcogen layer,
  \begin{equation}
    \begin{gathered}
        \vect{\tau}^{}_{A} = d_\perp \hat{\vect{e}}_z, \qquad
        \vect{\tau}^{}_{B} = \vect{\delta} + d_\perp \hat{\vect{e}}_z, \\
        \vect{\tau}^{}_{A'} = -\frac{d_\text{X--X}}{2} \, \hat{\vect{e}}_z, 
        \quad
        \vect{\tau}^{}_{B'_1} = \vect{\delta}', \quad
        \vect{\tau}^{}_{B'_2} = \vect{\delta}' - 
          d_\text{X--X}\,\hat{\vect{e}}_z,
    \end{gathered}
  \end{equation}
  with $d_\text{X--X}$ the TMDC chalcogen-chalcogen distance.
  
  \section{Slater-Koster tunneling coefficients and their Fourier transform}\label{sec:slaterkoster}
  
  We are interested in the tunneling between the $p_z$ orbitals of the carbon atoms in graphene and the $p$ orbitals of the closest TMDC chalcogen layer. Using the two-center approximation, the real space tunneling matrix elements $T_{XX'} (\vect{R})$ can be written in terms of Slater-Koster parameters \cite{slater_simplified_1954},
  \begin{subequations}\label{eqn:slaterkoster}
    \begin{align}
      T_{p_z, p_z} (\vect{R}) & =
        n_z^2 V_{pp\sigma} (R) + (1 - n_z^2) V_{pp\pi} (R), \\
      T_{p_z, p_x\,(p_y)} (\vect{R}) & =
        n_{x\,(y)} n_z (V_{pp\sigma} (R) - V_{pp\pi} (R)),
    \end{align}
  \end{subequations}
  with $R = |\vect{R}|$ and $(n_x, n_y, n_z) = \vect{R} / R$. Since $X=A,B$ refers always to the $p_z$ orbitals of the carbon atoms in graphene, there is no real dependence on $X$ and we omit it in the following, $T_{XX'} = T_{X'}$. 
  
  In cylindrical coordinates $(r, \varphi, z)$ we have $\vect{r} = r\cos\varphi \, \hat{\vect{e}}_x + r\sin\varphi \, \hat{\vect{e}}_y$, $\vect{R} = \vect{r} + z \hat{\vect{e}}_z$, $R = \sqrt{r^2 + z^2}$ and
  \begin{equation*}
    n_x = \frac{r \cos \varphi}{\sqrt{r^2 + z^2}}, \quad
    n_y = \frac{r \sin \varphi}{\sqrt{r^2 + z^2}}, \quad
    n_z = \frac{z}{\sqrt{r^2 + z^2}}.
  \end{equation*}
  We can separate the radial part from the angular part in Eqs.~\eqref{eqn:slaterkoster},
  \begin{subequations}\label{eqn:separateangle}
    \begin{align}
      T_{p_z} (r, \varphi, z) = & f_z (r, z), \\
      T_{p_x} (r, \varphi, z) = & \cos\varphi \, f_x (r, z), \\
      T_{p_y} (r, \varphi, z) = & \sin\varphi \, f_x (r, z),
    \end{align}
  \end{subequations}
  where
  \begin{equation}
    \begin{gathered}
      f_z (r, z) = \frac{1}{R^2} [z^2 V_{pp\sigma}(R) + r^2 V_{pp\pi}(R)], \\
      f_x (r, z) = f_y (r, z) =
        \frac{rz}{R^2} [V_{pp\sigma} (R) - V_{pp\pi} (R)].
    \end{gathered}
  \end{equation}
  In Eqs.~\eqref{eqn:separateangle}, we refer to the $\varphi$-dependent parts as $a_{X'} (\varphi)$, with $a_z (\varphi) = 1$, $a_x (\varphi) = \cos\varphi$ and $a_y (\varphi) = \sin\varphi$. Hence, we can write $T_{X'} (r, \varphi, z) = a_{X'} (\varphi) f_{X'} (r, z)$. Then, we take the Fourier trasform of Eq.~\eqref{eqn:slaterkoster} \cite{koshino_interlayer_2015},
  \begin{multline}\label{eqn:tunnft}
    t_{X'} (\vect{q}) = \frac{1}{\sqrt{SS'}} \int T_{X'} (\vect{r} + z \hat{\vect{e}}_z)
      e^{-i \vect{q}\cdot\vect{r}} \mathrm{d}^2 r, \\
    = \frac{1}{\sqrt{SS'}} \int_0^\infty \mathrm{d} r \; r f_{X'} (r, z) \\
      \times \int_{-\pi}^\pi \mathrm{d} \varphi \; a_{X'} (\varphi) e^{-iqr\cos(\varphi - \varphi_q)},
  \end{multline}
  where $\vect{q} = (q \cos \varphi_q, q \sin \varphi_q)$ and $S$ ($S'$) is the unit cell size of graphene (TMDC). The integral over the angle can be solved using the Jacobi-Anger expansion \cite{colton_inverse_1998, cuyt_handbook_2008},
  \begin{subequations}
    \begin{align}
      \int_{-\pi}^\pi \mathrm{d}\varphi \; e^{-iqr\cos(\varphi-\varphi_q)} & =
      2\pi J_0 (qr), \\
      \int_{-\pi}^\pi \mathrm{d}\varphi \, \cos \varphi \,
        e^{-iqr\cos(\varphi-\varphi_q)} & = - 2\pi i J_1 (qr) \cos \varphi_q, \\
      \int_{-\pi}^\pi \mathrm{d}\varphi \, \sin \varphi \,
        e^{-iqr\cos(\varphi-\varphi_q)} & = - 2\pi i J_1 (qr) \sin \varphi_q,
    \end{align}
  \end{subequations}
  where $J_m (x)$ is the $m$-th order Bessel function of the first kind. We see that the angular dependence of the tunneling matrix elements is preserved when switching from real space to momentum space. One may write
  \begin{equation}
    t_{X'} (q, \varphi_q, z) = (-i)^m a_{X'} (\varphi_q)
      P_{X'} (q, z),
  \end{equation}
  where $P_{X'} (q, z)$ is real and equal to the integral of the radial part,
  \begin{equation}
    P_{X'} (q, z) = \frac{2\pi}{\sqrt{SS'}}
      \int_0^\infty \mathrm{d} r \; r f_{X'} (r, z) J_m (qr),
  \end{equation}
  with $m = 0$ for $X' = p_z$, while $m = 1$ for $X' = p_{x}, p_y$.
  
  We define the tunneling strength from graphene to a band of the TMDC as
  \begin{equation}\label{eqn:tunnelingstrengthband}
    t_{b} (\vect{k}' + \vect{G}') = \sum_{X'} c_{bX'} (\vect{k}')
      t_{X'} (\vect{k}' + \vect{G}'),
  \end{equation}
  where $\vect{k}'$ is a vector inside the first TMDC BZ, $\vect{G}'$ is a reciprocal lattice vector of the TMDC and $c_{bX'} (\vect{k}')$ is the amplitude of orbital $X'$ in band $b$. We derive here the form of Eq.~\eqref{eqn:tunnelingstrengthband} for the points $\tau(\vect{k}'_j + \vect{G}'_j)$ of Eq.~\eqref{eqn:backfolded}, with $\vect{G}'_1 = \vect{b}'_1$, $\vect{G}'_2 = \vect{b}'_2$ and $\vect{G}'_3 = - \vect{b}'_1 - \vect{b}'_2$. Using the quasi-momentum conservation we have $\tau(\vect{k}'_j + \vect{G}'_j) = \tau(\vect{K}^\theta + \vect{G}_j^\theta) =: \tau\vect{K}_j^\theta$, with $\vect{G}_1 = 0$, $\vect{G}_2 = \vect{b}_2$ and $\vect{G}_3 = -\vect{b}_1$ (see Fig.~\ref{fig:geometry}). We remark here that all vectors $\tau\vect{K}_j^\theta$ have the same magnitude $K$. Renaming the in-plane integral as $-P_x (K, z_1) \equiv -P_y (K, z_1) \equiv t_\parallel$ and the out-of-plane integral as $P_z (K, z_1) \equiv t_\perp$, with $z_1 = d_\perp$, we have then
  \begin{multline}\label{eqn:tunnstrengthlong}
    t_{b} (\tau\vect{K}_j^\theta) =
      i [c_{bx} (\tau\vect{k}'_j) \cos\varphi_{\tau\vect{K}_j^\theta}
        + c_{by} (\tau\vect{k}'_j) \sin\varphi_{\tau\vect{K}_j^\theta}] \;
        t_\parallel \\
      + c_{bz} (\tau\vect{k}'_j) \; t_\perp,
  \end{multline}
  where $\varphi_{\tau\vect{K}_j^\theta}$ is the polar angle of $\tau\vect{K}_j^\theta$. One may write $\varphi_{\tau\vect{K}_j^\theta} = \varphi_{\tau\vect{K}_j} + \theta$ with $\varphi_{\vect{K}_1} = \varphi_{\vect{K}} = 0$, $\varphi_{\vect{K}_2} = 2\pi/3$ and $\varphi_{\vect{K}_3} = -2\pi/3$, while $\varphi_{-\vect{K}_j} = \varphi_{\vect{K}_j} + \pi$. We treat $t_\parallel$ and $t_\perp$ as two real parameters to be determined from experiments, \emph{ab initio} calculations or tight binding models.
  
  \section{Symmetry of orbital amplitudes in a TMDC band}\label{sec:symmetryorbitalweights}
  
  To define the tunneling strength in Eq.~\eqref{eqn:tunnelingstrengthband}, we have expanded the state of an electron in band $b$ of the TMDC as a linear combination of single orbital Bloch states,
  \begin{equation}
    \ket{b, \vect{k}'} = \sum_{X'} c_{bX'} (\vect{k}') \ket{X', \vect{k}'}.
  \end{equation}
  The properties of the coefficients $c_{bX'} (\vect{k}')$ therefore play an important role in the form of the bilayer Hamiltonian, Eq.~\eqref{eqn:effham}. These coefficients are constrained by the TMDC lattice symmetry and the coordinate transformations of the orbitals and of the Bloch states. We prove a useful relation focusing on $c_{b,x} (\vect{k}')$ and $c_{b,y} (\vect{k}')$, the coefficients of orbitals $p_x$ and $p_y$ respectively. For the sake of clarity we indicate $\ket{X', \vect{k}'} \equiv \ket{\psi_{X'}, \vect{k}'}$, where we made the orbital wavefunction $\psi_{X'}$ explicit, $\braket{\vect{r} | \psi_{X'}} = \psi_{X'} (\vect{r})$, with $\vect{r} = (x, y, z)^T$.
  
  Consider two wavevectors $\vect{k}'$ and $R(\alpha) \vect{k}'$ where $R(\alpha)$ is a rotation of the point group of the TMDC crystal, i.e. $\alpha = \pm 2\pi/3$. Following Ref.~\citenum{dresselhaus_group_2010}, we know that
  \begin{equation}\label{eqn:grouptheory}
    \begin{aligned}
      \ket{b, R(\alpha)\vect{k}'} & = R(\alpha) \ket{b, \vect{k}'} \\
        & = \sum_{X'} c_{bX'} (\vect{k}') R(\alpha) \ket{\psi_{X'}, \vect{k}'}.
    \end{aligned}
  \end{equation}
  For a single orbital Bloch state, $\ket{\psi_{X'}, \vect{k}'}$, the transformation under rotation results in a rotation of the orbital wavefunction,
  \begin{multline}
    \braket{\vect{r} | R(\alpha) | \psi_{X'}, \vect{k}'} =
      \braket{R(-\alpha) \vect{r} | \psi_{X'}, \vect{k}'} \\
    \begin{aligned}
        & = \frac{1}{\sqrt{N}} \sum_{\vect{R}_{X'}}
          e^{i \vect{k}' \cdot \vect{R}_{X'}}
          \psi_{X'} (R(-\alpha)\vect{r} - \vect{R}_{X'}) \\
        & = \frac{1}{\sqrt{N}} \sum_{\vect{R}_{X'}}
          e^{i \vect{k}' \cdot \vect{R}_{X'}}
          \psi_{X'} (R(-\alpha)(\vect{r} - R(\alpha)\vect{R}_{X'})) \\
        & = \frac{1}{\sqrt{N}} \sum_{\widetilde{\vect{R}}_{X'}}
          e^{i \vect{k}' \cdot R(-\alpha)\widetilde{\vect{R}}_{X'}}
          (R(\alpha) \psi_{X'}) (\vect{r} - \widetilde{\vect{R}}_{X'}) \\
        & = \frac{1}{\sqrt{N}} \sum_{\widetilde{\vect{R}}_{X'}}
          e^{i R(\alpha)\vect{k}' \cdot \widetilde{\vect{R}}_{X'}}
          (R(\alpha) \psi_{X'}) (\vect{r} - \widetilde{\vect{R}}_{X'}) \\
        & = \braket{\vect{r} | R(\alpha) \psi_{X'}, R(\alpha)\vect{k}'},
    \end{aligned}
  \end{multline}
  therefore
  \begin{equation}
    R(\alpha) \ket{\psi_{X'}, \vect{k}'} =
      \ket{R(\alpha) \psi_{X'}, R(\alpha) \vect{k}'}.
  \end{equation}
  Due to the linear dependence of $p_x(\vect{r})$ and $p_y(\vect{r})$ on $x$ and $y$ respectively, we have the following transformations for $\psi_{X'} = p_x, p_y$,
  \begin{equation}
    \begin{aligned}
      (R(\alpha) p_x) (\vect{r}) & = p_x (R(-\alpha) \vect{r}) = 
        \cos\alpha \, p_x (\vect{r}) + \sin\alpha \, p_y (\vect{r}), \\
      (R(\alpha) p_y) (\vect{r}) & = p_y (R(-\alpha) \vect{r}) = 
        -\sin\alpha \, p_x (\vect{r}) + \cos\alpha \, p_y (\vect{r}),
    \end{aligned}
  \end{equation}
  which is reflected then in the Bloch states,
  \begin{equation}
    \begin{aligned}
      \ket{R(\alpha) p_x, \vect{k}'} & = 
        \cos\alpha \, \ket{p_x, \vect{k}'}
        + \sin\alpha \, \ket{p_y, \vect{k}'}, \\
      \ket{R(\alpha) p_y, \vect{k}'} & =
        - \sin\alpha \, \ket{p_x, \vect{k}'}
        + \cos\alpha \, \ket{p_y, \vect{k}'}.
    \end{aligned}
  \end{equation}
  Finally, multiplying the left and the right hand side of Eq.~\eqref{eqn:grouptheory} by $\bra{\psi_{\widetilde{X}'}, R(\alpha)\vect{k}'}$ and using the orthogonality between $p_x$ and $p_y$ orbitals, we obtain
  \begin{equation}
    \begin{aligned}
      c_{b,x} (R(\alpha)\vect{k}') & =
        \cos\alpha \, c_{b,x} (\vect{k}')
        - \sin\alpha \, c_{b,y} (\vect{k}'), \\
      c_{b,y} (R(\alpha)\vect{k}') & =
        \sin\alpha \, c_{b,x} (\vect{k}')
        + \cos\alpha \, c_{b,y} (\vect{k}'),
    \end{aligned}
  \end{equation}
  which can be written in short form as
  \begin{equation}\label{eqn:amplituderotation}
    \vect{c}_b (R(\alpha)\vect{k}') = R(\alpha) \vect{c}_b (\vect{k}'),
  \end{equation}
  with $\vect{c}_b (\vect{k}') = (c_{b,x} (\vect{k}'), c_{b,y} (\vect{k}'))^T$.
  
  We need Eq.~\eqref{eqn:amplituderotation} to prove that the band tunneling strength in Eq.~\eqref{eqn:tunnstrengthlong} has the same value for all the three backfolded vectors $\tau\vect{k}'_j$ in Eq.~\eqref{eqn:backfolded}. Eq.~\eqref{eqn:tunnstrengthlong} can be rewritten as
  \begin{equation}
    t_b (\tau\vect{K}_j^\theta) = \vect{c}_b (\tau\vect{k}'_j) \cdot
      R(\varphi_{\tau\vect{K}_j^\theta}) \vect{t},
  \end{equation}
  where $\vect{t} = (i t_\parallel, 0, t_\perp)$. Here we have included the $p_z$ coefficient $c_{b,z} (\tau\vect{k}'_j)$ in the vector $\vect{c}_b (\tau\vect{k}'_j)$ and the rotation operator $R(\varphi_{\tau\vect{K}_j^\theta})$ is a $3 \times 3$ matrix rotating only the first two components of $\vect{t}$ while leaving the third one unchanged. We show that $t_b (\vect{K}_2^\theta) = t_b (\vect{K}_1^\theta)$ and one can obtain similar results for $\vect{K}_3^\theta$ and for the opposite Dirac point ($\tau=-$). We remark that $\varphi_{\vect{K}_2^\theta} = \varphi_{\vect{K}_1^\theta} + 2\pi/3$. Then,
  \begin{equation}
    \begin{aligned}
      t_b (\vect{K}_2^\theta) & = \vect{c}_b (\vect{k}'_2) \cdot
          R(\varphi_{\vect{K}_2^\theta}) \vect{t} \\
        & = \vect{c}_b (R(2\pi/3)\vect{k}'_1) \cdot
          R(\varphi_{\vect{K}_1^\theta} + 2\pi/3) \vect{t} \\
        & = \vect{c}_b (\vect{k}'_1) \cdot R(\varphi_{\vect{K}_1^\theta})
          \vect{t} = t_b (\vect{K}_1^\theta),
    \end{aligned}
  \end{equation}
  where we have used Eq.~\eqref{eqn:amplituderotation}. It follows that we need to compute the band tunneling strength only for $\tau\vect{K}_1^\theta = \tau\vect{K}^\theta$. Since $\varphi_{\vect{K}^\theta} = \theta$ and $\varphi_{-\vect{K}^\theta} = \theta + \pi$, we can write Eq.~\eqref{eqn:tunnstrengthlong} as
  \begin{multline}
    t_b (\tau\vect{K}^\theta) =
      i \tau [c_{bx} (\tau\vect{k}'_1) \cos\theta
        + c_{by} (\tau\vect{k}'_1) \sin\theta] \;
        t_\parallel \\
      + c_{bz} (\tau\vect{k}'_1) \; t_\perp.
  \end{multline}
  
  \section{Second order Schrieffer-Wolff transformation}\label{sec:schriefferwolff}
  
  Here we derive Eq.~\eqref{eqn:diagonal} and Eq.~\eqref{eqn:offdiagonal}. The second order Schrieffer-Wolff matrix elements are given by
  \begin{equation}
    \delta H^{\text{gr},\tau}_{Xs,X's'} = \sum_{j,b,s''}
      \frac{(T_{\tau\vect{k}'_j})_{Xs,bs''}
        (T_{\tau\vect{k}'_j}^\dagger)_{bs'',X's'}}
        {E_D^\text{gr} - E_{bs''}^\text{tmdc}(\tau\vect{k}'_j+\delta\vect{k})}.
  \end{equation}
  In the following we treat diagonal and off-diagonal elements separately. We also expand the numerator using Eq.~\eqref{eqn:tunnelingmatrix} and we obtain for the diagonal elements
  \begin{equation}
    \delta H_{Xs, Xs}^{\text{gr},\tau} =
      -\sum_{j,b} \frac{|t_b(\tau\vect{K}^\theta)|^2}{E_{bs}^\text{tmdc} (\tau\vect{k}'_j+\delta\vect{k})}.
  \end{equation}
  Since the tunneling matrices in Eq.~\eqref{eqn:tunnelingmatrix} preserve the spin, we have $\delta H_{Xs, X's'}^{\text{gr},\tau} = 0$ for $s \neq s'$. Hence only two independent off-diagonal elements are non-zero,
  \begin{equation}\label{eqn:SWOffDiag}
    \delta H_{As, Bs}^{\text{gr},\tau} =
      -\sum_{j,b} \frac{| t_b(\tau\vect{K}^\theta) |^2 e^{-i\tau\phi_j}}
        {E_{bs}^\text{tmdc} (\tau\vect{k}'_j+\delta\vect{k})},
  \end{equation}
  for $s = \uparrow, \downarrow$. As one can see, the diagonal elements are obtained from the off-diagonal ones by setting $\phi_j = 0$.
  
  We expand the $\delta \vect{k}$-dependence of $E_{bs}^\text{tmdc} (\tau\vect{k}'_j+\delta\vect{k})$ using $k \cdot p$ theory \cite{kormanyos_k_2015}. For a general $\vect{k}'_1$ point in the TMDC BZ,
  \begin{multline}\label{eqn:kdotp}
    E_{bs}^\text{tmdc} (\tau\vect{k}'_1 + \delta\vect{k}) =
      E_b(\vect{k}'_1) + s\tau \Delta_{0,b}(\vect{k}'_1) \\
      + (w_{x,b}(\vect{k}'_1) + s\tau \Delta_{1x,b}(\vect{k}'_1))
        \tau \delta \vect{k}_x \\
      + (w_{y,b}(\vect{k}'_1) + s\tau \Delta_{1y,b}(\vect{k}'_1))
        \tau \delta \vect{k}_y \\
      + \frac{\hbar^2 \delta\vect{k}_x^2}{2m_x^{\tau,s} (\vect{k}'_1)}
      + \frac{\hbar^2 \delta\vect{k}_y^2}{2m_y^{\tau,s} (\vect{k}'_1)}
      + \frac{\hbar^2 \delta\vect{k}_x \delta\vect{k}_y}
             {2m_{xy}^{\tau,s} (\vect{k}'_1)}
      + \mathcal{O}(\delta\vect{k}^3),
  \end{multline}
  where $E_b$, $\Delta_{0,b}$, $w_{x,b}$, $w_{y,b}$, $\Delta_{1x,b}$, $\Delta_{1y,b}$, $m_x^{\tau,s}$, $m_y^{\tau,s}$, $m_{xy}^{\tau,s}$ are material parameters for band $b$ locally dependent on the BZ point. They can be extracted from experiments, \emph{ab initio} calculations or tight-binding models. In particular, $E_b$ is the energy of band $b$ (ignoring SOC) with respect to the Dirac point of graphene, $\Delta_{0,b}$ is the local spin-splitting, $w_{x,b}$, $w_{y,b}$, $\Delta_{1x,b}$, $\Delta_{1y,b}$ describe the local slope of the band and $m_x^{\tau,s}$, $m_y^{\tau,s}$, $m_{xy}^{\tau,s}$ are the effective masses of the quadratic dispersion. The $k \cdot p$ expansion close to $\vect{k}'_{2,3}$ is obtained from $E_{bs}^\text{tmdc} (\tau\vect{k}'_1 + \delta\vect{k})$ by rotating $\delta\vect{k}$ according to Eq.~\eqref{eqn:tmdsymmetry}. One may write
  \begin{equation}\label{eqn:kdotpsymm}
    E_{bs}^\text{tmdc} (\tau\vect{k}'_j + \delta\vect{k}) = 
      E_{bs}^\text{tmdc} (\tau\vect{k}'_1 + \delta\vect{k}^{-\varphi_j}),
  \end{equation}
  with $\varphi_j=0,2\pi/3,-2\pi/3$ for $j=1,2,3$. We expand the denominator of Eq.~\eqref{eqn:SWOffDiag} with Eqs.~\eqref{eqn:kdotp},~\eqref{eqn:kdotpsymm} and we retain up to the linear terms in $\delta\vect{k}$,
  \begin{equation}\label{eqn:denominator}
    \frac{1}{E_{bs}^\text{tmdc} (\tau\vect{k}'_j + \delta\vect{k})} \approx
      \frac{1}{E_{bs\tau} (\vect{k}'_1)}
      - \frac{\vect{w}_{bs\tau}(\vect{k}'_1) \cdot 
          \tau\delta\vect{k}^{-\varphi_j}}{E_{bs\tau} (\vect{k}'_1)^2},
  \end{equation}
  where $E_{bs\tau} (\vect{k}'_1) = E_b(\vect{k}'_1) + s\tau \Delta_{0,b}(\vect{k}'_1)$ and $\vect{w}_{bs\tau}(\vect{k}'_1) = (w_{x,b}(\vect{k}'_1) + s\tau \Delta_{1x,b}(\vect{k}'_1), w_{y,b}(\vect{k}'_1) + s\tau \Delta_{1y,b}(\vect{k}'_1))^T$. This holds under the condition that $|\vect{w}_{bs\tau}(\vect{k}'_1) \cdot \delta\vect{k}^{-\varphi_j}| \ll E_{bs\tau} (\vect{k}'_1)$ and terms containing higher powers of $\delta\vect{k}$ are therefore negligible. Substituting Eq.~\eqref{eqn:denominator} in Eq.~\eqref{eqn:SWOffDiag} we have
  \begin{equation}
    \delta H_{As, Bs}^{\text{gr},\tau} = A_{bs\tau}
      + B_{bs\tau,x} \tau\delta\vect{k}_x + B_{bs\tau,y} \tau\delta\vect{k}_y,
  \end{equation}
  which is a sum of a $\delta\vect{k}$-independent part,
  \begin{equation}\label{eqn:Abstau}
    A_{bs\tau} = -\sum_{j,b}
      \frac{|t_b(\tau\vect{K}^\theta)|^2 e^{-i\tau\phi_j}}
        {E_{bs\tau} (\vect{k}'_1)},
  \end{equation}
  and a $\delta\vect{k}$-dependent part whose coefficients are given by
  \begin{equation}\label{eqn:Bbstau}
    B_{bs\tau,\xi} =
      \sum_b \frac{|t_b(\tau\vect{K}^\theta)|^2}{E_{bs\tau} (\vect{k}'_1)^2}
      \sum_j e^{-i\tau\phi_j}
        \left ( R(\varphi_j) \vect{w}_{bs\tau}(\vect{k}'_1) \right )_\xi,
  \end{equation}
  for $\xi = x, y$. The two sets of angles $\phi_j$ and $\varphi_j$ have the same values (0, $2\pi/3$, $-2\pi/3$ for $j=1,2,3$), but different origin. The angles $\phi_j$ come from the tunneling matrix elements in Eq.~\eqref{eqn:tunnelingmatrix}, while the angles $\varphi_j$ are connected to the $C_3$ symmetry of the TMDC crystal and they come from Eq.~\eqref{eqn:kdotpsymm}. In order to carry out the sum over index $j$ in Eq.~\eqref{eqn:Bbstau} we compute
  \begin{multline}\label{eqn:BxPMBy}
    B_{bs\tau,x} \pm i B_{bs\tau,y} = \\
      \sum_b \frac{|t_b(\tau\vect{K}^\theta)|^2}{E_{bs\tau} (\vect{k}'_1)^2}
        w_{bs\tau,\pm} (\vect{k}'_1) \sum_j e^{-i\tau\phi_j \pm i\varphi_j},
  \end{multline}
  with $w_{bs\tau,\pm} (\vect{k}'_1) = w_{x,b} (\vect{k}'_1) + s\tau \Delta_{1x,b} (\vect{k}'_1) \pm i (w_{y,b} (\vect{k}'_1) + s\tau \Delta_{1y,b} (\vect{k}'_1))$.
  
  At this point we have again to distinguish between the case of diagonal and off-diagonal elements. For the diagonal elements we have $\phi_j = 0$, therefore $\sum_j e^{-i\tau\phi_j} = 3$ in Eq.~\eqref{eqn:Abstau}, while $\sum_j e^{-i\tau\phi_j \pm i\varphi_j} = \sum_j e^{\pm i\varphi_j} = 0$ in Eq.~\eqref{eqn:BxPMBy} because $e^{i\varphi_j}$ are the complex cube roots of the unity and sum to zero. We have then $B_{bs\tau,x} = B_{bs\tau,y} = 0$. The diagonal elements are therefore $\delta\vect{k}$-independent,
  \begin{equation}
    \delta H_{Xs, Xs}^{\text{gr},\tau} = -3 \sum_b
      \frac{|t_b(\tau\vect{K}^\theta)|^2}
        {E_{bs\tau} (\vect{k}'_1)}.
  \end{equation}
  The off-diagonal elements have instead $\phi_j \in \{0, 2\pi/3, -2\pi/3\}$ and consequently $A_{bs\tau} = 0$. Looking at Eq.~\eqref{eqn:BxPMBy}, the sum $\sum_j e^{-i\tau\phi_j + i\varphi_j}$ is equal to 3 for $\tau=+$ and it is equal to 0 for $\tau=-$. On the other hand $\sum_j e^{-i\tau\phi_j - i\varphi_j} = 0$ for $\tau=+$ and is equal to 3 for $\tau=-$. We conclude then that $B_{bs\tau,x} - i\tau B_{bs\tau,y} = 0$ and $B_{bs\tau,y} = -i\tau B_{bs\tau,x}$, while $B_{bs\tau,x} + i\tau B_{bs\tau,y} = 2 B_{bs\tau,x}$. Therefore
  \begin{equation}
    B_{bs\tau,x} = \frac{3}{2}
      \sum_b \frac{|t_b(\tau\vect{K}^\theta)|^2}{E_{bs\tau} (\vect{k}'_1)^2}
        w_{bs\tau,\tau} (\vect{k}'_1)
  \end{equation}
  and $\delta H_{As, Bs}^{\text{gr},\tau} = B_{bs\tau,x} \tau\delta\vect{k}_x + B_{bs\tau,y} \tau\delta\vect{k}_y = B_{bs\tau,x} (\tau\delta\vect{k}_x - i \delta\vect{k}_y)$ as reported in Eq.~\eqref{eqn:offdiagonal}.
  
  \section{Estimation of $t_\parallel$ and $t_\perp$}\label{sec:estimation}
  
  \begin{figure}
    \centering
    \includegraphics[width=\columnwidth]{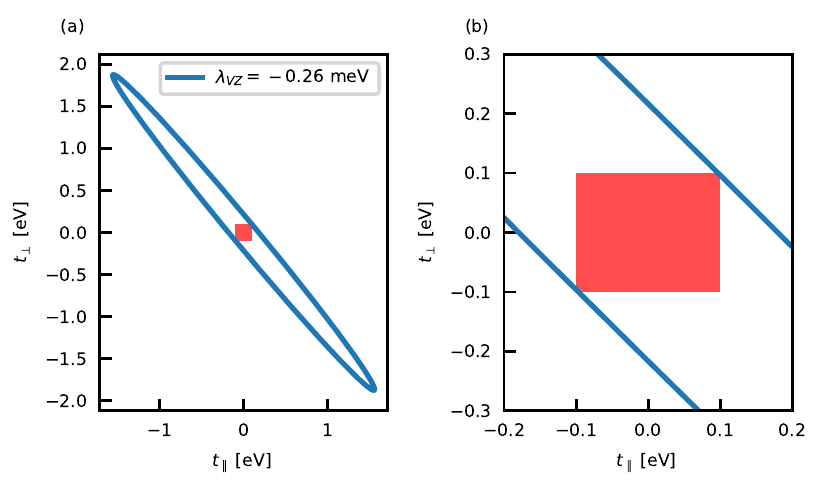}
    \caption{Estimation of $t_\parallel$ and $t_\perp$. \emph{(a)}. The blue ellipse indicates the possible values of $t_\parallel$ and $t_\perp$ that give a valley Zeeman spin-orbit strength of $-0.26$ meV at $\theta = 0^\circ$ for a corresponding value of $f_G = 0.95$. \emph{(b)}. Magnification of (a). The red rectangle indicates the window of values where $|t_\parallel|, |t_\perp| \le 100$ meV.}
    \label{fig:estimate}
  \end{figure}
  
  According to Ref.~\citenum{bistritzer_moire_2011} the value of $t_\perp$ for bilayer graphene is 110 meV. We expect $t_\perp$ for graphene/TMDC bilayers to be of the same order of magnitude because the distance between graphene and the closest chalcogen layer is $d_\perp = 3.4$ Å \cite{pierucci_band_2016} and happens to be equal to the distance reported between graphene layers \cite{castro_neto_electronic_2009}. For further comparison and in order to obtain the relative value of $t_\parallel$, we look at DFT calculations for graphene/TMDC heterostructures. Ref.~\citenum{gmitra_trivial_2016} reports an induced valley Zeeman spin-orbit splitting in graphene of $-0.26$ meV from the MoS$_2$ TMDC compound. This does not reveal immediately the values of $t_\parallel$ and $t_\perp$, but we can extract information about them using Eq.~\eqref{eqn:lambdavalleyzeeman}. Substituting Eq.~\eqref{eqn:bandtunn} in Eq.~\eqref{eqn:lambdavalleyzeeman}, we expand the dependence of $|t_b|^2$ in $t_\parallel$ and $t_\perp$,
  \begin{equation}\label{eqn:lambdaVZdependence}
    \lambda_\text{VZ} = \alpha t_\parallel^2 + \beta t_\perp^2 + 2\gamma t_\parallel t_\perp,
  \end{equation}
  where
  \begin{equation}
    \begin{gathered}
      \alpha = 3 \sum_b \frac{\widetilde{\alpha} \Delta_{0,b} (\vect{k}'_1)}{E_b^2 (\vect{k}'_1) - \Delta_{0,b}^2 (\vect{k}'_1)}, \\
      \beta = 3 \sum_b \frac{\widetilde{\beta} \Delta_{0,b} (\vect{k}'_1)}{E_b^2 (\vect{k}'_1) - \Delta_{0,b}^2 (\vect{k}'_1)}, \\
      \gamma = 3 \sum_b \frac{\widetilde{\gamma} \Delta_{0,b} (\vect{k}'_1)}{E_b^2 (\vect{k}'_1) - \Delta_{0,b}^2 (\vect{k}'_1)}. \\
    \end{gathered}
  \end{equation}
  and
  \begin{equation}
    \begin{gathered}
      \widetilde{\alpha} = |c_{bx} (\tau\vect{k}'_1) \cos\theta + c_{by} (\tau\vect{k}'_1) \sin\theta|^2, \\
      \widetilde{\beta} = |c_{bz} (\tau\vect{k}'_1)|^2, \\
      \widetilde{\gamma} = -\mathrm{Im}[(c_{bx} (\tau\vect{k}'_1) \cos\theta + c_{by} (\tau\vect{k}'_1) \sin\theta) c_{bz}^* (\tau\vect{k}'_1)].
    \end{gathered}
  \end{equation}
  We see that $\alpha$, $\beta$ and $\gamma$ depend on the orbital amplitudes $c_{b,x,y,z} (\tau\vect{k}'_1)$, the band dispersion $E_b (\vect{k}'_1)$ and the spin splitting $\Delta_{0,b} (\vect{k}'_1)$ which are intrinsic properties of the isolated TMDC layer and therefore can be readily calculated using the TB model of Ref.~\citenum{fang_ab_2015}. The only missing external parameter is the value of $f_G$ which defines the distance of $E_b (\vect{k}'_1)$ from the Dirac point. From Ref.~\citenum{gmitra_trivial_2016}, the Dirac point is very close to the conduction band of the TMDC and we set $f_G = 0.95$, meaning that the Dirac point of graphene has an energy distance from the TMDC conduction band edge equal to 5\% of the TMDC band gap. We plug the resulting $\alpha$, $\beta$, $\gamma$ and the value of $\lambda_\text{VZ} = -0.26$ meV in Eq.~\eqref{eqn:lambdaVZdependence} and the solutions for $t_\parallel$ and $t_\perp$ form an ellipse in the $(t_\parallel, t_\perp)$-plane (see Fig.~\ref{fig:estimate}). This ellipse is elongated and inclined by an angle of $\sim -40^\circ$. In principle all the points $(t_\parallel, t_\perp)$ on this ellipse give $\lambda_\text{VZ} = -0.26$ meV, but some values are unphysically large. Zooming closely to the center, see Fig.~\ref{fig:estimate}(b), the ellipse touches the point $(t_\parallel, t_\perp) = (100, 100)$ meV. Since this is the order of magnitude that we expect, we estimate $t_\parallel \approx t_\perp \approx 100$ meV.
  
  \section{Rashba type induced spin-orbit coupling}\label{sec:RashbaSOC}
  
  In this section we will show that the induced Rashba-like SOC in graphene can be understood by taking into account spin-flip processes between even ($e$) and odd ($o$) bands of the TMDC. The energy bands of monolayer TMDCs can be classified as $e$ or $o$ under $\sigma_h$, which is the reflection with respect to the horizontal mirror plane of the TMDC.

  Consider the following term in the effective low energy Hamiltonian of graphene that can be obtained in third order perturbation theory \cite{winkler_spin--orbit_2003},
  \begin{multline}\label{eqn:Rashba-gen-appendix}
     (\delta H^{\text{gr},\tau}_R)_{Xs,X's'} = \\
     \sum_{j,b,b',s'',s'''}
      \frac{(T_{\tau\vect{k}'_j})_{Xs,bs''}
      (H_\text{soc})_{bs'',b's'''}
        (T_{\tau\vect{k}'_j}^\dagger)_{b's''',X's'}}
        {[E_D^\text{gr} - E_{b}^\text{tmdc}(\tau\vect{k}'_j)][E_D^\text{gr} - E_{b'}^\text{tmdc}(\tau\vect{k}'_j)]}.
  \end{multline}
  Here $b \neq b'$ are band indices, and in the denominator we have neglected the dependence of the TMDC band energies $E_{b}^\text{tmdc}(\tau\vect{k}'_j)$ on the intrinsic SOC (c.f., Eq.~\eqref{eqn:deltaHgr}) because it would lead to higher order effects. Here, $(H_\text{soc})_{bs'',b's'''}$ are matrix elements of the SOC operator
  \begin{equation}
    \hat{H}_\text{soc} = \gamma_d \hat{\vect{L}} \cdot \hat{\vect{S}} =
      \gamma_d \left ( \hat{L}_z \hat{S}_z + \frac{1}{2}
        (\hat{L}_{+} \hat{S}_{-} + \hat{L}_{-} \hat{S}_{+}) \right ),
  \end{equation}
  which are non-zero only between $e$ and $o$ bands of the TMDC. Moreover $\gamma_d$ is the atomic SOC strength of the metal atoms' $d$ orbitals, $\hat{L}_{\pm}= \hat{L}_{x}\pm i \hat{L}_{y}$, $\hat{L}_z$ are angular momentum operators and $\hat{\vect{S}} = (\hat{S}_x, \hat{S}_y, \hat{S}_z)^T$, $\hat{S}_{\pm} = \hat{S}_{x} \pm i \hat{S}_{y}$ are spin operators, i.e. $\hat{\vect{S}} = (\hbar/2) \vect{s}$, where $\vect{s} = (s_x, s_y, s_z)^T$ are Pauli matrices. In order to show that Eq.~\eqref{eqn:Rashba-gen-appendix} describes Rashba-like induced SOC, we focus, as a first step, on the matrix element between an even ($b = e$) and an odd ($b' = o$) band. At a general point $\mathbf{k}'$ of the BZ the Bloch wavefunction of these bands can be written as 
  \begin{subequations}
  \begin{multline}
    \ket{e, \vect{k}'} = c_{e,x^2-y^2} (\vect{k}') \ket{d_{x^2-y^2}, \vect{k}'}
      + c_{e,xy} (\vect{k}') \ket{d_{xy}, \vect{k}'} \\
      + c_{e,z^2} (\vect{k}') \ket{d_{z^2}, \vect{k}'},
  \end{multline}
  \begin{equation}
    \ket{o, \vect{k}'} = c_{o,xz} (\vect{k}') \ket{d_{xz}, \vect{k}'}
      + c_{o,yz} (\vect{k}') \ket{d_{yz}, \vect{k}'},
  \end{equation}
  \label{eq:Psi-even-odd}
  \end{subequations}
  where $\ket{d_\mu, \vect{k}'}$ are the usual Bloch wavefunctions formed using the $d$ atomic orbitals of the metal atoms, $\mu \in \{x^2-y^2, xy, z^2, xz, yz\}$, and $c_{e\,(o),\mu} (\vect{k}')$ are complex amplitudes giving the weight of each type of atomic orbital at a given k-space point. Other Bloch wavefunctions formed from the atomic orbitals $\{p_z, p_x, p_{y}\}$ of the chalcogen atoms have also finite weight in $\ket{e\,(o), \vect{k}'}$ and as argued in previous sections, they are crucial to understand band-to-band tunneling. However, they are less important in the calculation of interband SOC matrix elements and therefore we do not take them into account explicitly in Eq.~\eqref{eq:Psi-even-odd}. The inter-band spin matrices of $\hat{H}_\text{soc}$ between these $e$ and $o$ bands can be written as 
  \begin{multline}
    [H_\text{soc} (\vect{k}')]_{e,o} =
      \bra{e, \vect{k}'} \hat{H}_\text{soc} \ket{o, \vect{k}'} \\
    = i \gamma_d \left [ \alpha_{e,o}^{(x)}(\vect{k}') \hat{S}_x
      + \alpha_{e,o}^{(y)}(\vect{k}') \hat{S}_y \right ],
  \label{eq:SOC-off-diag}
  \end{multline}
  where $\alpha_{e,o}^{(x)} = (c_{e,x^2-y^2})^{*} c_{o,yz} - (c_{e,xy})^{*} c_{o,xz} + \sqrt{3} (c_{e,z^2})^{*} c_{o,yz}$ and $\alpha_{e,o}^{(y)} = (c_{e,x^2-y^2})^{*} c_{o,xz} + (c_{e,xy})^{*} c_{o,yz} - \sqrt{3} (c_{e,z^2})^{*} c_{o,xz}$ (for simplicity, we have suppressed the dependence of $\alpha_{e,o}^{(x,y)}$ on $\vect{k}'$, which will be restored later). Eq.~\eqref{eq:SOC-off-diag} can be easily obtained by taking into account Table~\ref{tbl:SOC-matrix}. Note that $(H_\text{soc})_{e,o}$ in Eq.~\eqref{eq:SOC-off-diag} has only off-diagonal non-zero elements in spin-space $\uparrow$, $\downarrow$, i.e., it describes spin-flip processes between the two bands. The term that would be $\sim \hat{S}_z$ vanishes between $e$ and $o$ bands by symmetry.
  
  \begin{table}[ht]
  \begin{tabular}{ccc}
  \hline  
   \hline
   Orbital        &   $d_{xz}$  & $d_{yz}$ \\
   \hline
   $d_{z^{2}}$    &   $-i \sqrt{3} \hat{S}_y$   &    $i \sqrt{3} \hat{S}_x$  \\
   $d_{xy}$       &   $-i \hat{S}_x$  & $i \hat{S}_y$ \\
   $d_{x^2-y^2}$  &    $i \hat{S}_y$       & $i \hat{S}_x$\\
    \hline
    \hline
  \end{tabular}
  \caption{Matrix elements of the SOC operator in the basis of $\{d_{x^2-y^2}, d_{xy}, d_{z^2}, d_{xz}, d_{yz}\}$ atomic orbitals.}
  \label{tbl:SOC-matrix}
  \end{table}
  
  As one can see from Eq.~\eqref{eqn:Rashba-gen-appendix}, one needs to calculate $(H_\text{soc})_{e s'',o s'''}$ at the three $\vect{k}'_j$ BZ points of the TMDC defined in Eq.~\eqref{eqn:backfolded} that satisfy the quasimomentum conservation for interlayer tunneling. These points are related to each other by a $2\pi/3$ rotation. Following Ref.~\citenum{dresselhaus_group_2010}, we may write $\ket{e\,(o), R_{\pm 2\pi/3}\vect{k}'_1} = R_{\pm 2\pi/3}\ket{e\,(o), \vect{k}'_1}$, where $R_{\pm 2\pi/3}$ denotes rotation by $\pm 2\pi/3$. Therefore, given $\bra{e, \vect{k}'_1} \hat{H}_\text{soc} \ket{o, \vect{k}'_1}$, one needs to evaluate
  \begin{subequations}
    \begin{multline}
    \bra{e, R_{2\pi/3} \vect{k}'_1} \hat{H}_\text{soc}
      \ket{o, R_{2\pi/3} \vect{k}'_1} = \\
    \bra{e, \vect{k}'_1}
      (R_{2\pi/3})^\dagger \, \hat{H}_\text{soc} \, R_{2\pi/3}
      \ket{o, \vect{k}'_1},
    \end{multline}
    \begin{multline}
    \bra{e, R_{-2\pi/3} \vect{k}'_1} \hat{H}_\text{soc}
      \ket{o, R_{-2\pi/3} \vect{k}'_1} = \\
    \bra{e, \vect{k}'_1}
      (R_{-2\pi/3})^\dagger \, \hat{H}_\text{soc} \, R_{-2\pi/3}
      \ket{o, \vect{k}'_1},
    \end{multline}
    \label{eq:rotated-matrix-elements}
  \end{subequations}%
  This means that the necessary matrix elements can be calculated using $\ket{e, \vect{k}'_1}$ and $\ket{o, \vect{k}'_1}$ and a rotated $\hat{H}_\text{soc}$. The transformed operators $(R_{\pm 2\pi/3})^{\dagger} \hat{H}_\text{soc} R_{\pm 2\pi/3}$ can be easily calculated by noticing that
  \begin{subequations}
    \begin{gather}
      (R_{\pm 2\pi/3})^{\dagger} \, \hat{L}_z \, R_{\pm 2\pi/3} = \hat{L}_z, \\
      R_{2\pi/3} \, \hat{L}_{\pm} \, (R_{2\pi/3})^{\dagger} =
        e^{\mp i 2\pi/3}\hat{L}_{\pm}, \\
      R_{-2\pi/3} \, \hat{L}_{\pm} \, (R_{-2\pi/3})^{\dagger} =
        e^{\pm i 2\pi/3}\hat{L}_{\pm}.
    \end{gather}
  \end{subequations}
  Let us define the vectors $\vect{n}_{e,o} (\vect{k}'_1) = (\alpha_{e,o}^{(x)} (\vect{k}'_1), \alpha_{e,o}^{(y)}(\vect{k}'_1))^{T}$, 
$\vect{S} = (S_x, S_y)^{T}$. Then one finds that 
  \begin{subequations}
    \begin{align}
      [H_\text{soc} (\vect{k}'_1)]_{e,o} & =
        i \gamma_d \, \vect{n}_{e,o} (\vect{k}'_1) \cdot \vect{S}, \\ 
      [H_\text{soc} (R_{2\pi/3} \vect{k}'_1)]_{e,o} & =
        i \gamma_d (R_{2\pi/3} \vect{n}_{e,o} (\vect{k}'_1))
          \cdot \vect{S}, \\
      [H_\text{soc} (R_{-2\pi/3} \vect{k}'_1)]_{e,o} & =
        i \gamma_d (R_{-2\pi/3} \vect{n}_{e,o} (\vect{k}'_1))
          \cdot \vect{S}.
    \end{align}
  \label{eqn:H_soc-kj}
  \end{subequations}
  Note that $\vect{n}_{e,o} (\vect{k}'_1)$ in Eqs.~\eqref{eqn:H_soc-kj} is in general a complex vector because the weights $c_{e\,(o),\mu} (\vect{k}'_1)$ of the atomic orbitals in band $e$ ($o$) can be complex.
  
  We can compute now the contribution to $\delta H_R^{\text{gr},\tau}$ from the interaction of two bands of the TMDC (e.g., the conduction band which is $e$  and the first $o$ band above the conduction band). Then the indices $b$ and $b'$ in Eq.~\eqref{eqn:Rashba-gen-appendix} can take the values $(b, b')=(e, o)$ and $(b, b') = (o, e)$. For simplicity we focus on the Dirac point $\vect{K}$, i.e., $\tau=1$. Note that the energy differences $(E^\text{gr}_{D}-E_{b}^\text{tmdc}(\vect{k}'_j))$ and $(E^\text{gr}_{D}-E_{b'}^\text{tmdc}(\vect{k}'_j))$ appearing in Eq.~\eqref{eqn:Rashba-gen-appendix} are equal for all $\vect{k}'_j$ because of the threefold rotational ($C_3$) symmetry of the TMDC. Therefore the corresponding factor can be pulled out of the sum in  Eq.~\eqref{eqn:Rashba-gen-appendix}. Using Eq.~\eqref{eqn:tunnelingmatrix} one may write explicitly
  \begin{widetext}
    \begin{multline}
      \delta H^\text{gr}_{R} =
      \frac{1}{
        \left ( E^\text{gr}_{D} - E_{e}^\text{tmdc} (\vect{k}'_{1}) \right ) 
        \left ( E^\text{gr}_{D} - E_{o}^\text{tmdc} (\vect{k}'_{1})\right)}
      \left[
        \left(
        \begin{array}{cc}
          1 & 1 \\
          1 & 1
        \end{array}
        \right)
        \otimes
        \left[
            {T}_{e,o}^{} [H_\text{soc}^{}(\vect{k}'_1)]_{e,o}
          + {T}_{o,e}^{} [H_\text{soc}^{}(\vect{k}'_1)]_{o,e}
        \right]
      \right . \\
      +
      \left.
        \left(
        \begin{array}{cc}
          1 & e^{-2i \pi/3} \\
          e^{2i \pi/3} & 1
        \end{array}
        \right)
        \otimes
  \left[{T}_{e,o}^{} [H_\text{soc}^{}( R_{2\pi/3}\vect{k}'_1)]_{e,o}+{T}_{o,e}^{} [H_\text{soc}^{}( R_{2\pi/3}\vect{k}'_1)]_{o,e}\right]\right.\\
   + 
   \left.
  \left(
   \begin{array}{cc}
   1 & e^{2i \pi/3} \\
   e^{-2i \pi/3} & 1
   \end{array}
  \right)
  \otimes \left[{T}_{e,o}^{} [H_\text{soc}^{}(R_{-2\pi/3}\vect{k}'_1)]_{e,o}+{T}_{o,e}^{} [H_\text{soc}^{}(R_{-2\pi/3}\vect{k}'_1)]_{o,e}\right]\right].
  \label{eqn:Gr-Rashba-1}
  \end{multline}
  \end{widetext}
  Here ${T}_{e,o}^{}=t_e(\vect{K}^{\theta}) t_o^{*}(\vect{K}^{\theta})$ where $t_b(\vect{K}^{\theta})$ is given in Eq.~\eqref{eqn:bandtunn},  
 ${T}_{o,e}^{}={T}_{e,o}^{*}$ and  $[H_\text{soc}^{}(\vect{k}')]_{o,e}=[H_\text{soc}^{}(\vect{k}')]_{e,o}^{\dagger}$.
Let us write ${T}_{e,o}^{}=|{T}_{e,o}^{}|e^{i\eta}$, then using Eqs.~\eqref{eqn:H_soc-kj}
\begin{widetext}
 \begin{subequations}
\begin{align}
{T}_{e,o}^{} \left [ H_\text{soc}^{}(\vect{k}'_1)\right ]_{e,o}+{T}_{e,o}^{*}\left [ H_\text{soc}^{}(\vect{k}'_1)\right ]_{e,o}^{\dagger} & = 
 i \gamma_d |{T}_{e,o}^{}|(e^{i\eta} \vect{n}_{e,o}(\vect{k}'_1)-e^{-i\eta}(\vect{n}_{e,o})^{*}(\vect{k}'_1)) \cdot\vect{S} \nonumber \\
  & = -2  \gamma_d |{T}_{e,o}^{}| \left(\mathrm{Im} \left [ e^{i\eta} \vect{n}_{e,o}(\vect{k}'_1) \right ] \right ) \cdot \vect{S} \nonumber \\
  & = 2 i \gamma_d |{T}_{e,o}^{}|
  \left(
    \begin{array}{cc}
     0 & \Lambda_2(\vect{k}'_1) \\
    \Lambda_1(\vect{k}'_1) & 0
   \end{array}
  \right),
\end{align}
\begin{align}
{T}_{e,o}^{} \left [ H_\text{soc}^{}(R_{2\pi/3}\vect{k}'_1)\right ]_{e,o}+ 
{T}_{e,o}^{*} \left [ H_\text{soc}^{} (R_{2\pi/3}\vect{k}'_1)\right ]_{e,o}^{\dagger} & = 
  -2  \gamma_d |{T}_{e,o}^{}| \left ( R_{2\pi/3} \, \mathrm{Im} \left [ e^{i\eta} \vect{n}_{e,o}(\vect{k}'_1) \right ] \right ) \cdot \vect{S} \nonumber \\
    & = 2 i \gamma_d  |{T}_{e,o}^{}| \left(
   \begin{array}{cc}
     0 & e^{-2i\pi/3}\Lambda_2(\vect{k}'_1) \\
    e^{2i\pi/3}\Lambda_1(\vect{k}'_1) & 0
   \end{array}
  \right),
\end{align}
\begin{align}
 {T}_{e,o}^{} \left [ H_\text{soc}^{}(R_{-2\pi/3}\vect{k}'_1)\right ]_{e,o}+
 {T}_{e,o}^{*}  \left [ H_\text{soc}^{}(R_{-2\pi/3}\vect{k}'_1)\right ]_{e,o}^{\dagger} & = 
  -2  \gamma_d |{T}_{e,o}^{}| \left ( R_{-2\pi/3} \, \mathrm{Im} \left [ e^{i\eta} \vect{n}_{e,o}(\vect{k}'_1) \right ] \right )\cdot\vect{S} \nonumber \\
   & = 2 i \gamma_d |{T}_{e,o}^{}| \left(
    \begin{array}{cc}
     0 & e^{2i\pi/3}\Lambda_2(\vect{k}'_1) \\
    e^{-2i\pi/3}\Lambda_1(\vect{k}'_1) & 0
   \end{array}
  \right).
\end{align}
\label{eqn:H_soc-H_soc-dag}
\end{subequations}
\end{widetext}
Here $\Lambda_1 (\vect{k}'_1) = -\mathrm{Im}[e^{i\eta} \alpha_{e,o}^{(y)}(\vect{k}'_1)]+ i \mathrm{Im}[e^{i\eta} \alpha_{e,o}^{(x)}(\vect{k}'_1)]$ and 
$\Lambda_2(\vect{k}'_1)=\mathrm{Im}[e^{i\eta} \alpha_{e,o}^{(y)}(\vect{k}'_1)]+ i \mathrm{Im}[e^{i\eta} \alpha_{e,o}^{(x)}(\vect{k}'_1)]$. 
Note that one can write 
$\Lambda_1(\vect{k}'_1)=|\Lambda_1(\vect{k}'_1)|e^{i \vartheta(\vect{k}'_1)}$ and 
$\Lambda_2(\vect{k}'_1)=-|\Lambda_1(\vect{k}'_1)|e^{-i \vartheta(\vect{k}'_1)}$ where 
$\vartheta(\vect{k}'_1)=\mathrm{Arg}[\Lambda_1(\vect{k}'_1)]$.
Substituting now Eqs.~\eqref{eqn:H_soc-H_soc-dag} into Eq.~\eqref{eqn:Gr-Rashba-1} one finds
\begin{align}
\label{eqn:Gr-Rashba-2}
 \delta H_{R}^{gr}=
 \left(
 \begin{array}{cccc}
  0 & 0 & 0 & 0 \\
  0 & 0 & i\lambda_R(\vect{k}'_1) e^{i\vartheta(\vect{k}'_1)} & 0 \\
  0 & -i\lambda_R(\vect{k}'_1) e^{-i\vartheta(\vect{k}'_1)} & 0 & 0 \\
  0 & 0 & 0 & 0
 \end{array}
\right),
\end{align}
where
\begin{equation}
 \lambda_R(\vect{k}'_1)=\frac{6 \gamma_d |{T}_{e,o}^{}(\vect{k}'_1)| |\Lambda_1(\vect{k}'_1)|}
 {\left(E^\text{gr}_{D}-E_{e}^\text{tmdc}(\vect{k}'_1)\right) 
 \left(E^\text{gr}_{D}-E_{o}^\text{tmdc}(\vect{k}'_1)\right)}.
 \label{eqn:Rashba-strength-appendix}
\end{equation}
Eq.~\eqref{eqn:Rashba-strength-appendix} is the strength of the Rashba type SOC induced in graphene by each pair of $e$ and $o$ bands. As Eq.~\eqref{eqn:Rashba-gen-appendix} shows, in order to calculate the total SOC coupling $\lambda_R(\vect{k}'_1)$ one needs to sum up the contributions coming from all pairs of even and odd bands with the correct phase factors shown in Eq.~\eqref{eqn:Gr-Rashba-2}. A similar result to Eq.~\eqref{eqn:Gr-Rashba-2} can be obtained in an analogous way for the opposite Dirac point $-\vect{K}$.
  
  We conclude this Appendix commenting the technique used to produce Fig.~\ref{fig:induced-rashba}, which plots Eq.~\eqref{eqn:Rashba-strength-appendix} for three different pairs of $e$ and $o$ bands and their total sum. In Eq.~\eqref{eqn:Rashba-strength-appendix}, $\Lambda_1 (\vect{k}'_1)$ contains the SOC matrix elements $\alpha_{e,o}^{(x)}$ and $\alpha_{e,o}^{(y)}$ that we obtained with the TB model of Ref.~\citenum{fang_ab_2015}. These matrix elements are computed separately for each point of the TMDC BZ, but this procedure leads to several phase jumps of $\pm\pi$ in their complex value across the entire BZ. This indeed hinders the computation of $\lambda_R (\vect{k}'_1)$. We were able to partially smooth the phases of these matrix elements with the help of the NumPy function \texttt{unwrap} \cite{oliphant_guide_2006}. This function is designed to work on one dimensional data and its generalization to two dimensional arrays, as we would need in this case, is non-trivial. Nevertheless the result is satisfactory between twist angles $\theta = 0^\circ$ and $\theta = 30^\circ$. Instead, between $\theta = 30^\circ$ and $\theta = 60^\circ$ the surviving phase jumps cause the values of $\lambda_R$ to also change abruptly. Analysing Eq.~\eqref{eqn:Rashba-strength-appendix} one notices that the values of $\lambda_R$ for $\theta \in [0^\circ, 30^\circ]$ must be equal to those for $60^\circ - \theta$. This comes from the fact that the tunneling $|T_{e,o}|$, the TMDC band dispersion and the SOC matrix elements in $\Lambda_1$ have this same symmetry (see Fig.~\ref{fig:valleyzeemanone}(f) and Ref.~\citenum{kormanyos_k_2015}). Therefore, in Fig.~\ref{fig:induced-rashba} we have used for $\theta \in [30^\circ, 60^\circ]$ the same values of $\lambda_R$ as for $\theta \in [0^\circ, 30^\circ]$ but mirrored with respect to $\theta = 30^\circ$.
  
  \bibliography{bibliography}

\begin{thebibliography}{59}%
\makeatletter
\providecommand \@ifxundefined [1]{%
 \@ifx{#1\undefined}
}%
\providecommand \@ifnum [1]{%
 \ifnum #1\expandafter \@firstoftwo
 \else \expandafter \@secondoftwo
 \fi
}%
\providecommand \@ifx [1]{%
 \ifx #1\expandafter \@firstoftwo
 \else \expandafter \@secondoftwo
 \fi
}%
\providecommand \natexlab [1]{#1}%
\providecommand \enquote  [1]{``#1''}%
\providecommand \bibnamefont  [1]{#1}%
\providecommand \bibfnamefont [1]{#1}%
\providecommand \citenamefont [1]{#1}%
\providecommand \href@noop [0]{\@secondoftwo}%
\providecommand \href [0]{\begingroup \@sanitize@url \@href}%
\providecommand \@href[1]{\@@startlink{#1}\@@href}%
\providecommand \@@href[1]{\endgroup#1\@@endlink}%
\providecommand \@sanitize@url [0]{\catcode `\\12\catcode `\$12\catcode
  `\&12\catcode `\#12\catcode `\^12\catcode `\_12\catcode `\%12\relax}%
\providecommand \@@startlink[1]{}%
\providecommand \@@endlink[0]{}%
\providecommand \url  [0]{\begingroup\@sanitize@url \@url }%
\providecommand \@url [1]{\endgroup\@href {#1}{\urlprefix }}%
\providecommand \urlprefix  [0]{URL }%
\providecommand \Eprint [0]{\href }%
\providecommand \doibase [0]{http://dx.doi.org/}%
\providecommand \selectlanguage [0]{\@gobble}%
\providecommand \bibinfo  [0]{\@secondoftwo}%
\providecommand \bibfield  [0]{\@secondoftwo}%
\providecommand \translation [1]{[#1]}%
\providecommand \BibitemOpen [0]{}%
\providecommand \bibitemStop [0]{}%
\providecommand \bibitemNoStop [0]{.\EOS\space}%
\providecommand \EOS [0]{\spacefactor3000\relax}%
\providecommand \BibitemShut  [1]{\csname bibitem#1\endcsname}%
\let\auto@bib@innerbib\@empty
\bibitem [{\citenamefont {Novoselov}\ \emph {et~al.}(2004)\citenamefont
  {Novoselov}, \citenamefont {Geim}, \citenamefont {Morozov}, \citenamefont
  {Jiang}, \citenamefont {Zhang}, \citenamefont {Dubonos}, \citenamefont
  {Grigorieva},\ and\ \citenamefont {Firsov}}]{novoselov_electric_2004}%
  \BibitemOpen
  \bibfield  {author} {\bibinfo {author} {\bibfnamefont {K.~S.}\ \bibnamefont
  {Novoselov}}, \bibinfo {author} {\bibfnamefont {A.~K.}\ \bibnamefont {Geim}},
  \bibinfo {author} {\bibfnamefont {S.~V.}\ \bibnamefont {Morozov}}, \bibinfo
  {author} {\bibfnamefont {D.}~\bibnamefont {Jiang}}, \bibinfo {author}
  {\bibfnamefont {Y.}~\bibnamefont {Zhang}}, \bibinfo {author} {\bibfnamefont
  {S.~V.}\ \bibnamefont {Dubonos}}, \bibinfo {author} {\bibfnamefont {I.~V.}\
  \bibnamefont {Grigorieva}}, \ and\ \bibinfo {author} {\bibfnamefont {A.~A.}\
  \bibnamefont {Firsov}},\ }\href {\doibase 10.1126/science.1102896} {\bibfield
   {journal} {\bibinfo  {journal} {Science}\ }\textbf {\bibinfo {volume}
  {306}},\ \bibinfo {pages} {666} (\bibinfo {year} {2004})}\BibitemShut
  {NoStop}%
\bibitem [{\citenamefont {Novoselov}\ \emph {et~al.}(2005)\citenamefont
  {Novoselov}, \citenamefont {Geim}, \citenamefont {Morozov}, \citenamefont
  {Jiang}, \citenamefont {Katsnelson}, \citenamefont {Grigorieva},
  \citenamefont {Dubonos},\ and\ \citenamefont
  {Firsov}}]{novoselov_two-dimensional_2005}%
  \BibitemOpen
  \bibfield  {author} {\bibinfo {author} {\bibfnamefont {K.~S.}\ \bibnamefont
  {Novoselov}}, \bibinfo {author} {\bibfnamefont {A.~K.}\ \bibnamefont {Geim}},
  \bibinfo {author} {\bibfnamefont {S.~V.}\ \bibnamefont {Morozov}}, \bibinfo
  {author} {\bibfnamefont {D.}~\bibnamefont {Jiang}}, \bibinfo {author}
  {\bibfnamefont {M.~I.}\ \bibnamefont {Katsnelson}}, \bibinfo {author}
  {\bibfnamefont {I.~V.}\ \bibnamefont {Grigorieva}}, \bibinfo {author}
  {\bibfnamefont {S.~V.}\ \bibnamefont {Dubonos}}, \ and\ \bibinfo {author}
  {\bibfnamefont {A.~A.}\ \bibnamefont {Firsov}},\ }\href {\doibase
  10.1038/nature04233} {\bibfield  {journal} {\bibinfo  {journal} {Nature}\
  }\textbf {\bibinfo {volume} {438}},\ \bibinfo {pages} {197} (\bibinfo {year}
  {2005})}\BibitemShut {NoStop}%
\bibitem [{\citenamefont {Castro~Neto}\ \emph {et~al.}(2009)\citenamefont
  {Castro~Neto}, \citenamefont {Guinea}, \citenamefont {Peres}, \citenamefont
  {Novoselov},\ and\ \citenamefont {Geim}}]{castro_neto_electronic_2009}%
  \BibitemOpen
  \bibfield  {author} {\bibinfo {author} {\bibfnamefont {A.~H.}\ \bibnamefont
  {Castro~Neto}}, \bibinfo {author} {\bibfnamefont {F.}~\bibnamefont {Guinea}},
  \bibinfo {author} {\bibfnamefont {N.~M.~R.}\ \bibnamefont {Peres}}, \bibinfo
  {author} {\bibfnamefont {K.~S.}\ \bibnamefont {Novoselov}}, \ and\ \bibinfo
  {author} {\bibfnamefont {A.~K.}\ \bibnamefont {Geim}},\ }\href {\doibase
  10.1103/RevModPhys.81.109} {\bibfield  {journal} {\bibinfo  {journal} {Rev.
  Mod. Phys.}\ }\textbf {\bibinfo {volume} {81}},\ \bibinfo {pages} {109}
  (\bibinfo {year} {2009})}\BibitemShut {NoStop}%
\bibitem [{\citenamefont {Drögeler}\ \emph {et~al.}(2016)\citenamefont
  {Drögeler}, \citenamefont {Franzen}, \citenamefont {Volmer}, \citenamefont
  {Pohlmann}, \citenamefont {Banszerus}, \citenamefont {Wolter}, \citenamefont
  {Watanabe}, \citenamefont {Taniguchi}, \citenamefont {Stampfer},\ and\
  \citenamefont {Beschoten}}]{drogeler_spin_2016}%
  \BibitemOpen
  \bibfield  {author} {\bibinfo {author} {\bibfnamefont {M.}~\bibnamefont
  {Drögeler}}, \bibinfo {author} {\bibfnamefont {C.}~\bibnamefont {Franzen}},
  \bibinfo {author} {\bibfnamefont {F.}~\bibnamefont {Volmer}}, \bibinfo
  {author} {\bibfnamefont {T.}~\bibnamefont {Pohlmann}}, \bibinfo {author}
  {\bibfnamefont {L.}~\bibnamefont {Banszerus}}, \bibinfo {author}
  {\bibfnamefont {M.}~\bibnamefont {Wolter}}, \bibinfo {author} {\bibfnamefont
  {K.}~\bibnamefont {Watanabe}}, \bibinfo {author} {\bibfnamefont
  {T.}~\bibnamefont {Taniguchi}}, \bibinfo {author} {\bibfnamefont
  {C.}~\bibnamefont {Stampfer}}, \ and\ \bibinfo {author} {\bibfnamefont
  {B.}~\bibnamefont {Beschoten}},\ }\href {\doibase
  10.1021/acs.nanolett.6b00497} {\bibfield  {journal} {\bibinfo  {journal}
  {Nano Lett.}\ }\textbf {\bibinfo {volume} {16}},\ \bibinfo {pages} {3533}
  (\bibinfo {year} {2016})}\BibitemShut {NoStop}%
\bibitem [{\citenamefont {Singh}\ \emph {et~al.}(2016)\citenamefont {Singh},
  \citenamefont {Katoch}, \citenamefont {Xu}, \citenamefont {Tan},
  \citenamefont {Zhu}, \citenamefont {Amamou}, \citenamefont {Hone},\ and\
  \citenamefont {Kawakami}}]{singh_nanosecond_2016}%
  \BibitemOpen
  \bibfield  {author} {\bibinfo {author} {\bibfnamefont {S.}~\bibnamefont
  {Singh}}, \bibinfo {author} {\bibfnamefont {J.}~\bibnamefont {Katoch}},
  \bibinfo {author} {\bibfnamefont {J.}~\bibnamefont {Xu}}, \bibinfo {author}
  {\bibfnamefont {C.}~\bibnamefont {Tan}}, \bibinfo {author} {\bibfnamefont
  {T.}~\bibnamefont {Zhu}}, \bibinfo {author} {\bibfnamefont {W.}~\bibnamefont
  {Amamou}}, \bibinfo {author} {\bibfnamefont {J.}~\bibnamefont {Hone}}, \ and\
  \bibinfo {author} {\bibfnamefont {R.}~\bibnamefont {Kawakami}},\ }\href
  {\doibase 10.1063/1.4962635} {\bibfield  {journal} {\bibinfo  {journal}
  {Appl. Phys. Lett.}\ }\textbf {\bibinfo {volume} {109}},\ \bibinfo {pages}
  {122411} (\bibinfo {year} {2016})}\BibitemShut {NoStop}%
\bibitem [{\citenamefont {Ingla-Ayn{\'e}s}\ \emph {et~al.}(2015)\citenamefont
  {Ingla-Ayn{\'e}s}, \citenamefont {Guimar{\~ a}es}, \citenamefont {Meijerink},
  \citenamefont {Zomer},\ and\ \citenamefont {van
  Wees}}]{ingla-aynes_$24ensuremath-ensuremathmumathrmm$_2015}%
  \BibitemOpen
  \bibfield  {author} {\bibinfo {author} {\bibfnamefont {J.}~\bibnamefont
  {Ingla-Ayn{\'e}s}}, \bibinfo {author} {\bibfnamefont {M.~H.~D.}\ \bibnamefont
  {Guimar{\~ a}es}}, \bibinfo {author} {\bibfnamefont {R.~J.}\ \bibnamefont
  {Meijerink}}, \bibinfo {author} {\bibfnamefont {P.~J.}\ \bibnamefont
  {Zomer}}, \ and\ \bibinfo {author} {\bibfnamefont {B.~J.}\ \bibnamefont {van
  Wees}},\ }\href {\doibase 10.1103/PhysRevB.92.201410} {\bibfield  {journal}
  {\bibinfo  {journal} {Phys. Rev. B}\ }\textbf {\bibinfo {volume} {92}},\
  \bibinfo {pages} {201410(R)} (\bibinfo {year} {2015})}\BibitemShut {NoStop}%
\bibitem [{\citenamefont {Han}\ \emph {et~al.}(2014)\citenamefont {Han},
  \citenamefont {Kawakami}, \citenamefont {Gmitra},\ and\ \citenamefont
  {Fabian}}]{han_graphene_2014}%
  \BibitemOpen
  \bibfield  {author} {\bibinfo {author} {\bibfnamefont {W.}~\bibnamefont
  {Han}}, \bibinfo {author} {\bibfnamefont {R.~K.}\ \bibnamefont {Kawakami}},
  \bibinfo {author} {\bibfnamefont {M.}~\bibnamefont {Gmitra}}, \ and\ \bibinfo
  {author} {\bibfnamefont {J.}~\bibnamefont {Fabian}},\ }\href {\doibase
  10.1038/nnano.2014.214} {\bibfield  {journal} {\bibinfo  {journal} {Nature
  Nanotechnology}\ }\textbf {\bibinfo {volume} {9}},\ \bibinfo {pages} {794}
  (\bibinfo {year} {2014})}\BibitemShut {NoStop}%
\bibitem [{\citenamefont {Kane}\ and\ \citenamefont
  {Mele}(2005)}]{kane_quantum_2005}%
  \BibitemOpen
  \bibfield  {author} {\bibinfo {author} {\bibfnamefont {C.~L.}\ \bibnamefont
  {Kane}}\ and\ \bibinfo {author} {\bibfnamefont {E.~J.}\ \bibnamefont
  {Mele}},\ }\href {\doibase 10.1103/PhysRevLett.95.226801} {\bibfield
  {journal} {\bibinfo  {journal} {Phys. Rev. Lett.}\ }\textbf {\bibinfo
  {volume} {95}},\ \bibinfo {pages} {226801} (\bibinfo {year}
  {2005})}\BibitemShut {NoStop}%
\bibitem [{\citenamefont {Gmitra}\ \emph {et~al.}(2009)\citenamefont {Gmitra},
  \citenamefont {Konschuh}, \citenamefont {Ertler}, \citenamefont
  {Ambrosch-Draxl},\ and\ \citenamefont {Fabian}}]{gmitra_band-structure_2009}%
  \BibitemOpen
  \bibfield  {author} {\bibinfo {author} {\bibfnamefont {M.}~\bibnamefont
  {Gmitra}}, \bibinfo {author} {\bibfnamefont {S.}~\bibnamefont {Konschuh}},
  \bibinfo {author} {\bibfnamefont {C.}~\bibnamefont {Ertler}}, \bibinfo
  {author} {\bibfnamefont {C.}~\bibnamefont {Ambrosch-Draxl}}, \ and\ \bibinfo
  {author} {\bibfnamefont {J.}~\bibnamefont {Fabian}},\ }\href {\doibase
  10.1103/PhysRevB.80.235431} {\bibfield  {journal} {\bibinfo  {journal} {Phys.
  Rev. B}\ }\textbf {\bibinfo {volume} {80}},\ \bibinfo {pages} {235431}
  (\bibinfo {year} {2009})}\BibitemShut {NoStop}%
\bibitem [{\citenamefont {Geim}\ and\ \citenamefont
  {Grigorieva}(2013)}]{geim_van_2013}%
  \BibitemOpen
  \bibfield  {author} {\bibinfo {author} {\bibfnamefont {A.~K.}\ \bibnamefont
  {Geim}}\ and\ \bibinfo {author} {\bibfnamefont {I.~V.}\ \bibnamefont
  {Grigorieva}},\ }\href {\doibase 10.1038/nature12385} {\bibfield  {journal}
  {\bibinfo  {journal} {Nature}\ }\textbf {\bibinfo {volume} {499}},\ \bibinfo
  {pages} {419} (\bibinfo {year} {2013})}\BibitemShut {NoStop}%
\bibitem [{\citenamefont {Avsar}\ \emph {et~al.}(2014)\citenamefont {Avsar},
  \citenamefont {Tan}, \citenamefont {Taychatanapat}, \citenamefont
  {Balakrishnan}, \citenamefont {Koon}, \citenamefont {Yeo}, \citenamefont
  {Lahiri}, \citenamefont {Carvalho}, \citenamefont {Rodin}, \citenamefont
  {O’Farrell}, \citenamefont {Eda}, \citenamefont {Castro~Neto},\ and\
  \citenamefont {Özyilmaz}}]{avsar_spinorbit_2014}%
  \BibitemOpen
  \bibfield  {author} {\bibinfo {author} {\bibfnamefont {A.}~\bibnamefont
  {Avsar}}, \bibinfo {author} {\bibfnamefont {J.~Y.}\ \bibnamefont {Tan}},
  \bibinfo {author} {\bibfnamefont {T.}~\bibnamefont {Taychatanapat}}, \bibinfo
  {author} {\bibfnamefont {J.}~\bibnamefont {Balakrishnan}}, \bibinfo {author}
  {\bibfnamefont {G.~K.~W.}\ \bibnamefont {Koon}}, \bibinfo {author}
  {\bibfnamefont {Y.}~\bibnamefont {Yeo}}, \bibinfo {author} {\bibfnamefont
  {J.}~\bibnamefont {Lahiri}}, \bibinfo {author} {\bibfnamefont
  {A.}~\bibnamefont {Carvalho}}, \bibinfo {author} {\bibfnamefont {A.~S.}\
  \bibnamefont {Rodin}}, \bibinfo {author} {\bibfnamefont {E.~C.~T.}\
  \bibnamefont {O’Farrell}}, \bibinfo {author} {\bibfnamefont
  {G.}~\bibnamefont {Eda}}, \bibinfo {author} {\bibfnamefont {A.~H.}\
  \bibnamefont {Castro~Neto}}, \ and\ \bibinfo {author} {\bibfnamefont
  {B.}~\bibnamefont {Özyilmaz}},\ }\href {\doibase 10.1038/ncomms5875}
  {\bibfield  {journal} {\bibinfo  {journal} {Nat. Comm.}\ }\textbf {\bibinfo
  {volume} {5}},\ \bibinfo {pages} {4875} (\bibinfo {year} {2014})}\BibitemShut
  {NoStop}%
\bibitem [{\citenamefont {Wang}\ \emph
  {et~al.}(2015{\natexlab{a}})\citenamefont {Wang}, \citenamefont {Ki},
  \citenamefont {Chen}, \citenamefont {Berger}, \citenamefont {MacDonald},\
  and\ \citenamefont {Morpurgo}}]{wang_strong_2015}%
  \BibitemOpen
  \bibfield  {author} {\bibinfo {author} {\bibfnamefont {Z.}~\bibnamefont
  {Wang}}, \bibinfo {author} {\bibfnamefont {D.-K.}\ \bibnamefont {Ki}},
  \bibinfo {author} {\bibfnamefont {H.}~\bibnamefont {Chen}}, \bibinfo {author}
  {\bibfnamefont {H.}~\bibnamefont {Berger}}, \bibinfo {author} {\bibfnamefont
  {A.~H.}\ \bibnamefont {MacDonald}}, \ and\ \bibinfo {author} {\bibfnamefont
  {A.~F.}\ \bibnamefont {Morpurgo}},\ }\href {\doibase 10.1038/ncomms9339}
  {\bibfield  {journal} {\bibinfo  {journal} {Nat. Comm.}\ }\textbf {\bibinfo
  {volume} {6}},\ \bibinfo {pages} {8339} (\bibinfo {year}
  {2015}{\natexlab{a}})}\BibitemShut {NoStop}%
\bibitem [{\citenamefont {Wang}\ \emph {et~al.}(2016)\citenamefont {Wang},
  \citenamefont {Ki}, \citenamefont {Khoo}, \citenamefont {Mauro},
  \citenamefont {Berger}, \citenamefont {Levitov},\ and\ \citenamefont
  {Morpurgo}}]{wang_origin_2016}%
  \BibitemOpen
  \bibfield  {author} {\bibinfo {author} {\bibfnamefont {Z.}~\bibnamefont
  {Wang}}, \bibinfo {author} {\bibfnamefont {D.-K.}\ \bibnamefont {Ki}},
  \bibinfo {author} {\bibfnamefont {J.~Y.}\ \bibnamefont {Khoo}}, \bibinfo
  {author} {\bibfnamefont {D.}~\bibnamefont {Mauro}}, \bibinfo {author}
  {\bibfnamefont {H.}~\bibnamefont {Berger}}, \bibinfo {author} {\bibfnamefont
  {L.~S.}\ \bibnamefont {Levitov}}, \ and\ \bibinfo {author} {\bibfnamefont
  {A.~F.}\ \bibnamefont {Morpurgo}},\ }\href {\doibase
  10.1103/PhysRevX.6.041020} {\bibfield  {journal} {\bibinfo  {journal} {Phys.
  Rev. X}\ }\textbf {\bibinfo {volume} {6}},\ \bibinfo {pages} {041020}
  (\bibinfo {year} {2016})}\BibitemShut {NoStop}%
\bibitem [{\citenamefont {Yang}\ \emph {et~al.}(2016)\citenamefont {Yang},
  \citenamefont {Tu}, \citenamefont {Kim}, \citenamefont {Wu}, \citenamefont
  {Wang}, \citenamefont {Alicea}, \citenamefont {Wu}, \citenamefont
  {Bockrath},\ and\ \citenamefont {Shi}}]{yang_tunable_2016}%
  \BibitemOpen
  \bibfield  {author} {\bibinfo {author} {\bibfnamefont {B.}~\bibnamefont
  {Yang}}, \bibinfo {author} {\bibfnamefont {M.-F.}\ \bibnamefont {Tu}},
  \bibinfo {author} {\bibfnamefont {J.}~\bibnamefont {Kim}}, \bibinfo {author}
  {\bibfnamefont {Y.}~\bibnamefont {Wu}}, \bibinfo {author} {\bibfnamefont
  {H.}~\bibnamefont {Wang}}, \bibinfo {author} {\bibfnamefont {J.}~\bibnamefont
  {Alicea}}, \bibinfo {author} {\bibfnamefont {R.}~\bibnamefont {Wu}}, \bibinfo
  {author} {\bibfnamefont {M.}~\bibnamefont {Bockrath}}, \ and\ \bibinfo
  {author} {\bibfnamefont {J.}~\bibnamefont {Shi}},\ }\href {\doibase
  10.1088/2053-1583/3/3/031012} {\bibfield  {journal} {\bibinfo  {journal} {2D
  Mater.}\ }\textbf {\bibinfo {volume} {3}},\ \bibinfo {pages} {031012}
  (\bibinfo {year} {2016})}\BibitemShut {NoStop}%
\bibitem [{\citenamefont {Yan}\ \emph {et~al.}(2016)\citenamefont {Yan},
  \citenamefont {Txoperena}, \citenamefont {Llopis}, \citenamefont {Dery},
  \citenamefont {Hueso},\ and\ \citenamefont
  {Casanova}}]{yan_two-dimensional_2016}%
  \BibitemOpen
  \bibfield  {author} {\bibinfo {author} {\bibfnamefont {W.}~\bibnamefont
  {Yan}}, \bibinfo {author} {\bibfnamefont {O.}~\bibnamefont {Txoperena}},
  \bibinfo {author} {\bibfnamefont {R.}~\bibnamefont {Llopis}}, \bibinfo
  {author} {\bibfnamefont {H.}~\bibnamefont {Dery}}, \bibinfo {author}
  {\bibfnamefont {L.~E.}\ \bibnamefont {Hueso}}, \ and\ \bibinfo {author}
  {\bibfnamefont {F.}~\bibnamefont {Casanova}},\ }\href {\doibase
  10.1038/ncomms13372} {\bibfield  {journal} {\bibinfo  {journal} {Nat. Comm.}\
  }\textbf {\bibinfo {volume} {7}},\ \bibinfo {pages} {13372} (\bibinfo {year}
  {2016})}\BibitemShut {NoStop}%
\bibitem [{\citenamefont {Yang}\ \emph {et~al.}(2017)\citenamefont {Yang},
  \citenamefont {Lohmann}, \citenamefont {Barroso}, \citenamefont {Liao},
  \citenamefont {Lin}, \citenamefont {Liu}, \citenamefont {Bartels},
  \citenamefont {Watanabe}, \citenamefont {Taniguchi},\ and\ \citenamefont
  {Shi}}]{yang_strong_2017}%
  \BibitemOpen
  \bibfield  {author} {\bibinfo {author} {\bibfnamefont {B.}~\bibnamefont
  {Yang}}, \bibinfo {author} {\bibfnamefont {M.}~\bibnamefont {Lohmann}},
  \bibinfo {author} {\bibfnamefont {D.}~\bibnamefont {Barroso}}, \bibinfo
  {author} {\bibfnamefont {I.}~\bibnamefont {Liao}}, \bibinfo {author}
  {\bibfnamefont {Z.}~\bibnamefont {Lin}}, \bibinfo {author} {\bibfnamefont
  {Y.}~\bibnamefont {Liu}}, \bibinfo {author} {\bibfnamefont {L.}~\bibnamefont
  {Bartels}}, \bibinfo {author} {\bibfnamefont {K.}~\bibnamefont {Watanabe}},
  \bibinfo {author} {\bibfnamefont {T.}~\bibnamefont {Taniguchi}}, \ and\
  \bibinfo {author} {\bibfnamefont {J.}~\bibnamefont {Shi}},\ }\href {\doibase
  10.1103/PhysRevB.96.041409} {\bibfield  {journal} {\bibinfo  {journal} {Phys.
  Rev. B}\ }\textbf {\bibinfo {volume} {96}},\ \bibinfo {pages} {041409(R)}
  (\bibinfo {year} {2017})}\BibitemShut {NoStop}%
\bibitem [{\citenamefont {Ghiasi}\ \emph {et~al.}(2017)\citenamefont {Ghiasi},
  \citenamefont {Ingla-Ayn{\'e}s}, \citenamefont {Kaverzin},\ and\
  \citenamefont {van Wees}}]{ghiasi_large_2017}%
  \BibitemOpen
  \bibfield  {author} {\bibinfo {author} {\bibfnamefont {T.~S.}\ \bibnamefont
  {Ghiasi}}, \bibinfo {author} {\bibfnamefont {J.}~\bibnamefont
  {Ingla-Ayn{\'e}s}}, \bibinfo {author} {\bibfnamefont {A.~A.}\ \bibnamefont
  {Kaverzin}}, \ and\ \bibinfo {author} {\bibfnamefont {B.~J.}\ \bibnamefont
  {van Wees}},\ }\href {\doibase 10.1021/acs.nanolett.7b03460} {\bibfield
  {journal} {\bibinfo  {journal} {Nano Lett.}\ }\textbf {\bibinfo {volume}
  {17}},\ \bibinfo {pages} {7528} (\bibinfo {year} {2017})}\BibitemShut
  {NoStop}%
\bibitem [{\citenamefont {Dankert}\ and\ \citenamefont
  {Dash}(2017)}]{dankert_electrical_2017}%
  \BibitemOpen
  \bibfield  {author} {\bibinfo {author} {\bibfnamefont {A.}~\bibnamefont
  {Dankert}}\ and\ \bibinfo {author} {\bibfnamefont {S.~P.}\ \bibnamefont
  {Dash}},\ }\href {\doibase 10.1038/ncomms16093} {\bibfield  {journal}
  {\bibinfo  {journal} {Nat. Comm.}\ }\textbf {\bibinfo {volume} {8}},\
  \bibinfo {pages} {16093} (\bibinfo {year} {2017})}\BibitemShut {NoStop}%
\bibitem [{\citenamefont {V{\"o}lkl}\ \emph {et~al.}(2017)\citenamefont
  {V{\"o}lkl}, \citenamefont {Rockinger}, \citenamefont {Drienovsky},
  \citenamefont {Watanabe}, \citenamefont {Taniguchi}, \citenamefont {Weiss},\
  and\ \citenamefont {Eroms}}]{volkl_magnetotransport_2017}%
  \BibitemOpen
  \bibfield  {author} {\bibinfo {author} {\bibfnamefont {T.}~\bibnamefont
  {V{\"o}lkl}}, \bibinfo {author} {\bibfnamefont {T.}~\bibnamefont
  {Rockinger}}, \bibinfo {author} {\bibfnamefont {M.}~\bibnamefont
  {Drienovsky}}, \bibinfo {author} {\bibfnamefont {K.}~\bibnamefont
  {Watanabe}}, \bibinfo {author} {\bibfnamefont {T.}~\bibnamefont {Taniguchi}},
  \bibinfo {author} {\bibfnamefont {D.}~\bibnamefont {Weiss}}, \ and\ \bibinfo
  {author} {\bibfnamefont {J.}~\bibnamefont {Eroms}},\ }\href {\doibase
  10.1103/PhysRevB.96.125405} {\bibfield  {journal} {\bibinfo  {journal} {Phys.
  Rev. B}\ }\textbf {\bibinfo {volume} {96}},\ \bibinfo {pages} {125405}
  (\bibinfo {year} {2017})}\BibitemShut {NoStop}%
\bibitem [{\citenamefont {Zihlmann}\ \emph {et~al.}(2018)\citenamefont
  {Zihlmann}, \citenamefont {Cummings}, \citenamefont {Garcia}, \citenamefont
  {Kedves}, \citenamefont {Watanabe}, \citenamefont {Taniguchi}, \citenamefont
  {Sch{\"o}nenberger},\ and\ \citenamefont {Makk}}]{zihlmann_large_2018}%
  \BibitemOpen
  \bibfield  {author} {\bibinfo {author} {\bibfnamefont {S.}~\bibnamefont
  {Zihlmann}}, \bibinfo {author} {\bibfnamefont {A.~W.}\ \bibnamefont
  {Cummings}}, \bibinfo {author} {\bibfnamefont {J.~H.}\ \bibnamefont
  {Garcia}}, \bibinfo {author} {\bibfnamefont {M.}~\bibnamefont {Kedves}},
  \bibinfo {author} {\bibfnamefont {K.}~\bibnamefont {Watanabe}}, \bibinfo
  {author} {\bibfnamefont {T.}~\bibnamefont {Taniguchi}}, \bibinfo {author}
  {\bibfnamefont {C.}~\bibnamefont {Sch{\"o}nenberger}}, \ and\ \bibinfo
  {author} {\bibfnamefont {P.}~\bibnamefont {Makk}},\ }\href {\doibase
  10.1103/PhysRevB.97.075434} {\bibfield  {journal} {\bibinfo  {journal} {Phys.
  Rev. B}\ }\textbf {\bibinfo {volume} {97}},\ \bibinfo {pages} {075434}
  (\bibinfo {year} {2018})}\BibitemShut {NoStop}%
\bibitem [{\citenamefont {Wakamura}\ \emph {et~al.}(2018)\citenamefont
  {Wakamura}, \citenamefont {Reale}, \citenamefont {Palczynski}, \citenamefont
  {Gu{\'e}ron}, \citenamefont {Mattevi},\ and\ \citenamefont
  {Bouchiat}}]{wakamura_strong_2018}%
  \BibitemOpen
  \bibfield  {author} {\bibinfo {author} {\bibfnamefont {T.}~\bibnamefont
  {Wakamura}}, \bibinfo {author} {\bibfnamefont {F.}~\bibnamefont {Reale}},
  \bibinfo {author} {\bibfnamefont {P.}~\bibnamefont {Palczynski}}, \bibinfo
  {author} {\bibfnamefont {S.}~\bibnamefont {Gu{\'e}ron}}, \bibinfo {author}
  {\bibfnamefont {C.}~\bibnamefont {Mattevi}}, \ and\ \bibinfo {author}
  {\bibfnamefont {H.}~\bibnamefont {Bouchiat}},\ }\href {\doibase
  10.1103/PhysRevLett.120.106802} {\bibfield  {journal} {\bibinfo  {journal}
  {Phys. Rev. Lett.}\ }\textbf {\bibinfo {volume} {120}},\ \bibinfo {pages}
  {106802} (\bibinfo {year} {2018})}\BibitemShut {NoStop}%
\bibitem [{\citenamefont {Leutenantsmeyer}\ \emph {et~al.}(2018)\citenamefont
  {Leutenantsmeyer}, \citenamefont {Ingla-Ayn{\'e}s}, \citenamefont {Fabian},\
  and\ \citenamefont {van Wees}}]{leutenantsmeyer_observation_2018}%
  \BibitemOpen
  \bibfield  {author} {\bibinfo {author} {\bibfnamefont {J.~C.}\ \bibnamefont
  {Leutenantsmeyer}}, \bibinfo {author} {\bibfnamefont {J.}~\bibnamefont
  {Ingla-Ayn{\'e}s}}, \bibinfo {author} {\bibfnamefont {J.}~\bibnamefont
  {Fabian}}, \ and\ \bibinfo {author} {\bibfnamefont {B.~J.}\ \bibnamefont {van
  Wees}},\ }\href {\doibase 10.1103/PhysRevLett.121.127702} {\bibfield
  {journal} {\bibinfo  {journal} {Phys. Rev. Lett.}\ }\textbf {\bibinfo
  {volume} {121}},\ \bibinfo {pages} {127702} (\bibinfo {year}
  {2018})}\BibitemShut {NoStop}%
\bibitem [{\citenamefont {Omar}\ and\ \citenamefont {van
  Wees}(2018)}]{omar_spin_2018}%
  \BibitemOpen
  \bibfield  {author} {\bibinfo {author} {\bibfnamefont {S.}~\bibnamefont
  {Omar}}\ and\ \bibinfo {author} {\bibfnamefont {B.~J.}\ \bibnamefont {van
  Wees}},\ }\href {\doibase 10.1103/PhysRevB.97.045414} {\bibfield  {journal}
  {\bibinfo  {journal} {Phys. Rev. B}\ }\textbf {\bibinfo {volume} {97}},\
  \bibinfo {pages} {045414} (\bibinfo {year} {2018})}\BibitemShut {NoStop}%
\bibitem [{\citenamefont {Benítez}\ \emph {et~al.}(2018)\citenamefont
  {Benítez}, \citenamefont {Sierra}, \citenamefont {Torres}, \citenamefont
  {Arrighi}, \citenamefont {Bonell}, \citenamefont {Costache},\ and\
  \citenamefont {Valenzuela}}]{benitez_strongly_2018}%
  \BibitemOpen
  \bibfield  {author} {\bibinfo {author} {\bibfnamefont {L.~A.}\ \bibnamefont
  {Benítez}}, \bibinfo {author} {\bibfnamefont {J.~F.}\ \bibnamefont
  {Sierra}}, \bibinfo {author} {\bibfnamefont {W.~S.}\ \bibnamefont {Torres}},
  \bibinfo {author} {\bibfnamefont {A.}~\bibnamefont {Arrighi}}, \bibinfo
  {author} {\bibfnamefont {F.}~\bibnamefont {Bonell}}, \bibinfo {author}
  {\bibfnamefont {M.~V.}\ \bibnamefont {Costache}}, \ and\ \bibinfo {author}
  {\bibfnamefont {S.~O.}\ \bibnamefont {Valenzuela}},\ }\href {\doibase
  10.1038/s41567-017-0019-2} {\bibfield  {journal} {\bibinfo  {journal} {Nature
  Physics}\ }\textbf {\bibinfo {volume} {14}},\ \bibinfo {pages} {303}
  (\bibinfo {year} {2018})}\BibitemShut {NoStop}%
\bibitem [{\citenamefont {Safeer}\ \emph {et~al.}(2019)\citenamefont {Safeer},
  \citenamefont {Ingla-Ayn{\'e}s}, \citenamefont {Herling}, \citenamefont
  {Garcia}, \citenamefont {Vila}, \citenamefont {Ontoso}, \citenamefont
  {Calvo}, \citenamefont {Roche}, \citenamefont {Hueso},\ and\ \citenamefont
  {Casanova}}]{safeer_room-temperature_2019}%
  \BibitemOpen
  \bibfield  {author} {\bibinfo {author} {\bibfnamefont {C.~K.}\ \bibnamefont
  {Safeer}}, \bibinfo {author} {\bibfnamefont {J.}~\bibnamefont
  {Ingla-Ayn{\'e}s}}, \bibinfo {author} {\bibfnamefont {F.}~\bibnamefont
  {Herling}}, \bibinfo {author} {\bibfnamefont {J.~H.}\ \bibnamefont {Garcia}},
  \bibinfo {author} {\bibfnamefont {M.}~\bibnamefont {Vila}}, \bibinfo {author}
  {\bibfnamefont {N.}~\bibnamefont {Ontoso}}, \bibinfo {author} {\bibfnamefont
  {M.~R.}\ \bibnamefont {Calvo}}, \bibinfo {author} {\bibfnamefont
  {S.}~\bibnamefont {Roche}}, \bibinfo {author} {\bibfnamefont {L.~E.}\
  \bibnamefont {Hueso}}, \ and\ \bibinfo {author} {\bibfnamefont
  {F.}~\bibnamefont {Casanova}},\ }\href {\doibase
  10.1021/acs.nanolett.8b04368} {\bibfield  {journal} {\bibinfo  {journal}
  {Nano Lett.}\ }\textbf {\bibinfo {volume} {19}},\ \bibinfo {pages} {1074}
  (\bibinfo {year} {2019})}\BibitemShut {NoStop}%
\bibitem [{\citenamefont {Kretinin}\ \emph {et~al.}(2014)\citenamefont
  {Kretinin}, \citenamefont {Cao}, \citenamefont {Tu}, \citenamefont {Yu},
  \citenamefont {Jalil}, \citenamefont {Novoselov}, \citenamefont {Haigh},
  \citenamefont {Gholinia}, \citenamefont {Mishchenko}, \citenamefont {Lozada},
  \citenamefont {Georgiou}, \citenamefont {Woods}, \citenamefont {Withers},
  \citenamefont {Blake}, \citenamefont {Eda}, \citenamefont {Wirsig},
  \citenamefont {Hucho}, \citenamefont {Watanabe}, \citenamefont {Taniguchi},
  \citenamefont {Geim},\ and\ \citenamefont
  {Gorbachev}}]{kretinin_electronic_2014}%
  \BibitemOpen
  \bibfield  {author} {\bibinfo {author} {\bibfnamefont {A.~V.}\ \bibnamefont
  {Kretinin}}, \bibinfo {author} {\bibfnamefont {Y.}~\bibnamefont {Cao}},
  \bibinfo {author} {\bibfnamefont {J.~S.}\ \bibnamefont {Tu}}, \bibinfo
  {author} {\bibfnamefont {G.~L.}\ \bibnamefont {Yu}}, \bibinfo {author}
  {\bibfnamefont {R.}~\bibnamefont {Jalil}}, \bibinfo {author} {\bibfnamefont
  {K.~S.}\ \bibnamefont {Novoselov}}, \bibinfo {author} {\bibfnamefont {S.~J.}\
  \bibnamefont {Haigh}}, \bibinfo {author} {\bibfnamefont {A.}~\bibnamefont
  {Gholinia}}, \bibinfo {author} {\bibfnamefont {A.}~\bibnamefont
  {Mishchenko}}, \bibinfo {author} {\bibfnamefont {M.}~\bibnamefont {Lozada}},
  \bibinfo {author} {\bibfnamefont {T.}~\bibnamefont {Georgiou}}, \bibinfo
  {author} {\bibfnamefont {C.~R.}\ \bibnamefont {Woods}}, \bibinfo {author}
  {\bibfnamefont {F.}~\bibnamefont {Withers}}, \bibinfo {author} {\bibfnamefont
  {P.}~\bibnamefont {Blake}}, \bibinfo {author} {\bibfnamefont
  {G.}~\bibnamefont {Eda}}, \bibinfo {author} {\bibfnamefont {A.}~\bibnamefont
  {Wirsig}}, \bibinfo {author} {\bibfnamefont {C.}~\bibnamefont {Hucho}},
  \bibinfo {author} {\bibfnamefont {K.}~\bibnamefont {Watanabe}}, \bibinfo
  {author} {\bibfnamefont {T.}~\bibnamefont {Taniguchi}}, \bibinfo {author}
  {\bibfnamefont {A.~K.}\ \bibnamefont {Geim}}, \ and\ \bibinfo {author}
  {\bibfnamefont {R.~V.}\ \bibnamefont {Gorbachev}},\ }\href {\doibase
  10.1021/nl5006542} {\bibfield  {journal} {\bibinfo  {journal} {Nano Lett.}\
  }\textbf {\bibinfo {volume} {14}},\ \bibinfo {pages} {3270} (\bibinfo {year}
  {2014})}\BibitemShut {NoStop}%
\bibitem [{\citenamefont {Korm{\'a}nyos}\ \emph {et~al.}(2015)\citenamefont
  {Korm{\'a}nyos}, \citenamefont {Burkard}, \citenamefont {Gmitra},
  \citenamefont {Fabian}, \citenamefont {Z{\'o}lyomi}, \citenamefont
  {Drummond},\ and\ \citenamefont {{Vladimir Fal’ko}}}]{kormanyos_k_2015}%
  \BibitemOpen
  \bibfield  {author} {\bibinfo {author} {\bibfnamefont {A.}~\bibnamefont
  {Korm{\'a}nyos}}, \bibinfo {author} {\bibfnamefont {G.}~\bibnamefont
  {Burkard}}, \bibinfo {author} {\bibfnamefont {M.}~\bibnamefont {Gmitra}},
  \bibinfo {author} {\bibfnamefont {J.}~\bibnamefont {Fabian}}, \bibinfo
  {author} {\bibfnamefont {V.}~\bibnamefont {Z{\'o}lyomi}}, \bibinfo {author}
  {\bibfnamefont {N.~D.}\ \bibnamefont {Drummond}}, \ and\ \bibinfo {author}
  {\bibnamefont {{Vladimir Fal’ko}}},\ }\href {\doibase
  10.1088/2053-1583/2/2/022001} {\bibfield  {journal} {\bibinfo  {journal} {2D
  Mater.}\ }\textbf {\bibinfo {volume} {2}},\ \bibinfo {pages} {022001}
  (\bibinfo {year} {2015})}\BibitemShut {NoStop}%
\bibitem [{\citenamefont {Cummings}\ \emph {et~al.}(2017)\citenamefont
  {Cummings}, \citenamefont {Garcia}, \citenamefont {Fabian},\ and\
  \citenamefont {Roche}}]{cummings_giant_2017}%
  \BibitemOpen
  \bibfield  {author} {\bibinfo {author} {\bibfnamefont {A.~W.}\ \bibnamefont
  {Cummings}}, \bibinfo {author} {\bibfnamefont {J.~H.}\ \bibnamefont
  {Garcia}}, \bibinfo {author} {\bibfnamefont {J.}~\bibnamefont {Fabian}}, \
  and\ \bibinfo {author} {\bibfnamefont {S.}~\bibnamefont {Roche}},\ }\href
  {\doibase 10.1103/PhysRevLett.119.206601} {\bibfield  {journal} {\bibinfo
  {journal} {Phys. Rev. Lett.}\ }\textbf {\bibinfo {volume} {119}},\ \bibinfo
  {pages} {206601} (\bibinfo {year} {2017})}\BibitemShut {NoStop}%
\bibitem [{\citenamefont {Kaloni}\ \emph {et~al.}(2014)\citenamefont {Kaloni},
  \citenamefont {Kou}, \citenamefont {Frauenheim},\ and\ \citenamefont
  {Schwingenschlögl}}]{kaloni_quantum_2014}%
  \BibitemOpen
  \bibfield  {author} {\bibinfo {author} {\bibfnamefont {T.~P.}\ \bibnamefont
  {Kaloni}}, \bibinfo {author} {\bibfnamefont {L.}~\bibnamefont {Kou}},
  \bibinfo {author} {\bibfnamefont {T.}~\bibnamefont {Frauenheim}}, \ and\
  \bibinfo {author} {\bibfnamefont {U.}~\bibnamefont {Schwingenschlögl}},\
  }\href {\doibase 10.1063/1.4903895} {\bibfield  {journal} {\bibinfo
  {journal} {Appl. Phys. Lett.}\ }\textbf {\bibinfo {volume} {105}},\ \bibinfo
  {pages} {233112} (\bibinfo {year} {2014})}\BibitemShut {NoStop}%
\bibitem [{\citenamefont {Gmitra}\ and\ \citenamefont
  {Fabian}(2015)}]{gmitra_graphene_2015}%
  \BibitemOpen
  \bibfield  {author} {\bibinfo {author} {\bibfnamefont {M.}~\bibnamefont
  {Gmitra}}\ and\ \bibinfo {author} {\bibfnamefont {J.}~\bibnamefont
  {Fabian}},\ }\href {\doibase 10.1103/PhysRevB.92.155403} {\bibfield
  {journal} {\bibinfo  {journal} {Phys. Rev. B}\ }\textbf {\bibinfo {volume}
  {92}},\ \bibinfo {pages} {155403} (\bibinfo {year} {2015})}\BibitemShut
  {NoStop}%
\bibitem [{\citenamefont {Gmitra}\ \emph {et~al.}(2016)\citenamefont {Gmitra},
  \citenamefont {Kochan}, \citenamefont {H{\"o}gl},\ and\ \citenamefont
  {Fabian}}]{gmitra_trivial_2016}%
  \BibitemOpen
  \bibfield  {author} {\bibinfo {author} {\bibfnamefont {M.}~\bibnamefont
  {Gmitra}}, \bibinfo {author} {\bibfnamefont {D.}~\bibnamefont {Kochan}},
  \bibinfo {author} {\bibfnamefont {P.}~\bibnamefont {H{\"o}gl}}, \ and\
  \bibinfo {author} {\bibfnamefont {J.}~\bibnamefont {Fabian}},\ }\href
  {\doibase 10.1103/PhysRevB.93.155104} {\bibfield  {journal} {\bibinfo
  {journal} {Phys. Rev. B}\ }\textbf {\bibinfo {volume} {93}},\ \bibinfo
  {pages} {155104} (\bibinfo {year} {2016})}\BibitemShut {NoStop}%
\bibitem [{\citenamefont {Singh}\ \emph {et~al.}(2018)\citenamefont {Singh},
  \citenamefont {Alsharari}, \citenamefont {Ulloa},\ and\ \citenamefont
  {Romero}}]{singh_proximity-induced_2018}%
  \BibitemOpen
  \bibfield  {author} {\bibinfo {author} {\bibfnamefont {S.}~\bibnamefont
  {Singh}}, \bibinfo {author} {\bibfnamefont {A.~M.}\ \bibnamefont
  {Alsharari}}, \bibinfo {author} {\bibfnamefont {S.~E.}\ \bibnamefont
  {Ulloa}}, \ and\ \bibinfo {author} {\bibfnamefont {A.~H.}\ \bibnamefont
  {Romero}},\ }\href {http://arxiv.org/abs/1806.11469} {\bibfield  {journal}
  {\bibinfo  {journal} {arXiv:1806.11469 [cond-mat]}\ } (\bibinfo {year}
  {2018})},\ \bibinfo {note} {arXiv: 1806.11469}\BibitemShut {NoStop}%
\bibitem [{\citenamefont {Pierucci}\ \emph {et~al.}(2016)\citenamefont
  {Pierucci}, \citenamefont {Henck}, \citenamefont {Avila}, \citenamefont
  {Balan}, \citenamefont {Naylor}, \citenamefont {Patriarche}, \citenamefont
  {Dappe}, \citenamefont {Silly}, \citenamefont {Sirotti}, \citenamefont
  {Johnson}, \citenamefont {Asensio},\ and\ \citenamefont
  {Ouerghi}}]{pierucci_band_2016}%
  \BibitemOpen
  \bibfield  {author} {\bibinfo {author} {\bibfnamefont {D.}~\bibnamefont
  {Pierucci}}, \bibinfo {author} {\bibfnamefont {H.}~\bibnamefont {Henck}},
  \bibinfo {author} {\bibfnamefont {J.}~\bibnamefont {Avila}}, \bibinfo
  {author} {\bibfnamefont {A.}~\bibnamefont {Balan}}, \bibinfo {author}
  {\bibfnamefont {C.~H.}\ \bibnamefont {Naylor}}, \bibinfo {author}
  {\bibfnamefont {G.}~\bibnamefont {Patriarche}}, \bibinfo {author}
  {\bibfnamefont {Y.~J.}\ \bibnamefont {Dappe}}, \bibinfo {author}
  {\bibfnamefont {M.~G.}\ \bibnamefont {Silly}}, \bibinfo {author}
  {\bibfnamefont {F.}~\bibnamefont {Sirotti}}, \bibinfo {author} {\bibfnamefont
  {A.~T.~C.}\ \bibnamefont {Johnson}}, \bibinfo {author} {\bibfnamefont
  {M.~C.}\ \bibnamefont {Asensio}}, \ and\ \bibinfo {author} {\bibfnamefont
  {A.}~\bibnamefont {Ouerghi}},\ }\href {\doibase 10.1021/acs.nanolett.6b00609}
  {\bibfield  {journal} {\bibinfo  {journal} {Nano Letters}\ }\textbf {\bibinfo
  {volume} {16}},\ \bibinfo {pages} {4054} (\bibinfo {year}
  {2016})}\BibitemShut {NoStop}%
\bibitem [{\citenamefont {Wang}\ \emph
  {et~al.}(2015{\natexlab{b}})\citenamefont {Wang}, \citenamefont {Chen},\ and\
  \citenamefont {Wang}}]{wang_electronic_2015}%
  \BibitemOpen
  \bibfield  {author} {\bibinfo {author} {\bibfnamefont {Z.}~\bibnamefont
  {Wang}}, \bibinfo {author} {\bibfnamefont {Q.}~\bibnamefont {Chen}}, \ and\
  \bibinfo {author} {\bibfnamefont {J.}~\bibnamefont {Wang}},\ }\href {\doibase
  10.1021/jp507751p} {\bibfield  {journal} {\bibinfo  {journal} {J. Phys. Chem.
  C}\ }\textbf {\bibinfo {volume} {119}},\ \bibinfo {pages} {4752} (\bibinfo
  {year} {2015}{\natexlab{b}})}\BibitemShut {NoStop}%
\bibitem [{\citenamefont {Felice}\ \emph {et~al.}(2017)\citenamefont {Felice},
  \citenamefont {Abad}, \citenamefont {González}, \citenamefont {Smogunov},\
  and\ \citenamefont {Dappe}}]{felice_angle_2017}%
  \BibitemOpen
  \bibfield  {author} {\bibinfo {author} {\bibfnamefont {D.~D.}\ \bibnamefont
  {Felice}}, \bibinfo {author} {\bibfnamefont {E.}~\bibnamefont {Abad}},
  \bibinfo {author} {\bibfnamefont {C.}~\bibnamefont {González}}, \bibinfo
  {author} {\bibfnamefont {A.}~\bibnamefont {Smogunov}}, \ and\ \bibinfo
  {author} {\bibfnamefont {Y.~J.}\ \bibnamefont {Dappe}},\ }\href {\doibase
  10.1088/1361-6463/aa64fe} {\bibfield  {journal} {\bibinfo  {journal} {J.
  Phys. D: Appl. Phys.}\ }\textbf {\bibinfo {volume} {50}},\ \bibinfo {pages}
  {17LT02} (\bibinfo {year} {2017})}\BibitemShut {NoStop}%
\bibitem [{\citenamefont {Alsharari}\ \emph {et~al.}(2016)\citenamefont
  {Alsharari}, \citenamefont {Asmar},\ and\ \citenamefont
  {Ulloa}}]{alsharari_mass_2016}%
  \BibitemOpen
  \bibfield  {author} {\bibinfo {author} {\bibfnamefont {A.~M.}\ \bibnamefont
  {Alsharari}}, \bibinfo {author} {\bibfnamefont {M.~M.}\ \bibnamefont
  {Asmar}}, \ and\ \bibinfo {author} {\bibfnamefont {S.~E.}\ \bibnamefont
  {Ulloa}},\ }\href {\doibase 10.1103/PhysRevB.94.241106} {\bibfield  {journal}
  {\bibinfo  {journal} {Phys. Rev. B}\ }\textbf {\bibinfo {volume} {94}},\
  \bibinfo {pages} {241106(R)} (\bibinfo {year} {2016})}\BibitemShut {NoStop}%
\bibitem [{\citenamefont {Alsharari}\ \emph {et~al.}(2018)\citenamefont
  {Alsharari}, \citenamefont {Asmar},\ and\ \citenamefont
  {Ulloa}}]{alsharari_topological_2018}%
  \BibitemOpen
  \bibfield  {author} {\bibinfo {author} {\bibfnamefont {A.~M.}\ \bibnamefont
  {Alsharari}}, \bibinfo {author} {\bibfnamefont {M.~M.}\ \bibnamefont
  {Asmar}}, \ and\ \bibinfo {author} {\bibfnamefont {S.~E.}\ \bibnamefont
  {Ulloa}},\ }\href {\doibase 10.1103/PhysRevB.98.195129} {\bibfield  {journal}
  {\bibinfo  {journal} {Phys. Rev. B}\ }\textbf {\bibinfo {volume} {98}},\
  \bibinfo {pages} {195129(R)} (\bibinfo {year} {2018})}\BibitemShut {NoStop}%
\bibitem [{\citenamefont {Li}\ and\ \citenamefont
  {Koshino}(2019)}]{li_twist-angle_2019}%
  \BibitemOpen
  \bibfield  {author} {\bibinfo {author} {\bibfnamefont {Y.}~\bibnamefont
  {Li}}\ and\ \bibinfo {author} {\bibfnamefont {M.}~\bibnamefont {Koshino}},\
  }\href {\doibase 10.1103/PhysRevB.99.075438} {\bibfield  {journal} {\bibinfo
  {journal} {Phys. Rev. B}\ }\textbf {\bibinfo {volume} {99}},\ \bibinfo
  {pages} {075438} (\bibinfo {year} {2019})}\BibitemShut {NoStop}%
\bibitem [{\citenamefont {Bistritzer}\ and\ \citenamefont
  {MacDonald}(2011)}]{bistritzer_moire_2011}%
  \BibitemOpen
  \bibfield  {author} {\bibinfo {author} {\bibfnamefont {R.}~\bibnamefont
  {Bistritzer}}\ and\ \bibinfo {author} {\bibfnamefont {A.~H.}\ \bibnamefont
  {MacDonald}},\ }\href {\doibase 10.1073/pnas.1108174108} {\bibfield
  {journal} {\bibinfo  {journal} {PNAS}\ }\textbf {\bibinfo {volume} {108}},\
  \bibinfo {pages} {12233} (\bibinfo {year} {2011})}\BibitemShut {NoStop}%
\bibitem [{\citenamefont {Carr}\ \emph {et~al.}(2017)\citenamefont {Carr},
  \citenamefont {Massatt}, \citenamefont {Fang}, \citenamefont {Cazeaux},
  \citenamefont {Luskin},\ and\ \citenamefont
  {Kaxiras}}]{carr_twistronics:_2017}%
  \BibitemOpen
  \bibfield  {author} {\bibinfo {author} {\bibfnamefont {S.}~\bibnamefont
  {Carr}}, \bibinfo {author} {\bibfnamefont {D.}~\bibnamefont {Massatt}},
  \bibinfo {author} {\bibfnamefont {S.}~\bibnamefont {Fang}}, \bibinfo {author}
  {\bibfnamefont {P.}~\bibnamefont {Cazeaux}}, \bibinfo {author} {\bibfnamefont
  {M.}~\bibnamefont {Luskin}}, \ and\ \bibinfo {author} {\bibfnamefont
  {E.}~\bibnamefont {Kaxiras}},\ }\href {\doibase 10.1103/PhysRevB.95.075420}
  {\bibfield  {journal} {\bibinfo  {journal} {Phys. Rev. B}\ }\textbf {\bibinfo
  {volume} {95}},\ \bibinfo {pages} {075420} (\bibinfo {year}
  {2017})}\BibitemShut {NoStop}%
\bibitem [{\citenamefont {Cao}\ \emph {et~al.}(2018)\citenamefont {Cao},
  \citenamefont {Fatemi}, \citenamefont {Fang}, \citenamefont {Watanabe},
  \citenamefont {Taniguchi}, \citenamefont {Kaxiras},\ and\ \citenamefont
  {Jarillo-Herrero}}]{cao_unconventional_2018}%
  \BibitemOpen
  \bibfield  {author} {\bibinfo {author} {\bibfnamefont {Y.}~\bibnamefont
  {Cao}}, \bibinfo {author} {\bibfnamefont {V.}~\bibnamefont {Fatemi}},
  \bibinfo {author} {\bibfnamefont {S.}~\bibnamefont {Fang}}, \bibinfo {author}
  {\bibfnamefont {K.}~\bibnamefont {Watanabe}}, \bibinfo {author}
  {\bibfnamefont {T.}~\bibnamefont {Taniguchi}}, \bibinfo {author}
  {\bibfnamefont {E.}~\bibnamefont {Kaxiras}}, \ and\ \bibinfo {author}
  {\bibfnamefont {P.}~\bibnamefont {Jarillo-Herrero}},\ }\href {\doibase
  10.1038/nature26160} {\bibfield  {journal} {\bibinfo  {journal} {Nature}\
  }\textbf {\bibinfo {volume} {556}},\ \bibinfo {pages} {43} (\bibinfo {year}
  {2018})}\BibitemShut {NoStop}%
\bibitem [{\citenamefont {Ribeiro-Palau}\ \emph {et~al.}(2018)\citenamefont
  {Ribeiro-Palau}, \citenamefont {Zhang}, \citenamefont {Watanabe},
  \citenamefont {Taniguchi}, \citenamefont {Hone},\ and\ \citenamefont
  {Dean}}]{ribeiro-palau_twistable_2018}%
  \BibitemOpen
  \bibfield  {author} {\bibinfo {author} {\bibfnamefont {R.}~\bibnamefont
  {Ribeiro-Palau}}, \bibinfo {author} {\bibfnamefont {C.}~\bibnamefont
  {Zhang}}, \bibinfo {author} {\bibfnamefont {K.}~\bibnamefont {Watanabe}},
  \bibinfo {author} {\bibfnamefont {T.}~\bibnamefont {Taniguchi}}, \bibinfo
  {author} {\bibfnamefont {J.}~\bibnamefont {Hone}}, \ and\ \bibinfo {author}
  {\bibfnamefont {C.~R.}\ \bibnamefont {Dean}},\ }\href {\doibase
  10.1126/science.aat6981} {\bibfield  {journal} {\bibinfo  {journal}
  {Science}\ }\textbf {\bibinfo {volume} {361}},\ \bibinfo {pages} {690}
  (\bibinfo {year} {2018})}\BibitemShut {NoStop}%
\bibitem [{\citenamefont {Mak}\ \emph {et~al.}(2010)\citenamefont {Mak},
  \citenamefont {Lee}, \citenamefont {Hone}, \citenamefont {Shan},\ and\
  \citenamefont {Heinz}}]{mak_atomically_2010}%
  \BibitemOpen
  \bibfield  {author} {\bibinfo {author} {\bibfnamefont {K.~F.}\ \bibnamefont
  {Mak}}, \bibinfo {author} {\bibfnamefont {C.}~\bibnamefont {Lee}}, \bibinfo
  {author} {\bibfnamefont {J.}~\bibnamefont {Hone}}, \bibinfo {author}
  {\bibfnamefont {J.}~\bibnamefont {Shan}}, \ and\ \bibinfo {author}
  {\bibfnamefont {T.~F.}\ \bibnamefont {Heinz}},\ }\href {\doibase
  10.1103/PhysRevLett.105.136805} {\bibfield  {journal} {\bibinfo  {journal}
  {Phys. Rev. Lett.}\ }\textbf {\bibinfo {volume} {105}},\ \bibinfo {pages}
  {136805} (\bibinfo {year} {2010})}\BibitemShut {NoStop}%
\bibitem [{\citenamefont {Splendiani}\ \emph {et~al.}(2010)\citenamefont
  {Splendiani}, \citenamefont {Sun}, \citenamefont {Zhang}, \citenamefont {Li},
  \citenamefont {Kim}, \citenamefont {Chim}, \citenamefont {Galli},\ and\
  \citenamefont {Wang}}]{splendiani_emerging_2010}%
  \BibitemOpen
  \bibfield  {author} {\bibinfo {author} {\bibfnamefont {A.}~\bibnamefont
  {Splendiani}}, \bibinfo {author} {\bibfnamefont {L.}~\bibnamefont {Sun}},
  \bibinfo {author} {\bibfnamefont {Y.}~\bibnamefont {Zhang}}, \bibinfo
  {author} {\bibfnamefont {T.}~\bibnamefont {Li}}, \bibinfo {author}
  {\bibfnamefont {J.}~\bibnamefont {Kim}}, \bibinfo {author} {\bibfnamefont
  {C.-Y.}\ \bibnamefont {Chim}}, \bibinfo {author} {\bibfnamefont
  {G.}~\bibnamefont {Galli}}, \ and\ \bibinfo {author} {\bibfnamefont
  {F.}~\bibnamefont {Wang}},\ }\href {\doibase 10.1021/nl903868w} {\bibfield
  {journal} {\bibinfo  {journal} {Nano Lett.}\ }\textbf {\bibinfo {volume}
  {10}},\ \bibinfo {pages} {1271} (\bibinfo {year} {2010})}\BibitemShut
  {NoStop}%
\bibitem [{\citenamefont {Koshino}(2015)}]{koshino_interlayer_2015}%
  \BibitemOpen
  \bibfield  {author} {\bibinfo {author} {\bibfnamefont {M.}~\bibnamefont
  {Koshino}},\ }\href {\doibase 10.1088/1367-2630/17/1/015014} {\bibfield
  {journal} {\bibinfo  {journal} {New J. Phys.}\ }\textbf {\bibinfo {volume}
  {17}},\ \bibinfo {pages} {015014} (\bibinfo {year} {2015})}\BibitemShut
  {NoStop}%
\bibitem [{\citenamefont {Bistritzer}\ and\ \citenamefont
  {MacDonald}(2010)}]{bistritzer_transport_2010}%
  \BibitemOpen
  \bibfield  {author} {\bibinfo {author} {\bibfnamefont {R.}~\bibnamefont
  {Bistritzer}}\ and\ \bibinfo {author} {\bibfnamefont {A.~H.}\ \bibnamefont
  {MacDonald}},\ }\href {\doibase 10.1103/PhysRevB.81.245412} {\bibfield
  {journal} {\bibinfo  {journal} {Phys. Rev. B}\ }\textbf {\bibinfo {volume}
  {81}},\ \bibinfo {pages} {245412} (\bibinfo {year} {2010})}\BibitemShut
  {NoStop}%
\bibitem [{\citenamefont {Slater}\ and\ \citenamefont
  {Koster}(1954)}]{slater_simplified_1954}%
  \BibitemOpen
  \bibfield  {author} {\bibinfo {author} {\bibfnamefont {J.~C.}\ \bibnamefont
  {Slater}}\ and\ \bibinfo {author} {\bibfnamefont {G.~F.}\ \bibnamefont
  {Koster}},\ }\href {\doibase 10.1103/PhysRev.94.1498} {\bibfield  {journal}
  {\bibinfo  {journal} {Phys. Rev.}\ }\textbf {\bibinfo {volume} {94}},\
  \bibinfo {pages} {1498} (\bibinfo {year} {1954})}\BibitemShut {NoStop}%
\bibitem [{\citenamefont {Fang}\ \emph {et~al.}(2015)\citenamefont {Fang},
  \citenamefont {Kuate~Defo}, \citenamefont {Shirodkar}, \citenamefont {Lieu},
  \citenamefont {Tritsaris},\ and\ \citenamefont {Kaxiras}}]{fang_ab_2015}%
  \BibitemOpen
  \bibfield  {author} {\bibinfo {author} {\bibfnamefont {S.}~\bibnamefont
  {Fang}}, \bibinfo {author} {\bibfnamefont {R.}~\bibnamefont {Kuate~Defo}},
  \bibinfo {author} {\bibfnamefont {S.~N.}\ \bibnamefont {Shirodkar}}, \bibinfo
  {author} {\bibfnamefont {S.}~\bibnamefont {Lieu}}, \bibinfo {author}
  {\bibfnamefont {G.~A.}\ \bibnamefont {Tritsaris}}, \ and\ \bibinfo {author}
  {\bibfnamefont {E.}~\bibnamefont {Kaxiras}},\ }\href {\doibase
  10.1103/PhysRevB.92.205108} {\bibfield  {journal} {\bibinfo  {journal} {Phys.
  Rev. B}\ }\textbf {\bibinfo {volume} {92}},\ \bibinfo {pages} {205108}
  (\bibinfo {year} {2015})}\BibitemShut {NoStop}%
\bibitem [{\citenamefont {Schrieffer}\ and\ \citenamefont
  {Wolff}(1966)}]{schrieffer_relation_1966}%
  \BibitemOpen
  \bibfield  {author} {\bibinfo {author} {\bibfnamefont {J.~R.}\ \bibnamefont
  {Schrieffer}}\ and\ \bibinfo {author} {\bibfnamefont {P.~A.}\ \bibnamefont
  {Wolff}},\ }\href {\doibase 10.1103/PhysRev.149.491} {\bibfield  {journal}
  {\bibinfo  {journal} {Phys. Rev.}\ }\textbf {\bibinfo {volume} {149}},\
  \bibinfo {pages} {491} (\bibinfo {year} {1966})}\BibitemShut {NoStop}%
\bibitem [{\citenamefont {Bravyi}\ \emph {et~al.}(2011)\citenamefont {Bravyi},
  \citenamefont {DiVincenzo},\ and\ \citenamefont
  {Loss}}]{bravyi_schriefferwolff_2011}%
  \BibitemOpen
  \bibfield  {author} {\bibinfo {author} {\bibfnamefont {S.}~\bibnamefont
  {Bravyi}}, \bibinfo {author} {\bibfnamefont {D.~P.}\ \bibnamefont
  {DiVincenzo}}, \ and\ \bibinfo {author} {\bibfnamefont {D.}~\bibnamefont
  {Loss}},\ }\href {\doibase 10.1016/j.aop.2011.06.004} {\bibfield  {journal}
  {\bibinfo  {journal} {Annals of Physics}\ }\textbf {\bibinfo {volume}
  {326}},\ \bibinfo {pages} {2793} (\bibinfo {year} {2011})}\BibitemShut
  {NoStop}%
\bibitem [{\citenamefont {Winkler}(2003)}]{winkler_spin--orbit_2003}%
  \BibitemOpen
  \bibfield  {author} {\bibinfo {author} {\bibfnamefont {R.}~\bibnamefont
  {Winkler}},\ }\href {\doibase 10.1007/b13586} {\emph {\bibinfo {title}
  {Spin--{Orbit} {Coupling} {Effects} in {Two}-{Dimensional} {Electron} and
  {Hole} {Systems}}}},\ \bibinfo {series} {Springer {Tracts} in {Modern}
  {Physics}}, Vol.\ \bibinfo {volume} {191}\ (\bibinfo  {publisher}
  {Springer},\ \bibinfo {address} {Berlin, Heidelberg},\ \bibinfo {year}
  {2003})\BibitemShut {NoStop}%
\bibitem [{\citenamefont {Hwang}\ \emph {et~al.}(2012)\citenamefont {Hwang},
  \citenamefont {Siegel}, \citenamefont {Mo}, \citenamefont {Regan},
  \citenamefont {Ismach}, \citenamefont {Zhang}, \citenamefont {Zettl},\ and\
  \citenamefont {Lanzara}}]{hwang_fermi_2012}%
  \BibitemOpen
  \bibfield  {author} {\bibinfo {author} {\bibfnamefont {C.}~\bibnamefont
  {Hwang}}, \bibinfo {author} {\bibfnamefont {D.~A.}\ \bibnamefont {Siegel}},
  \bibinfo {author} {\bibfnamefont {S.-K.}\ \bibnamefont {Mo}}, \bibinfo
  {author} {\bibfnamefont {W.}~\bibnamefont {Regan}}, \bibinfo {author}
  {\bibfnamefont {A.}~\bibnamefont {Ismach}}, \bibinfo {author} {\bibfnamefont
  {Y.}~\bibnamefont {Zhang}}, \bibinfo {author} {\bibfnamefont
  {A.}~\bibnamefont {Zettl}}, \ and\ \bibinfo {author} {\bibfnamefont
  {A.}~\bibnamefont {Lanzara}},\ }\href {\doibase 10.1038/srep00590} {\bibfield
   {journal} {\bibinfo  {journal} {Scientific Reports}\ }\textbf {\bibinfo
  {volume} {2}},\ \bibinfo {pages} {590} (\bibinfo {year} {2012})}\BibitemShut
  {NoStop}%
\bibitem [{\citenamefont {Min}\ \emph {et~al.}(2006)\citenamefont {Min},
  \citenamefont {Hill}, \citenamefont {Sinitsyn}, \citenamefont {Sahu},
  \citenamefont {Kleinman},\ and\ \citenamefont
  {MacDonald}}]{min_intrinsic_2006}%
  \BibitemOpen
  \bibfield  {author} {\bibinfo {author} {\bibfnamefont {H.}~\bibnamefont
  {Min}}, \bibinfo {author} {\bibfnamefont {J.~E.}\ \bibnamefont {Hill}},
  \bibinfo {author} {\bibfnamefont {N.~A.}\ \bibnamefont {Sinitsyn}}, \bibinfo
  {author} {\bibfnamefont {B.~R.}\ \bibnamefont {Sahu}}, \bibinfo {author}
  {\bibfnamefont {L.}~\bibnamefont {Kleinman}}, \ and\ \bibinfo {author}
  {\bibfnamefont {A.~H.}\ \bibnamefont {MacDonald}},\ }\href {\doibase
  10.1103/PhysRevB.74.165310} {\bibfield  {journal} {\bibinfo  {journal} {Phys.
  Rev. B}\ }\textbf {\bibinfo {volume} {74}},\ \bibinfo {pages} {165310}
  (\bibinfo {year} {2006})}\BibitemShut {NoStop}%
\bibitem [{\citenamefont {Korm{\'a}nyos}\ \emph {et~al.}(2014)\citenamefont
  {Korm{\'a}nyos}, \citenamefont {Z{\'o}lyomi}, \citenamefont {Drummond},\ and\
  \citenamefont {Burkard}}]{kormanyos_spin-orbit_2014}%
  \BibitemOpen
  \bibfield  {author} {\bibinfo {author} {\bibfnamefont {A.}~\bibnamefont
  {Korm{\'a}nyos}}, \bibinfo {author} {\bibfnamefont {V.}~\bibnamefont
  {Z{\'o}lyomi}}, \bibinfo {author} {\bibfnamefont {N.~D.}\ \bibnamefont
  {Drummond}}, \ and\ \bibinfo {author} {\bibfnamefont {G.}~\bibnamefont
  {Burkard}},\ }\href {\doibase 10.1103/PhysRevX.4.011034} {\bibfield
  {journal} {\bibinfo  {journal} {Phys. Rev. X}\ }\textbf {\bibinfo {volume}
  {4}},\ \bibinfo {pages} {011034} (\bibinfo {year} {2014})}\BibitemShut
  {NoStop}%
\bibitem [{\citenamefont {Ili{\'c}}\ \emph {et~al.}(2019)\citenamefont
  {Ili{\'c}}, \citenamefont {Meyer},\ and\ \citenamefont
  {Houzet}}]{ilic_weak_2019}%
  \BibitemOpen
  \bibfield  {author} {\bibinfo {author} {\bibfnamefont {S.}~\bibnamefont
  {Ili{\'c}}}, \bibinfo {author} {\bibfnamefont {J.~S.}\ \bibnamefont {Meyer}},
  \ and\ \bibinfo {author} {\bibfnamefont {M.}~\bibnamefont {Houzet}},\ }\href
  {\doibase 10.1103/PhysRevB.99.205407} {\bibfield  {journal} {\bibinfo
  {journal} {Phys. Rev. B}\ }\textbf {\bibinfo {volume} {99}},\ \bibinfo
  {pages} {205407} (\bibinfo {year} {2019})}\BibitemShut {NoStop}%
\bibitem [{\citenamefont {Colton}\ and\ \citenamefont
  {Kress}(1998)}]{colton_inverse_1998}%
  \BibitemOpen
  \bibfield  {author} {\bibinfo {author} {\bibfnamefont {D.}~\bibnamefont
  {Colton}}\ and\ \bibinfo {author} {\bibfnamefont {R.}~\bibnamefont {Kress}},\
  }\href {https://www.springer.com/us/book/9783662035375} {\emph {\bibinfo
  {title} {Inverse {Acoustic} and {Electromagnetic} {Scattering} {Theory}}}},\
  \bibinfo {edition} {2nd}\ ed.,\ Applied {Mathematical} {Sciences}\ (\bibinfo
  {publisher} {Springer-Verlag},\ \bibinfo {address} {Berlin Heidelberg},\
  \bibinfo {year} {1998})\BibitemShut {NoStop}%
\bibitem [{\citenamefont {Cuyt}\ \emph {et~al.}(2008)\citenamefont {Cuyt},
  \citenamefont {Petersen}, \citenamefont {Verdonk}, \citenamefont
  {Waadeland},\ and\ \citenamefont {Jones}}]{cuyt_handbook_2008}%
  \BibitemOpen
  \bibfield  {author} {\bibinfo {author} {\bibfnamefont {A.~A.~M.}\
  \bibnamefont {Cuyt}}, \bibinfo {author} {\bibfnamefont {V.}~\bibnamefont
  {Petersen}}, \bibinfo {author} {\bibfnamefont {B.}~\bibnamefont {Verdonk}},
  \bibinfo {author} {\bibfnamefont {H.}~\bibnamefont {Waadeland}}, \ and\
  \bibinfo {author} {\bibfnamefont {W.~B.}\ \bibnamefont {Jones}},\ }\href
  {https://www.springer.com/us/book/9781402069482} {\emph {\bibinfo {title}
  {Handbook of {Continued} {Fractions} for {Special} {Functions}}}}\ (\bibinfo
  {publisher} {Springer Netherlands},\ \bibinfo {year} {2008})\BibitemShut
  {NoStop}%
\bibitem [{\citenamefont {Dresselhaus}\ \emph {et~al.}(2010)\citenamefont
  {Dresselhaus}, \citenamefont {Dresselhaus},\ and\ \citenamefont
  {Jorio}}]{dresselhaus_group_2010}%
  \BibitemOpen
  \bibfield  {author} {\bibinfo {author} {\bibfnamefont {M.~S.}\ \bibnamefont
  {Dresselhaus}}, \bibinfo {author} {\bibfnamefont {G.}~\bibnamefont
  {Dresselhaus}}, \ and\ \bibinfo {author} {\bibfnamefont {A.}~\bibnamefont
  {Jorio}},\ }\href@noop {} {\emph {\bibinfo {title} {Group theory: application
  to the physics of condensed matter}}}\ (\bibinfo  {publisher}
  {Springer-Verlag},\ \bibinfo {address} {Berlin},\ \bibinfo {year} {2010})\
  \bibinfo {note} {oCLC: 692760083}\BibitemShut {NoStop}%
\bibitem [{\citenamefont {Oliphant}(2006)}]{oliphant_guide_2006}%
  \BibitemOpen
  \bibfield  {author} {\bibinfo {author} {\bibfnamefont {T.~E.}\ \bibnamefont
  {Oliphant}},\ }\href@noop {} {\emph {\bibinfo {title} {A guide to {NumPy}}}}\
  (\bibinfo  {publisher} {Trelgol Publishing},\ \bibinfo {address} {USA},\
  \bibinfo {year} {2006})\BibitemShut {NoStop}%
\end{thebibliography}%

\end{document}